\documentclass[journal=jctcce,manuscript=article]{achemso}

\usepackage[version=4]{mhchem}
\usepackage{bm,amssymb,mathtools,graphicx,xcolor}
\usepackage{booktabs,array,longtable,makecell,multirow}
\usepackage{siunitx,placeins,needspace,etoolbox}
\usepackage[normalem]{ulem}
\usepackage{hyperref}
\hypersetup{hidelinks}
\AtBeginDocument{\let\cite\cite}

\colorlet{reviewtext}{black}
\newif\ifshowreview
\showreviewfalse

\DeclareRobustCommand{\deleted}[1]{\ifshowreview{\sout{#1}}\else\ifhmode\unskip\fi\fi}
\newif\ifshowmanuscriptchanges
\showmanuscriptchangesfalse

\newcommand{\reviewmath}{\ifshowmanuscriptchanges\color{reviewtext}\fi}

\newcommand{\papertitle}{Learning to SHINE: Semiempirical Hamiltonians Informed by Machine-Learned Neighborhood Environments for Excited States}
\author{J. Diego Castro-Miyashiro}

\author{Lavanya Sankaranarayanan}

\author{Archit Gupta}
\author{Ismael A. Elayan}
\author{J. Terence Blaskovits}
\email{jtblaskovits@ualberta.ca}

\affiliation[UofA]
{Department of Chemistry, University of Alberta, Edmonton, Alberta, Canada, T6G 2G2}

\alsoaffiliation[Amii]
{Alberta Machine Intelligence Institute, Edmonton, Alberta, Canada, T5J 1S5}

\alsoaffiliation[CIFAR]
{Canada CIFAR AI Chair}

\title{\papertitle}
\SectionNumbersOn
\AtBeginDocument{
\newlabel{sec:si-readout-contractions}{{S1}{S5}{Readout Contractions}{section.1}{SI.pdf}}
\newlabel{sec:si-encoder-projections}{{S1.1}{S5}{Encoder Features and Learned Projections}{subsection.1.1}{SI.pdf}}
\newlabel{eq:source-19}{{S1}{S5}{Encoder Features and Learned Projections}{equation.1.1}{SI.pdf}}
\newlabel{fig:S1}{{S1}{S6}{Readout inputs}{figure.caption.4}{SI.pdf}}
\newlabel{sec:si-pair-inputs}{{S1.2}{S7}{Pair Inputs}{subsection.1.2}{SI.pdf}}
\newlabel{eq:source-21}{{S2}{S7}{Pair Inputs}{equation.1.2}{SI.pdf}}
\newlabel{tab:S1}{{S1}{S8}{Pair inputs for atoms i and j.\relax }{table.1}{SI.pdf}}
\newlabel{eq:source-22}{{S3}{S8}{Pair Inputs}{equation.1.3}{SI.pdf}}
\newlabel{eq:source-23}{{S4}{S8}{Pair Inputs}{equation.1.4}{SI.pdf}}
\newlabel{eq:source-24}{{S5}{S9}{Pair Inputs}{equation.1.5}{SI.pdf}}
\newlabel{eq:source-25}{{S6}{S9}{Pair Inputs}{equation.1.6}{SI.pdf}}
\newlabel{sec:si-site-inputs}{{S1.3}{S9}{Site Inputs}{subsection.1.3}{SI.pdf}}
\newlabel{eq:source-26}{{S7}{S9}{Site Inputs}{equation.1.7}{SI.pdf}}
\newlabel{tab:S2}{{S2}{S10}{Site inputs for atom i.\relax }{table.2}{SI.pdf}}
\newlabel{eq:source-27}{{S8}{S10}{Site Inputs}{equation.1.8}{SI.pdf}}
\newlabel{eq:source-28}{{S9}{S10}{Site Inputs}{equation.1.9}{SI.pdf}}
\newlabel{eq:source-29}{{S10}{S10}{Site Inputs}{equation.1.10}{SI.pdf}}
\newlabel{eq:source-30}{{S11}{S10}{Site Inputs}{equation.1.11}{SI.pdf}}
\newlabel{sec:si-kan-construction}{{S2}{S11}{Kolmogorov--Arnold Readout Construction}{section.2}{SI.pdf}}
\newlabel{sec:si-connection-functions}{{S2.1}{S11}{Learned Connection Functions}{subsection.2.1}{SI.pdf}}
\newlabel{eq:source-31}{{S12}{S11}{Learned Connection Functions}{equation.2.12}{SI.pdf}}
\newlabel{sec:si-screening-output}{{S2.2}{S11}{From Descriptors to a Screening Output}{subsection.2.2}{SI.pdf}}
\newlabel{eq:source-33}{{S13}{S11}{From Descriptors to a Screening Output}{equation.2.13}{SI.pdf}}
\newlabel{eq:source-34}{{S14}{S12}{From Descriptors to a Screening Output}{equation.2.14}{SI.pdf}}
\newlabel{fig:S2}{{S2}{S12}{KAN readout construction}{figure.caption.6}{SI.pdf}}
\newlabel{sec:si-spline-implementation}{{S2.3}{S13}{Spline Implementation and Initialization}{subsection.2.3}{SI.pdf}}
\newlabel{sec:si-intersite-rescaling}{{S2.4}{S13}{Rescaling and Screening of Intersite Coulomb Interactions}{subsection.2.4}{SI.pdf}}
\newlabel{eq:source-08}{{S15}{S13}{Rescaling and Screening of Intersite Coulomb Interactions}{equation.2.15}{SI.pdf}}
\newlabel{eq:source-09}{{S16}{S13}{Rescaling and Screening of Intersite Coulomb Interactions}{equation.2.16}{SI.pdf}}
\newlabel{sec:si-magnetic-moments}{{S3}{S14}{Magnetic Transition Moments About the Molecular Centre of Mass}{section.3}{SI.pdf}}
\newlabel{sec:si-magnetic-translation}{{S3.1}{S14}{Translation of Reference and PPP Moments}{subsection.3.1}{SI.pdf}}
\newlabel{eq:source-35}{{S17}{S14}{Translation of Reference and PPP Moments}{equation.3.17}{SI.pdf}}
\newlabel{sec:si-magnetic-resolution}{{S3.2}{S15}{Resolution of Magnetic Directions and Dissymmetry}{subsection.3.2}{SI.pdf}}
\newlabel{eq:source-39}{{S18}{S15}{Resolution of Magnetic Directions and Dissymmetry}{equation.3.18}{SI.pdf}}
\newlabel{fig:S3}{{S3}{S16}{Prescreening dissymmetry predictions}{figure.caption.7}{SI.pdf}}
\newlabel{sec:si-local-pi-score}{{S4}{S16}{\texorpdfstring {The S\textsubscript {1} Local-\ensuremath {\pi } Character Score}{The S1 Local-pi Character Score}}{section.4}{SI.pdf}}
\newlabel{eq:source-40}{{S19}{S17}{\texorpdfstring {The S\textsubscript {1} Local-\ensuremath {\pi } Character Score}{The S1 Local-pi Character Score}}{equation.4.19}{SI.pdf}}
\newlabel{eq:source-41}{{S20}{S17}{\texorpdfstring {The S\textsubscript {1} Local-\ensuremath {\pi } Character Score}{The S1 Local-pi Character Score}}{equation.4.20}{SI.pdf}}
\newlabel{eq:source-42}{{S21}{S18}{\texorpdfstring {The S\textsubscript {1} Local-\ensuremath {\pi } Character Score}{The S1 Local-pi Character Score}}{equation.4.21}{SI.pdf}}
\newlabel{tab:S3}{{S3}{S18}{The score of two FORMED molecules, configuration by configuration. Kept S\textsubscript {1} configurations in order of weight; c\textsuperscript {2} is the printed squared amplitude, the share is c\textsuperscript {2} over the kept total, and the contribution is c\textsuperscript {2}(f\textsubscript {occ} + f\textsubscript {virt})/2. The score is the sum of contributions divided by the sum of c\textsuperscript {2}. The kept configurations carry 92.4\% (KAVHUL) and 95.8\% (KEQHET) of the printed weight. The \ensuremath {\pi } fractions are rounded to three decimals, so recomputed contributions can differ in the last digit.\relax }{table.3}{SI.pdf}}
\newlabel{sec:si-structural-coverage}{{S5}{S19}{\texorpdfstring {Structural Coverage and Reference S\textsubscript {1} Local-\ensuremath {\pi } Character}{Structural Coverage and Reference S1 Local-pi Character}}{section.5}{SI.pdf}}
\newlabel{fig:S4}{{S4}{S19}{Structural coverage and excitation character}{figure.caption.8}{SI.pdf}}
\newlabel{sec:si-population-selection}{{S6}{S20}{Population Selection and Reference Criteria}{section.6}{SI.pdf}}
\newlabel{fig:S6}{{S5}{S20}{Population-selection counts}{figure.caption.9}{SI.pdf}}
\newlabel{fig:S7}{{S6}{S21}{Effects of the reference criteria}{figure.caption.10}{SI.pdf}}
\newlabel{sec:si-training-specification}{{S7}{S22}{Training and Checkpoint Selection}{section.7}{SI.pdf}}
\newlabel{sec:si-loss-normalization}{{S7.1}{S22}{Loss Normalization}{subsection.7.1}{SI.pdf}}
\newlabel{eq:source-43}{{S22}{S22}{Loss Normalization}{equation.7.22}{SI.pdf}}
\newlabel{sec:si-differentiation-optimization}{{S7.2}{S23}{Optimization}{subsection.7.2}{SI.pdf}}
\newlabel{tab:S5}{{S4}{S23}{\siadded {Shared optimization settings for all fourteen models.}\relax }{table.4}{SI.pdf}}
\newlabel{sec:si-validation-trajectories}{{S7.3}{S23}{Checkpoint Selection}{subsection.7.3}{SI.pdf}}
\newlabel{sec:si-failed-calculations}{{S8}{S24}{Failed Test Calculations and Comparison Sizes}{section.8}{SI.pdf}}
\newlabel{sec:si-comparison-sizes}{{S8.1}{S24}{Failures and Comparison Sizes}{subsection.8.1}{SI.pdf}}
\newlabel{tab:S7}{{S5}{S25}{Failures and Comparison Sizes}{table.5}{SI.pdf}}
\newlabel{tab:S7a}{{S5(a)}{S25}{Failures and Comparison Sizes}{tablepart.5.1}{SI.pdf}}
\newlabel{tab:S7b}{{S5(b)}{S26}{Failures and Comparison Sizes}{tablepart.5.2}{SI.pdf}}
\newlabel{sec:si-failed-case-structures}{{S8.2}{S26}{Molecular Structures of Failed Cases}{subsection.8.2}{SI.pdf}}
\newlabel{fig:S12}{{S7}{S27}{Structures of degenerate test cases}{figure.caption.11}{SI.pdf}}
\newlabel{sec:si-optical-only-training}{{S9}{S28}{Joint-Model Predictions and the Role of Gap Supervision}{section.9}{SI.pdf}}
\newlabel{fig:S13}{{S8}{S29}{Favorable joint optical–spin predictions}{figure.caption.12}{SI.pdf}}
\newlabel{fig:S14}{{S9}{S30}{Broad joint optical–spin predictions}{figure.caption.13}{SI.pdf}}
\newlabel{fig:S15}{{S10}{S31}{Favorable energy predictions}{figure.caption.14}{SI.pdf}}
\newlabel{fig:S16}{{S11}{S32}{Broad energy predictions}{figure.caption.15}{SI.pdf}}
\newlabel{sec:si-cross-property-comparisons}{{S10}{S33}{Full Cross-Property Comparisons}{section.10}{SI.pdf}}
\newlabel{tab:S8}{{S6}{S33}{Favorable all-model mean absolute errors. All models use the same 884 molecules, except native and logarithmic unsigned dissymmetry (694 molecules). Magnetic and derived quantities use the full-molecule centre of mass. Bold entries identify the lowest MAE in each column, using unrounded values.\relax }{table.6}{SI.pdf}}
\newlabel{tab:S9}{{S7}{S34}{Favorable all-model Pearson correlations. All models use the same 884 molecules, except native and logarithmic unsigned dissymmetry (694 molecules). Magnetic and derived quantities use the full-molecule centre of mass. Bold entries identify the highest correlation in each column, using unrounded values.\relax }{table.7}{SI.pdf}}
\newlabel{tab:S10}{{S8}{S34}{Broad all-model mean absolute errors. All models use the same 3,432 molecules, except native and logarithmic unsigned dissymmetry (2,991 molecules). Magnetic and derived quantities use the full-molecule centre of mass. Bold entries identify the lowest MAE in each column, using unrounded values.\relax }{table.8}{SI.pdf}}
\newlabel{tab:S11}{{S9}{S35}{Broad all-model Pearson correlations. All models use the same 3,432 molecules, except native and logarithmic unsigned dissymmetry (2,991 molecules). Magnetic and derived quantities use the full-molecule centre of mass. Bold entries identify the highest correlation in each column, using unrounded values.\relax }{table.9}{SI.pdf}}
\newlabel{sec:si-screening-errors}{{S11}{S35}{Prediction Errors Across Excitation Character}{section.11}{SI.pdf}}
\newlabel{fig:S17}{{S12}{S36}{Errors across excitation character}{figure.caption.16}{SI.pdf}}
\newlabel{sec:si-baseline-group-corrections}{{S12}{S37}{Learned Corrections by Baseline Group}{section.12}{SI.pdf}}
\newlabel{tab:S12}{{S10}{S37}{Single-site corrections by baseline site type. Changes in on-site repulsion and explicit site energy across 884 favorable molecules evaluated with the epoch-5 COM joint optical--spin model. Each site contributes one observation. Entries give the mean and 10th--90th percentiles. Site-type numerals denote \ensuremath {\pi }-electron contributions. The explicit site offset is bounded at \ensuremath {\pm }5 eV; the on-site fractional correction is bounded at \ensuremath {\pm }100\%.\relax }{table.10}{SI.pdf}}
\newlabel{fig:S22}{{S13}{S38}{Pair corrections by endpoint group}{figure.caption.17}{SI.pdf}}
}
\makeatletter
\newcommand{\mergedsubsectionanchor}{\phantomsection\protected@edef\@currentlabel{\thesubsection}}
\makeatother

\begin{document}
\maketitle
\begin{abstract}
\setstretch{1.0}
Despite advances in machine learning for molecular spectroscopy, predicting excitation energies and transition moments across unseen molecular scaffolds remains challenging. We introduce SHINE (\textbf{S}emiempirical \textbf{H}amiltonians \textbf{I}nformed by Machine-Learned \textbf{N}eighborhood Environments for \textbf{E}xcited States), a machine learning model which uses frozen MACE representations and rotation-invariant descriptors of local chemical environments to learn bounded corrections to semiempirical Hamiltonians. Using the simplified Pariser--Parr--Pople (PPP) \ensuremath{\pi}-electron Hamiltonian, we train corrections through restricted Hartree--Fock and configuration interaction singles calculations and evaluate predictions against Tamm--Dancoff time dependent density functional theory references on molecular scaffolds excluded from training. Joint training on optical and spin targets improves excitation energies, the singlet--triplet gap, oscillator strengths, and transition moment magnitudes relative to fixed PPP. {These gains hold in both the favorable population, drawn from the top 15\% by reference S\textsubscript{1} local \ensuremath{\pi} character, and the broader population without that cutoff. In the favorable population, the joint model reaches S\textsubscript{1}, T\textsubscript{1} and signed gap mean absolute errors of 0.127, 0.118 and 0.105 eV, respectively. Within the same Hamiltonian, electric and magnetic transition moment magnitude errors are 0.270 e~a\textsubscript{0} and 0.231 \(\mu_{\mathrm B}\). Single-target supervision more than halves the joint model's S\textsubscript{1} and T\textsubscript{1} mean absolute errors, bringing them to 0.060 and 0.041 eV, respectively. The electric and magnetic specialists also lower their respective transition moment magnitude errors to 0.256 e~a\textsubscript{0} and 0.224 \(\mu_{\mathrm B}\).} Together, these results show that learned corrections improve predictions within a reduced electronic description and motivate extending SHINE to richer semiempirical Hamiltonians.
\end{abstract}

\section{Introduction}\label{sec:introduction}

The lowest singlet excited state, S\textsubscript{1}, plays a central role in molecular photophysics. In many chromophores, excitation into a higher singlet state is followed by internal conversion down to S\textsubscript{1} before emission.\cite{ref01} The properties of S\textsubscript{1} help determine whether absorbed energy is released as light, transferred to another molecule or dissipated through nonradiative pathways. Therefore, to understand these processes and design organic optical materials, it is crucial to know the energy of S\textsubscript{1} and how strongly it couples to light. In particular, the excitation energy \(\Delta E\) sets the position of the vertical transition from the ground state, S\textsubscript{0}, to S\textsubscript{1}. The electric transition dipole moment \(\boldsymbol{\mu}\) governs the electric-dipole absorption strength, while its coupling with the magnetic transition dipole moment \(\mathbf{m}\) gives rise to differential absorption of left- and right-circularly polarized light. {Rotatory strength characterizes the signed differential absorption, and the absorption dissymmetry \(g_{\mathrm{abs}}\) expresses that difference relative to the mean absorption.\cite{caricato2024ecd} Together, these quantities describe the strength and relative selectivity of a molecule's interaction with circularly polarized light, making them relevant to the design of chiroptical materials.} Across a range of wavelengths, these differences in absorption give rise to the electronic circular dichroism (ECD) spectrum.

Recent machine-learning approaches,\cite{ref03,ref02} including ECDFormer and PhysECD, have concentrated on reconstructing electronic circular dichroism spectra. {However,~agreement between broadened spectra can coexist with errors in individual transition~properties. \cite{elayan2026tmc}} Therefore, assessing the underlying transition properties complements spectrum-level evaluation, although predicting them remains challenging. PhysECD reports that its per-state transition dipoles fail to generalize even under direct supervision,\cite{ref02} while DetaNet reports difficulty predicting oscillator strengths \(f\) from ground-state structures.\cite{ref04} Uceda et al.~predicted electric and magnetic transition dipole magnitudes and absorption dissymmetry from close structural neighbours within the [6]helicene family, leaving transfer across distinct scaffolds untested.\cite{uceda_helicene_2026}

{Beyond absorption, the pathways available after excitation depend on the relative energies of the excited states. In particular, the energy of the lowest triplet state, T\textsubscript{1}, relative to S\textsubscript{1} helps determine whether processes involving a change in spin are energetically accessible.\cite{ehrmaier2019heptazine}} The excitation energies of these states are denoted by \(\Delta E(\mathrm{S}_1)\) and \(\Delta E(\mathrm{T}_1)\). The signed singlet--triplet gap is \(\Delta E_{\mathrm{ST}}=\Delta E(\mathrm{S}_1)-\Delta E(\mathrm{T}_1)\), with both terms measured from the same ground state. For example, in organic light-emitting diodes (OLEDs), a small positive \(\Delta E_{\mathrm{ST}}\) can facilitate thermally activated reverse intersystem crossing from T\textsubscript{1} to S\textsubscript{1}, which enables delayed fluorescence.\cite{ref05} These two energies also help assess the energetic feasibility of singlet fission, which is a process relevant to solar-energy conversion. In this process, one singlet excitation generates two triplet excitations, so its energetic feasibility depends on how \(\Delta E(\mathrm{S}_1)\) compares with \(2\Delta E(\mathrm{T}_1)\).\cite{ref06} Therefore, identifying suitable chromophores requires reliable prediction of both levels.\cite{ref06}

Conjugated organic frameworks are  well suited to studying these properties because electron delocalization across the \ensuremath{\pi} network can support low-energy transitions with appreciable oscillator strengths.\cite{kuhn1949,garciaalvarez2024} The spatial arrangement of that network provides an additional means of controlling the optical response,  including when the conjugated structure adopts a chiral geometry. In helical conjugated families, for example, changes in molecular architecture substantially alter \(\mathbf{m}\).\cite{ref07} Furthermore, the extent and connectivity of the framework  influence \(\Delta E_{\mathrm{ST}}\), through changes in excitation localization and in the interactions between participating orbitals.\cite{ref06}{\cite{blaskovits2023symmetry,garner2023delocalization,blaskovits2024throughbond}} These links between conjugation, optical response and spin energetics motivate a shared \ensuremath{\pi}-electron model.

The Pariser--Parr--Pople (PPP) Hamiltonian\cite{pariser1953,pariser1953part2,pople1953} {provides a compact description of the excited states and optical absorption of conjugated molecules.\cite{ref08,fabian2026ppp} It treats the \(\pi\)-electron network explicitly, including hopping between sites and on-site and intersite electron repulsion, while representing the remaining electronic structure through effective parameters.\cite{pariser1953,pariser1953part2,pople1953}} When combined with configuration interaction (CI), the PPP model enables the calculation of singlet and triplet excitation energies and electric transition dipole moments within a reduced \(\pi\)-electron space.\cite{ref08,jorner2024} The Beveridge--Hinze parameterization serves as the baseline, hereafter referred to as fixed PPP.\cite{ref09} This compact \(\pi\)-electron representation requires its effective parameters both to describe the retained interactions and to compensate for some consequences of the model approximations. The extent of this compensation depends on the local atomic environment, which may limit the accuracy of assigning fixed parameters by atom or interaction type. For example, two carbon sites assigned the same baseline type may differ in their substituents. Likewise, two similarly assigned bonds may connect differently arranged \(\pi\) frameworks. Capturing these distinctions through manually defined parameter types becomes impractical as chemical and geometric possibilities multiply.

{Differentiable electronic-structure models provide a practical route to fitting such effective parameters. Vargas-Hernández et al.~used automatic differentiation to optimize Hückel parameters against excitation energies and molecular polarizabilities, demonstrating how property data can refine a compact electronic model.\cite{vargas2023huckel} Allowing the parameters to vary with the molecular environment extends this strategy beyond a shared parameter fit. Such environment-dependent Hamiltonians have improved ground-state predictions of semiempirical methods.\cite{ref19} For example,} NN-xTB\cite{ref10} learns bounded corrections to GFN2-xTB\cite{bannwarth2019gfn2xtb} parameters using energies, forces, and vibrational frequencies, while Suman et al.\cite{ref11} fitted effective single-particle Hamiltonians to orbital energies, dipole moments, and polarizabilities. Moving beyond ground-state properties, Cignoni et al.~learned an effective Hamiltonian by fitting orbital energies and atomic charges from seven hydrocarbon species, then calculated singlet excitations through sTDA and demonstrated transfer to selected larger hydrocarbons.\cite{ref12} Juergens et al.~learned electronic-state Hamiltonians and transition operators using nine QeMFi species and, separately, azobenzene, demonstrating excited-state surfaces and conical intersections.\cite{ref13}

A separate line of work has used \(\Delta\)-learning to improve ZINDO predictions of S\textsubscript{1} energies across large molecular sets.\cite{ref14} Using the same strategy with PPP+CIS, Nikhitha and Mondal\cite{nikhitha2025} learned additive corrections to S\textsubscript{1} and T\textsubscript{1} energies, the \(\Delta E_{\mathrm{ST}}\), and \(f\) using ADC(2) reference values. {These approaches improve the predicted properties without updating the underlying electronic states. Learning corrections to the Hamiltonian instead allows both excitation energies and transition moments to respond through the resulting wavefunctions. This shared electronic description also provides a route for supervision of one property to affect others, although improvement in unsupervised properties is not guaranteed.}

In this work, excited-state properties are used as training targets to learn bounded corrections to the parameters of the PPP Hamiltonian themselves, rather than the target observable. The~corrections~depend on the local molecular environment, which we describe using latent embeddings from MACE-OFF23.\cite{ref17} This representation allows atoms outside the PPP \(\pi\)-electron space to inform the interactions within it. The adapted Hamiltonian is used in restricted Hartree--Fock (RHF) calculations followed by configuration-interaction singles (CIS) calculations to obtain singlet and triplet excitation energies and the electric and magnetic transition dipole moments for the \(\mathrm{S}_0 \to \mathrm{S}_1\) transition. We call this framework SHINE (Semiempirical Hamiltonians Informed by Machine-Learned Neighborhood Environments for Excited States) and refer to the PPP model developed in this work as SHINE-PPP. Learning the Hamiltonian parameters makes the adaptation accessible for interpretation in terms of changes to site energies, hopping, and electron--electron interactions. Following established hybrid differentiation approaches \cite{zhang2022pyscfad,stohr2026differentiable}, training combines implicit differentiation \cite{zhang2022pyscfad,stohr2026differentiable,ref20}, first-order eigenpair perturbation theory \cite{ref21} and automatic differentiation in PyTorch \cite{paszke2019pytorch} to propagate excited-state property errors through RHF and CIS and update the learned corrections. Because the same corrected interactions generate \(\Delta E\) and transition moments, we can examine how supervision of one property affects the others.

{To test whether environment-dependent corrections improve PPP predictions for molecules with scaffolds excluded from training, we compare fixed and adapted PPP with reference calculations from the FORMED database,\cite{ref16} which provides calculated vertical \(\Delta E\) and transition properties for molecular geometries derived from the Cambridge Structural Database (CSD).~Our study concerns closed-shell molecules containing conjugated frameworks and exhibiting optically accessible S\textsubscript{1} transitions. The involvement of p orbitals perpendicular to the local molecular framework in each reference S\textsubscript{1} excitation is estimated from orbital coefficients and excitation configurations. This measure defines a \textbf{favorable population} drawn from the top 15\% of calculation-eligible molecules ranked by S\textsubscript{1} local-\(\pi\) character. To assess performance across a wider range of excitation character, we define a second cohort without this S\textsubscript{1}-character cutoff, hereafter referred to as the \textbf{broad population}. Separate models are trained and evaluated for each population using scaffold-group splits, so that each model is tested on scaffolds excluded from its training set.} 

{Across both populations, SHINE-PPP substantially reduces errors in excitation energies and the singlet--triplet gap relative to fixed PPP on molecular scaffolds excluded from training, and joint optical--spin supervision also improves electric and magnetic transition moment magnitudes within the same adapted Hamiltonian. Dedicated specialists generally lower errors further on their own targets, although the joint model is slightly more accurate for magnetic transition moment magnitude in the broad population.}

\Needspace{7\baselineskip}
\section{Methods}\label{sec:methods}

\begin{figure}[htbp]
\centering
\includegraphics[width=\linewidth,height=.76\textheight,keepaspectratio]{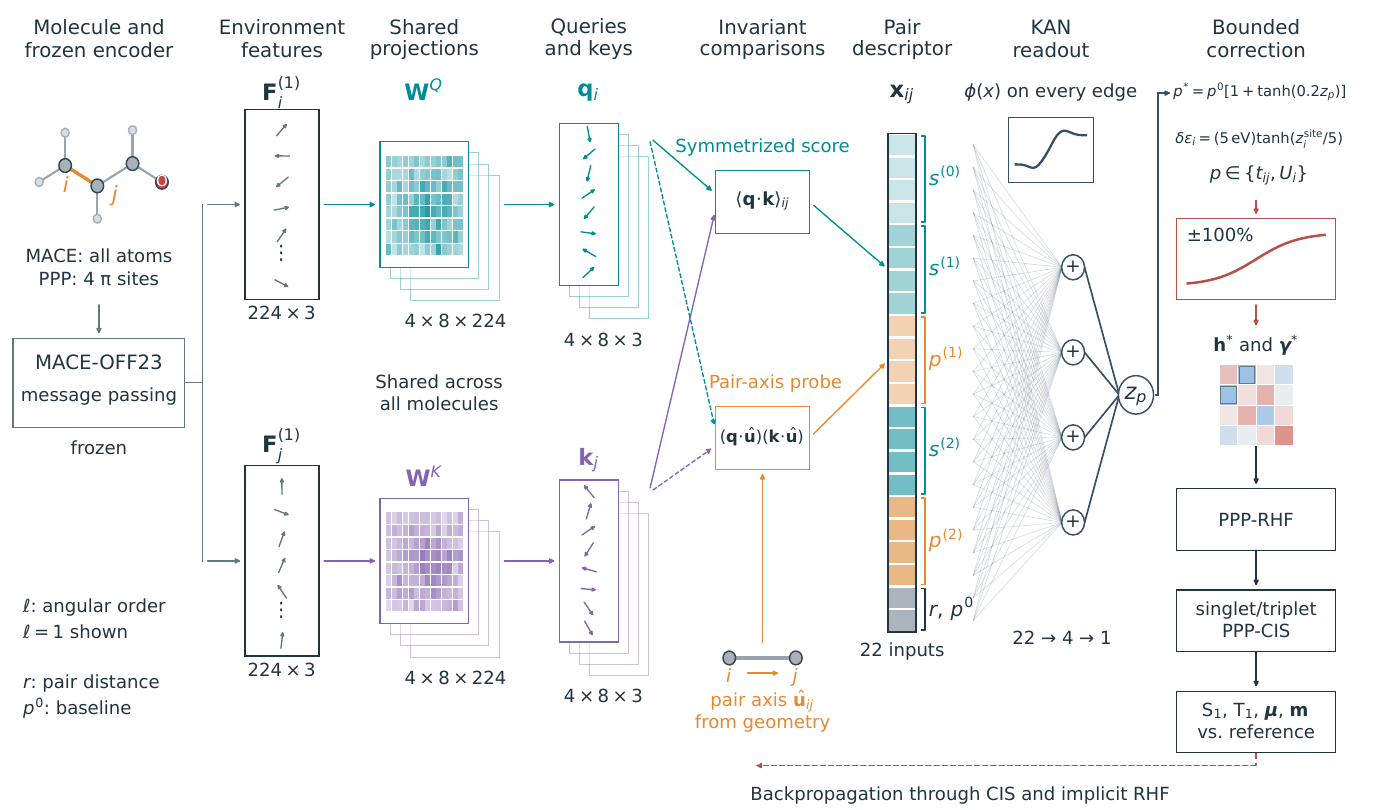}
\caption{Environmental features become a bounded Hamiltonian correction. The diagram follows one interacting pair along the \(\ell\) = 1 branch. Stacked cards represent four independent projection heads, each combining 224 encoder channels into eight query and eight key channels per angular order. Query--key~symmetric~contractions compare the environments, while pair-axis probes relate them to the connecting direction.~Across all the channels twelve environment-comparison scores, eight pair-axis probes, interatomic distance and the baseline interaction form the 22 inputs to the pair KAN.~Its scalar output passes through the bounded map to update the Hamiltonian parameter. The resulting core and Coulomb matrices enter PPP-RHF and singlet/triplet PPP-CIS to calculate the supervised properties. Gradients through these calculations train the projections and readouts while MACE remains frozen. Site readouts use 25 inputs, including four additional scalar features describing the individual environment and omitting the pair distance.~}
\label{fig:1}
\end{figure}

In SHINE-PPP, the molecular geometry determines how the fixed PPP parameters are corrected. MACE extracts features that describe the local atomic environments, and descriptors built from these features are fed into four Kolmogorov--Arnold networks (KANs).\cite{ref18} The network outputs are mapped to bounded corrections, which are applied to the fixed PPP parameters before calculating \(\Delta E\) and transition properties (Fig.~\ref{fig:1}).

The KANs are trained to minimize the deviations of these predictions from the corresponding FORMED reference values, which were calculated using the Tamm--Dancoff approximation (TDA) at the \ensuremath{\omega}B97X-D/6-31G(d) level of theory.\cite{ref16}

\subsection{Environment-Dependent PPP Hamiltonian}\label{sec:ppp-hamiltonian}

\subsubsection{Molecular Environment Features}\label{sec:molecular-environment}
A baseline PPP assignment fixes each parameter by atom type and geometry, so it cannot distinguish two sites of the same type whose substituents or neighboring atoms differ. We supply this environmental information using the frozen MACE-OFF23 large model,\cite{ref17} based on the higher-order equivariant message-passing architecture developed by Batatia et al.\cite{ref15} We retain its first interaction-layer features, which encode local atomic environments within a 5 Å cutoff radius. The readouts use these features at the retained \(\pi\) sites to generate bounded parameter corrections, allowing nearby atoms outside the PPP orbital space to influence the adapted interactions.

The MACE-OFF23 large checkpoint used here provides 224 channels for each angular order, \ensuremath{\ell} = 0, 1 and 2, in its first feature block. Each channel contains a scalar in \ensuremath{\mathbb{R}} for \ensuremath{\ell} = 0, a vector in \ensuremath{\mathbb{R}}\textsuperscript{3} for \ensuremath{\ell} = 1, or a five-component feature in \ensuremath{\mathbb{R}}\textsuperscript{5} for \ensuremath{\ell} = 2. The angular order specifies how these components transform when the molecule rotates. Scalars remain unchanged, the \ensuremath{\ell} = 1 features rotate as 3D vectors, and the \ensuremath{\ell} = 2 features transform as second-rank tensors. Retaining \ensuremath{\ell} = 2 allows the readouts to compare angular shape and relative alignment that vector summaries can lose. For example, an environment can have a pronounced axis or plane even when opposing directional contributions cancel in its vector summary. The Hamiltonian parameters must remain unchanged under molecular rotation, so the readouts use invariant scalar comparisons constructed from these directional features.

\subsubsection{Rotation-Invariant Descriptors}\label{sec:rotation-invariant-descriptors}

To compare local environments, we follow the SE(3)-Transformer principle of constructing rotation-invariant scores from inner products of equivariant queries and keys.\cite{fuchs2020se3} Within each angular order, learned projections produce queries and keys. For each pair, we average the scores obtained by assigning the query to either atom, giving a descriptor that is unchanged when the atoms are exchanged.

The query--key scores compare the two environments. To add their orientation relative to the direction joining the atoms, we project the vector features onto the pair axis and the tensor features onto its second-order spherical harmonics. We combine the projections from both atoms to preserve exchange symmetry.

Four heads provide twelve environment-comparison scores and eight pair-axis probes; the scaled interatomic distance and baseline interaction complete the 22-component pair descriptor supplied to the KAN readouts (Fig.~\ref{fig:1}). Site descriptors instead compare an atom's queries with its own keys and average directional probes over its bonds to other \ensuremath{\pi} sites. Together with scalar features and the baseline parameter, these give 25 inputs per site. Section~\ref{sec:si-readout-contractions} of the Supporting Information (SI) defines the complete descriptors and follows their construction in detail.

\subsubsection{Kolmogorov--Arnold Network (KAN) Readouts}\label{sec:kan-readouts}

The descriptors encode comparisons between local atomic environments and serve as inputs to the readouts (Fig.~\ref{fig:1}). The readouts comprise four separate Kolmogorov--Arnold networks (KANs), each with its own learned feature projections.\cite{ref18} The hopping and intersite screening KANs take atom-pair descriptors as inputs, while the site energy and on-site repulsion KANs take single-site descriptors.

In a conventional multilayer perceptron, each connection multiplies its input by a learned weight, and each neuron applies a fixed activation function to the weighted sum. A KAN instead learns a one-dimensional function (\ensuremath{\phi}) on each connection and sums the resulting contributions at each node.  In this implementation, each connection function combines cubic B-splines with a weighted SiLU term. Both the spline coefficients and the SiLU weight are learned.

Each readout applies these connection functions in two layers. The first layer computes each hidden value \(u_a^{(b)}\) by summing the transformed descriptor components and adding a learned bias, giving
\begin{equation}
u_a^{(b)}=\sum_d\phi_{ad}^{(1,b)}\!\left(x_d^{(b)}\right)+\beta_a^{(1,b)},
\qquad a=1,\ldots,4.
\label{eq:source-03}
\end{equation}
The second layer then transforms the four hidden values in Eq.~\eqref{eq:source-03}, sums their contributions, and adds an output bias to obtain
\begin{equation}
z^{(b)}=\sum_{a=1}^{4}\phi_a^{(2,b)}\!\left(u_a^{(b)}\right)+\beta^{(2,b)}.
\label{eq:source-04}
\end{equation}
Here, \(b\in\{\mathrm{hop},\mathrm{screen},\mathrm{site},U\}\) identifies the parameter channel, \(d\) indexes its descriptor components, and \(a\) indexes the four hidden units. Each channel has its own learned connection functions and biases. Each readout applies the same learned functions across molecules, so its scalar output \(z^{(b)}\) varies with the descriptor of the site or atom pair. This output is then mapped to a bounded correction to the corresponding PPP parameter (Fig.~\ref{fig:1}). 

{A chemist may recognize the sum over learned functions in Eq.~\eqref{eq:source-04} as reminiscent of a linear combination of atomic orbitals (LCAO), in which a molecular orbital is written as a weighted sum of atomic basis functions. The resemblance lies in the use of a sum, since the KAN functions act on scalar inputs and have no direct orbital interpretation. A further motivation is the potential for interpretability discussed by Liu et al.\cite{liu2025kanscience} Each learned function in the first readout layer acts on a single descriptor component, making it possible to inspect how that component is transformed before its contribution is combined with the others.}

\subsubsection{Bounded Hamiltonian Corrections}\label{sec:bounded-corrections}

{We initialize the Beveridge--Hinze PPP Hamiltonian\cite{ref09} using the Coulson software library.\cite{jorner2024}} Following the bounded parameter adaptation strategy of NN-xTB,\cite{ref10} we constrain the magnitude of the learned corrections to the PPP baseline. We impose these bounds through smooth \(\tanh\) maps. Hopping \(t_{ij}\) and on-site repulsion \(U_i\) share a multiplicative update, while the site energy receives an additive offset.

\begingroup\reviewmath
\begin{equation}
p^*=p^0\bigl[1+\tanh(0.2\,z_p)\bigr],
\qquad
\delta\varepsilon_i=(5\,\mathrm{eV})\tanh\!\left(\frac{z_i^{\mathrm{site}}}{5}\right).
\label{eq:source-05}
\end{equation}
\endgroup
Here \(p\) denotes either \(t_{ij}\) or \(U_i\), and \(z_p\) is its corresponding readout output. For intersite Coulomb interactions, we rescale the baseline to account for the updated on-site repulsions, then apply distance-dependent screening. These steps give the provisional interaction \(\widetilde{\gamma}_{ij}\), whose combined deviation from the baseline we bound using,

\begingroup
\begin{equation}
\gamma^*_{ij}=\gamma^0_{ij}\left[1+\tanh\left(\frac{\widetilde{\gamma}_{ij}-\gamma^0_{ij}}{\gamma^0_{ij}}\right)\right].
\label{eq:source-10}
\end{equation}
\endgroup

\begin{figure}[htbp]
\centering
\includegraphics[width=\linewidth,height=.76\textheight,keepaspectratio]{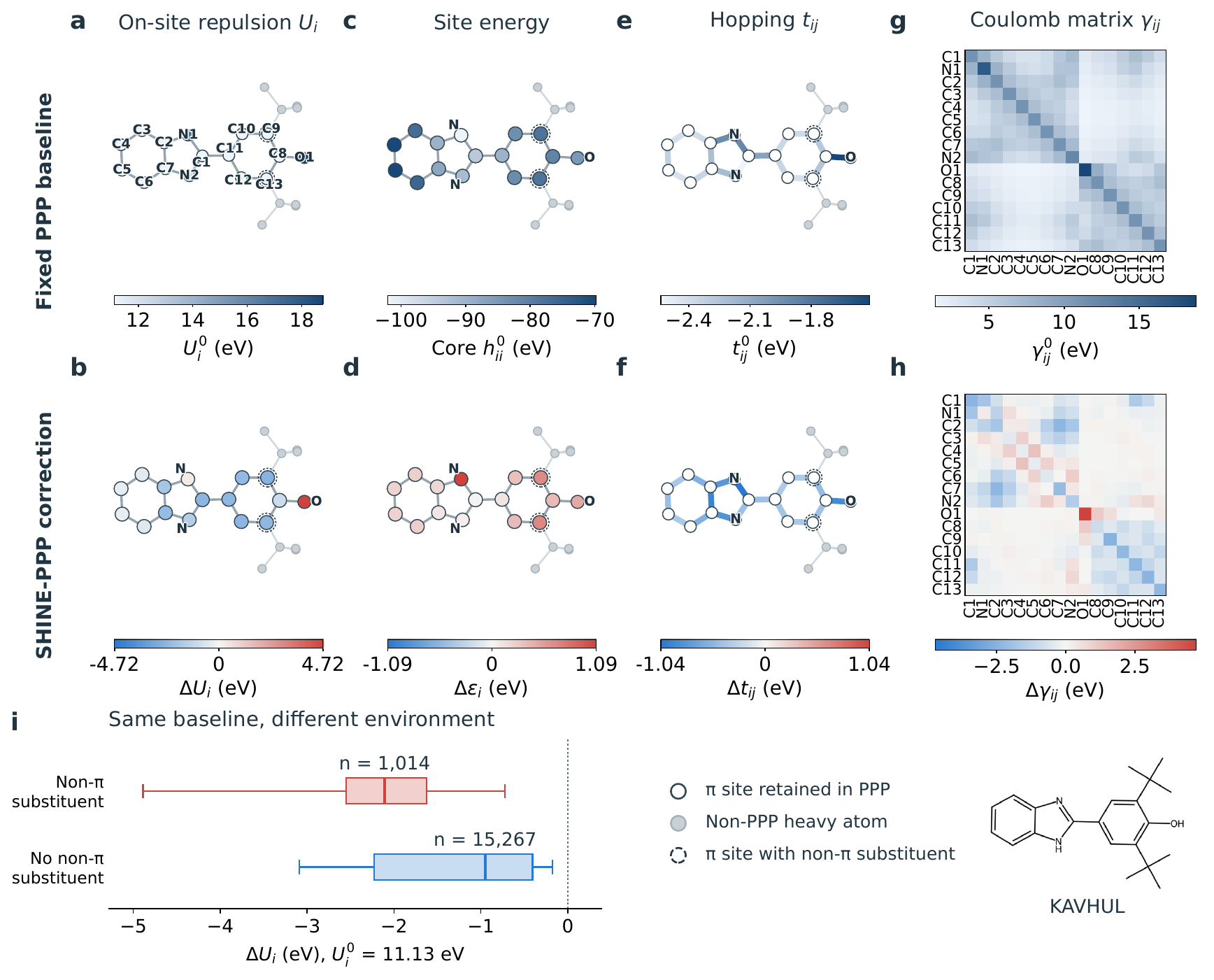}
\caption{{Baseline PPP values and local corrections. For the held-out molecule KAVHUL, the top row (a, c, e, g) shows fixed PPP values and the bottom row (b, d, f, h) shows learned changes from the favorable joint optical--spin model. From left to right, the columns show on-site repulsion, the core diagonal, hopping, and intersite Coulomb interactions. Panel d isolates the explicit site-energy offset \(\Delta\varepsilon_i\). Panel i pools carbon sites with \(U_i^0=11.13\,\mathrm{eV}\) across the favorable test set. Sites bearing a non-\(\pi\) heavy-atom substituent have a more negative median \(\Delta U_i\). Boxes show the interquartile range and median, with whiskers at the 5th and 95th percentiles.}}
\label{fig:2}
\end{figure}

 The maps recover the baseline exactly when all readout outputs are zero. The fractional changes in hopping, on-site repulsion and intersite Coulomb interactions are bounded by \(\pm 100\%\), and the explicit site-energy offset by \(\pm 5\) eV. Figure~\ref{fig:2} illustrates these updates for the held-out FORMED molecule KAVHUL, showing the baseline parameters alongside the corrections produced by~SHINE-PPP trained in the favorable regime for joint optical and singlet-triplet gap targets~. Readout details and the rescaling and screening formulas are provided in SI Section~\ref{sec:si-kan-construction}.  

up 

For each geometry, the readouts produce one set of environment-dependent PPP parameters, which remains fixed throughout the RHF iterations. We solve the resulting Hamiltonian with PPP-RHF and use the converged orbitals in singlet and triplet PPP-CIS. The lowest singlet and triplet CIS roots give \(\Delta E(\mathrm{S}_1)\) and \(\Delta E(\mathrm{T}_1)\) relative to the RHF ground state, from which \(\Delta E_{\mathrm{ST}}\) is obtained, while the S\textsubscript{1} transition density yields \(\boldsymbol{\mu}\) and \(\mathbf{m}\).

\subsection{Optical Observables and Spin-Energy Conventions}\label{sec:optical-spin-conventions}

For each S\textsubscript{1} transition, the optical observables are \(\Delta E(\mathrm{S}_1)\), \(f\), \(\boldsymbol{\mu}\), \(\mathbf{m}\), and \(\lvert g_{\mathrm{abs}}\rvert\), where \(f\) depends on \(\Delta E(\mathrm{S}_1)\) and \(\lVert\boldsymbol{\mu}\rVert^2\). The magnitudes \(\lVert\boldsymbol{\mu}\rVert\) and \(\lVert\mathbf{m}\rVert\) are reported in \(ea_0\) and \(\mu_{\mathrm B}\), respectively, where \(e\) is the elementary charge, \(a_0\) is the Bohr radius, and \(\mu_{\mathrm B}\) is the Bohr magneton.

The electric-equivalent rotatory strength \(R_{\mathrm{eq}}\) and the absorption dissymmetry \(g_{\mathrm{abs}}\) combine \(\boldsymbol{\mu}\) with the magnetic moment in electric-equivalent units, \(\widetilde{\mathbf{m}}\), where \(\alpha\) is the fine-structure constant:

\begin{equation}
\widetilde{\mathbf{m}}=\frac{\alpha}{2}\mathbf{m}
\label{eq:source-11}
\end{equation}

\begin{equation}
\left|R_{\mathrm{eq}}\right|=\left|\boldsymbol{\mu}\cdot\widetilde{\mathbf{m}}\right|
\label{eq:source-12}
\end{equation}

\begin{equation}
\left|g_{\mathrm{abs}}\right|=\frac{4\left|\boldsymbol{\mu}\cdot\widetilde{\mathbf{m}}\right|}{\left\|\boldsymbol{\mu}\right\|^2+\left\|\widetilde{\mathbf{m}}\right\|^2}=\frac{4\left|R_{\mathrm{eq}}\right|}{\left\|\boldsymbol{\mu}\right\|^2+\left\|\widetilde{\mathbf{m}}\right\|^2}.
\label{eq:source-13}
\end{equation}
 The joint objectives include \(\lvert R_{\mathrm{eq}}\rvert\) to constrain the electric--magnetic coupling in the dissymmetry numerator, where we report \(\lvert g_{\mathrm{abs}}\rvert\),  to quantify the absolute difference between absorption of left- and right-circularly polarized light relative to their mean absorption.

{Agreement with the reference direction is assessed separately for \(\boldsymbol{\mu}\) and \(\mathbf{m}\) using the absolute cosine of the angle between the predicted and reference vectors. For \(\boldsymbol{\mu}\), this gives}

\begin{equation}
{c_{\mu}=\frac{\left|\boldsymbol{\mu}^{\mathrm{pred}}\cdot\boldsymbol{\mu}^{\mathrm{ref}}\right|}{\left\lVert\boldsymbol{\mu}^{\mathrm{pred}}\right\rVert\,\left\lVert\boldsymbol{\mu}^{\mathrm{ref}}\right\rVert}.}
\label{eq:source-14}
\end{equation}

The same expression defines \(c_m\) for \(\mathbf{m}\). The arbitrary sign of the excited-state wavefunction can make the predicted and reference vectors point in opposite directions along the same axis. The absolute cosine gives a value of one in either case. Within a given calculation, changing the wavefunction sign multiplies both \(\boldsymbol{\mu}\) and \(\mathbf{m}\) by \(-1\), leaving the angle between them unchanged. The signed cosine below therefore describes their relative orientation independently of the wavefunction sign.

\begin{equation}
{\cos\theta_{\mu m}=\frac{\boldsymbol{\mu}\cdot\mathbf{m}}{\lVert\boldsymbol{\mu}\rVert\,\lVert\mathbf{m}\rVert}.}
\label{eq:source-15}
\end{equation}

Equation~\eqref{eq:source-13} shows that \(\lvert g_{\mathrm{abs}}\rvert\) depends on both dipole magnitudes and \(\lvert\cos\theta_{\mu m}\rvert\). The absolute value retains the magnitude of the chiroptical response while omitting its sign. When \(\boldsymbol{\mu}\) and \(\mathbf{m}\) are nearly perpendicular, even small angular errors can cause large relative errors in \(\lvert g_{\mathrm{abs}}\rvert\) despite close agreement with the reference axes. The calculated \(\mathbf{m}\) also depends on the coordinate origin. Therefore, the archived velocity transition vectors are used to translate the reference \(\mathbf{m}\) to the full-molecule center of mass, and the PPP \(\mathbf{m}\) is calculated about that same origin. Section~\ref{sec:si-magnetic-moments} provides the translation equations and reference-precision criteria. During training and validation, denominators in the direction and dissymmetry metrics are floored at \(\delta=10^{-10}\) in their respective units to stabilize the ratios near zero.

For each method, the lowest singlet and triplet roots are selected by energy at the supplied ground-state geometry and used to calculate \(\Delta E_{\mathrm{ST}}\). The comparison therefore follows energy ordering without requiring the states to have matching excitation character between methods. Nonpositive predictions of \(\Delta E(\mathrm{T}_1)\) are included in the error calculations and tallied separately.

\subsection{FORMED Curation and Study Populations}\label{sec:formed-curation}

To construct the study populations, we identify FORMED molecules that support the PPP calculation, then restrict the evaluation to bright, well-resolved S\textsubscript{1} transitions with usable triplet references. {These shared requirements define the broad population. The favorable population additionally requires} high reference S\textsubscript{1} local-\(\pi\) character.

\mergedsubsectionanchor\label{sec:calculation-eligibility}

We begin with~the full set of~raw~~FORMED metadata records.\cite{ref16} To construct the PPP Hamiltonian, we require a readable, single-fragment molecular graph without formal charges or radicals, containing only the supported elements H, C, N, O, F, S and Cl. Each molecule needs a conjugated or aromatic component of at least six atoms, supported types for its \(\pi\) sites, and a closed-shell \(\pi\)-electron count with both occupied and virtual orbitals. {We also require an available reference archive, defined local \(\pi\) axes, and a successfully constructed baseline Hamiltonian whose RHF calculation converges. The axis requirement is local, so helical and noncoplanar conjugated frameworks remain eligible. Among these calculation-eligible molecules, we retain reference S\textsubscript{1} transitions with \(f\geq 0.01\). This restricts the assessment to optically accessible transitions and avoids vanishingly small electric transition moments, supporting the direction and dissymmetry comparisons introduced in Section~\ref{sec:optical-spin-conventions}.}

Since FORMED prints each transition-moment component to four decimal places, we require each archived moment norm to exceed 3\(\sqrt{3}\) \(\times\) 5 \(\times\) 10\textsuperscript{-5} in its printed units, which helps define the study populations. We apply the same principle in the directional analysis, using the propagated rounding uncertainty to determine whether a magnetic direction is resolved (Section~\ref{sec:si-magnetic-resolution}). Where the direction is unresolved, eligible scalar values still enter the analysis. Furthermore, to distinguish close-lying singlet states, \emph{e.g.}, \(\mathrm{S}_1\) from \(\mathrm{S}_2\), we require a minimum energy separation of \(0.02\,\mathrm{eV}\), such that \(\Delta E(\mathrm{S}_2)-\Delta E(\mathrm{S}_1)\geq 0.02\,\mathrm{eV}\). We also require the reference \(\Delta E(\mathrm{T}_1)\) to be finite and positive. These requirements do not use S\textsubscript{1} local-\(\pi\) character or PPP prediction error, but they affect the chemistry represented. The median fraction of heavy atoms belonging to the largest conjugated network is 42\% across FORMED and 64\% in the broad population after calculation eligibility and the reference requirements are applied. The selection counts and resulting distribution shifts are documented in~detail in~SI Section~\ref{sec:si-population-selection}.

\subsubsection{\texorpdfstring{S\textsubscript{1} Local-\ensuremath{\pi} Character and the Favorable Population}{S1 Local-pi Character and the Favorable Population}}\label{sec:local-pi-character}

To assess how closely each reference S\textsubscript{1} excitation matches PPP's \ensuremath{\pi}-electron description, we calculate its  local-\ensuremath{\pi} character.  For each retained occupied-to-virtual excitation, we average the fractions of squared orbital coefficients associated with p functions perpendicular to the local plane. We then average over configurations, weighting each by its squared excitation amplitude. Section~\ref{sec:si-local-pi-score} gives the full construction and worked examples.

The selection retains the top 15\% of calculation-eligible molecules with the highest scores, corresponding to a character of at least 89.3\%. This defines a pool of 5,628 molecules before the shared reference criteria are applied. {Applying the shared reference requirements gives the favorable population of 4,608 molecules,} compared with 17,961 in the broad population, which has no character cutoff (Fig.~S5b; Table~1). Because the favorable selection concerns S\textsubscript{1}, it does not by itself establish that T\textsubscript{1} has the same orbital character.

\begingroup
\footnotesize
\setlength{\tabcolsep}{3pt}
\renewcommand{\arraystretch}{1.18}
\begin{longtable}{@{}>{\raggedright\arraybackslash}p{0.2170\linewidth}>{\raggedright\arraybackslash}p{0.2170\linewidth}>{\raggedright\arraybackslash}p{0.2170\linewidth}>{\raggedright\arraybackslash}p{0.2170\linewidth}@{}}
\caption{Training, validation and test memberships of the two populations.}\label{tab:1}\\
\toprule
Population & Training & Validation & Test \\
\midrule
\endfirsthead
\toprule
Population & Training & Validation & Test \\
\midrule
\endhead
\bottomrule
\endfoot
Favorable & 2,974 & 748 & 886 \\
Broad & 11,527 & 2,998 & 3,436 \\
\end{longtable}
\endgroup

\mergedsubsectionanchor\label{sec:structural-descriptors}

{Reference S\textsubscript{1} local-\(\pi\) character can help identify where the model performs well, but computing it already requires the excited-state calculation that a surrogate model is intended to avoid. Therefore, Section~\ref{sec:si-structural-coverage} in the SI offers a complementary analysis relating two structural measures of conjugated-network coverage to the S\textsubscript{1} local-\(\pi\) score. Overall, although there is substantial scatter among molecules of similar character, both rise with the score, particularly at high values.}

\mergedsubsectionanchor\label{sec:scaffold-splits}

To assess generalization to unseen molecular scaffolds, we group molecules by their RDKit Bemis--Murcko scaffold,\cite{bemis1996} serialized without stereochemistry. The scaffold consists of the ring systems and the linkers that connect them. Acyclic molecules are grouped by connectivity. Each group is assigned wholly to training, validation or test, preventing scaffold overlap between partitions within each population.

The broad and favorable splits were assigned independently over their calculation-eligible pools, before applying the reference requirements. Those requirements then removed molecules within each partition, giving the memberships in Table~\ref{tab:1}. Each model is trained, selected and evaluated only within its own population's split, and the two populations are reported separately.

\subsection{Learning Objective and Model Selection}\label{sec:learning-objective}

To test how the choice of supervision affects the adapted Hamiltonian, we train five specialists and two joint models in each population. {The specialists target \(\Delta E(\mathrm{S}_1)\), \(\Delta E(\mathrm{T}_1)\), \(f\), the electric transition-moment magnitude and direction, or the magnetic transition-moment magnitude and direction.} Their supervised properties establish target-specific accuracy, while their predictions on the remaining properties test how the learned corrections transfer, with PPP providing the common untrained baseline.

To combine targets with different units, we divide each residual by a scale derived from the training partition before squaring it,

\begin{equation}
e_k=\left(\frac{\widehat{y}_k-y_k}{s_k}\right)^2,
\label{eq:source-17}
\end{equation}
{where \(\widehat y_k\) and \(y_k\) are the predicted and reference values, and \(s_k\)} is fixed PPP's training-set root-mean-square (RMS) error for target k, subject to a small numerical floor. Each scalar specialist minimizes its target's normalized error. The electric and magnetic specialists weight magnitude and direction errors equally when both terms are available. The direction term is included only when the reference axis is resolved. Direction errors are normalized by fixed PPP's training-set RMS deviation from perfect axis agreement.

Joint supervision asks whether one adapted Hamiltonian can describe several properties together. The joint optical--spin objective averages six groups: \(\Delta E(\mathrm{S}_1)\), \(f\), \(\lVert\boldsymbol{\mu}\rVert\) and \(\lVert\mathbf{m}\rVert\), \(\lvert R_{\mathrm{eq}}\rvert\), \(\lvert g_{\mathrm{abs}}\rvert\) and \(\Delta E_{\mathrm{ST}}\):

\begin{equation}
\mathcal{L}_{\mathrm{joint}}=\frac{1}{6}\left[e_{\Delta E(\mathrm{S}_1)}+e_f+\frac{e_{\lVert\boldsymbol{\mu}\rVert}+e_{\lVert\mathbf{m}\rVert}}{2}+e_{|R_{\mathrm{eq}}|}+e_{|g_{\mathrm{abs}}|}+e_{\Delta E_{\mathrm{ST}}}\right],
\label{eq:source-18}
\end{equation}
\(\lVert\boldsymbol{\mu}\rVert\) and \(\lVert\mathbf{m}\rVert\) share one group's weight. {The joint optical model, referred to below as the \textbf{optical-only joint model},} omits the \(\Delta E_{\mathrm{ST}}\) group and averages the remaining five. Supervising \(\Delta E_{\mathrm{ST}}\) constrains \(\Delta E(\mathrm{T}_1)\) without introducing an independently weighted T\textsubscript{1} term; only the T\textsubscript{1} specialist supervises that energy directly.

Gradients pass through the electronic calculation to the learned Hamiltonian corrections. The converged RHF self-consistency condition is differentiated implicitly \cite{zhang2022pyscfad,stohr2026differentiable,ref20}. The CIS excitation energies and selected singlet eigenvector are differentiated using first-order perturbation theory \cite{ref21}. The singlet eigenpair derivatives enter PyTorch automatic differentiation \cite{paszke2019pytorch} through a custom backward operation \cite{xie2020eigensolver}. Including the eigenvector response allows transition-property errors, as well as errors in \(\Delta E\), to inform the Hamiltonian corrections. The MACE encoder remains frozen; training updates the feature projections and readouts. We use Adam at a constant learning rate of 10\textsuperscript{-3} for twenty epochs. An epoch is one pass through the training partition; a checkpoint stores the learned parameters at a given stage of training. All fourteen models are evaluated  with magnetic transition moments taken about the full-molecule center of mass. The magnetic specialist and both joint models use reference targets evaluated at that origin. The other specialist objectives are independent of this choice of origin.

We evaluate the full validation partition after every epoch, including initialization, and select the checkpoint with the lowest mean score. Specialists use their normalized validation objective. Joint models retain the training objective's property groups but replace native-unit error in \(\lvert g_{\mathrm{abs}}\rvert\) with a normalized squared logarithmic residual, flooring predicted and reference magnitudes at 10\textsuperscript{-5} before taking logarithms. This assesses proportional agreement across the wide range of dissymmetry magnitudes.

For each metric, we compare models on the same molecules and report MAE in native units unless a logarithmic scale is stated. We report Pearson \(r\) alongside MAE. Finite nonpositive T\textsubscript{1} predictions remain in the error statistics and are counted separately. Sections~\ref{sec:si-training-specification}--\ref{sec:si-failed-calculations} describe training, checkpoint selection and failure handling.

\section{Results}\label{sec:results}

\FloatBarrier
\subsection{\texorpdfstring{Favorable Regime: \(\pi\)-Like S\textsubscript{1} Transitions}{Favorable Regime: pi-Like S1 Transitions}}\label{sec:favorable-regime}

The favorable population tests the adapted Hamiltonian where the reference S\textsubscript{1} transition has high local-\(\pi\) character. To compare all models consistently, the comparison is restricted to molecules for which every model yields accepted predictions for all properties being compared. A calculation is rejected if the self-consistent-field (SCF) procedure fails to converge or if the two lowest excited states are too close in energy within either the singlet or the triplet sector. In the favorable population, all rejections are due to near-degeneracy. Thus, if a fixed PPP calculation is rejected, the molecule is excluded from the common set even if the adapted calculation succeeds. This retains 884 of the 886 favorable test molecules (99.8\%).

Among the scalar properties, only \(\lvert g_{\mathrm{abs}}\rvert\) is evaluated on a subset of the common test set. Near zero, rounding in the printed transition vectors can account for the entire discrepancy between the prediction and the reference. Therefore, only molecules whose reference \(\lvert g_{\mathrm{abs}}\rvert\) is at least three times the propagated rounding uncertainty are retained. This criterion retains 694 of the 884 molecules for dissymmetry comparisons. Comparisons of transition-moment directions apply the additional resolution criteria described in Section~2.3, and the corresponding molecule counts are reported separately. SI Table~\ref{tab:S7} gives a breakdown of calculation failures and the number of molecules included in each comparison.

\begin{figure}[!htbp]
\centering
\includegraphics[width=\linewidth,height=.68\textheight,keepaspectratio]{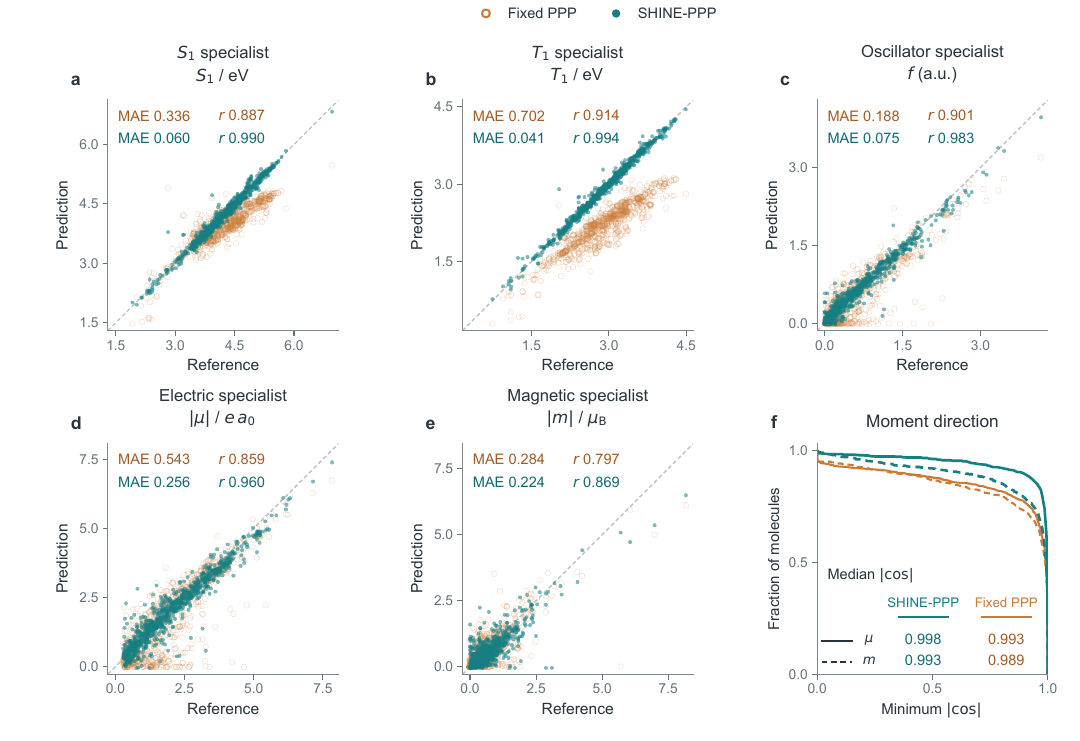}
\caption{Target-specific predictions of the five favorable population specialists. Panels show \(\Delta E(\mathrm{S}_1)\), \(\Delta E(\mathrm{T}_1)\), \(f\) and \(\lVert\boldsymbol{\mu}\rVert\) and \(\lVert\mathbf{m}\rVert\) against their references. Teal filled points show SHINE-PPP specialist predictions and orange open circles show fixed PPP on the same molecules. Matching scalar-panel annotations give each method's MAE and Pearson r; n is the evaluation count. {Panel f compares electric and magnetic direction agreement: teal curves show the specialists and orange curves show fixed PPP on the same molecules; solid lines indicate \(\boldsymbol{\mu}\) and dashed lines indicate \(\mathbf{m}\).} The horizontal axis is the minimum absolute direction cosine with the reference, and the vertical axis is the fraction of molecules reaching that value. The table gives specialist and fixed-PPP median absolute cosines. Scalar panels use the common 884-molecule evaluation set; direction curves retain 884 electric and 771 magnetic comparisons after the resolution masks.}
\label{fig:3}
\end{figure}

{Each specialist is trained on a single target to assess how accurately the adapted Hamiltonian can predict that property. All five specialists improve on fixed PPP for their supervised targets (Fig.~\ref{fig:3}), with the largest gains in excitation energies. The S\textsubscript{1} specialist reduces the MAE from 0.336 to 0.060 eV, and the T\textsubscript{1} specialist reduces it from 0.702 to 0.041 eV. Because the remaining properties receive no direct supervision, they provide a test of whether these gains transfer across properties.}

\begin{figure}[!htbp]
\centering
\includegraphics[width=\linewidth,height=.68\textheight,keepaspectratio]{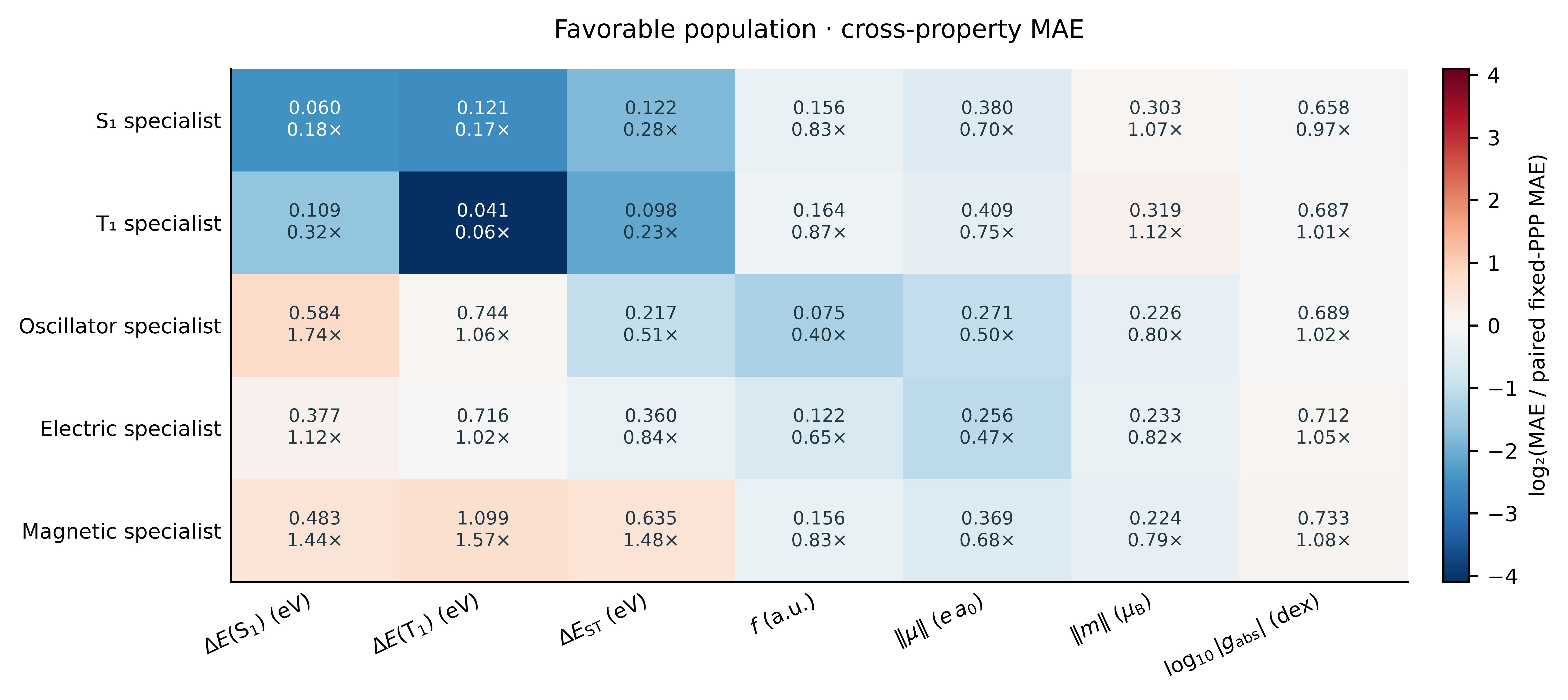}
\caption{Favorable specialist accuracy transfers differently across properties. Rows show separately supervised models and columns show their supervised and unsupervised properties. Cells give MAE and its ratio to fixed PPP on the same molecules. All columns use the common 884-molecule favorable evaluation set, except \(\log_{10}|g_{\mathrm{abs}}|\) (694 molecules). Accepted finite zero or negative triplet predictions remain included.}
\label{fig:4}
\end{figure}

Each correction changes the Hamiltonian used to calculate all properties. Therefore, improving the supervised target may also affect properties excluded from the training objective. Fig.~\ref{fig:4} evaluates each specialist across both its supervised and unsupervised properties. This comparison shows that supervision of either excitation energy also improves the other. The T\textsubscript{1} specialist lowers MAE in \(\Delta E(\mathrm{S}_1)\) from 0.336 to 0.109 eV, and the S\textsubscript{1} specialist lowers MAE in \(\Delta E(\mathrm{T}_1)\) from 0.702 to 0.121 eV. Both nevertheless worsen \(\lVert\mathbf{m}\rVert\) error, from 0.284 \(\mu\)B to 0.319 and 0.303 \(\mu\)B, respectively. In contrast, electric and \(f\) supervision lower the error in \(\lVert\mathbf{m}\rVert\) but increase both errors in \(\Delta E\), as does supervision of \(\lVert\mathbf{m}\rVert\). Thus, target accuracy does not imply an improvement in every property of the shared Hamiltonian.

\begin{figure}[!htbp]
\centering
\includegraphics[width=\linewidth,height=.78\textheight,keepaspectratio]{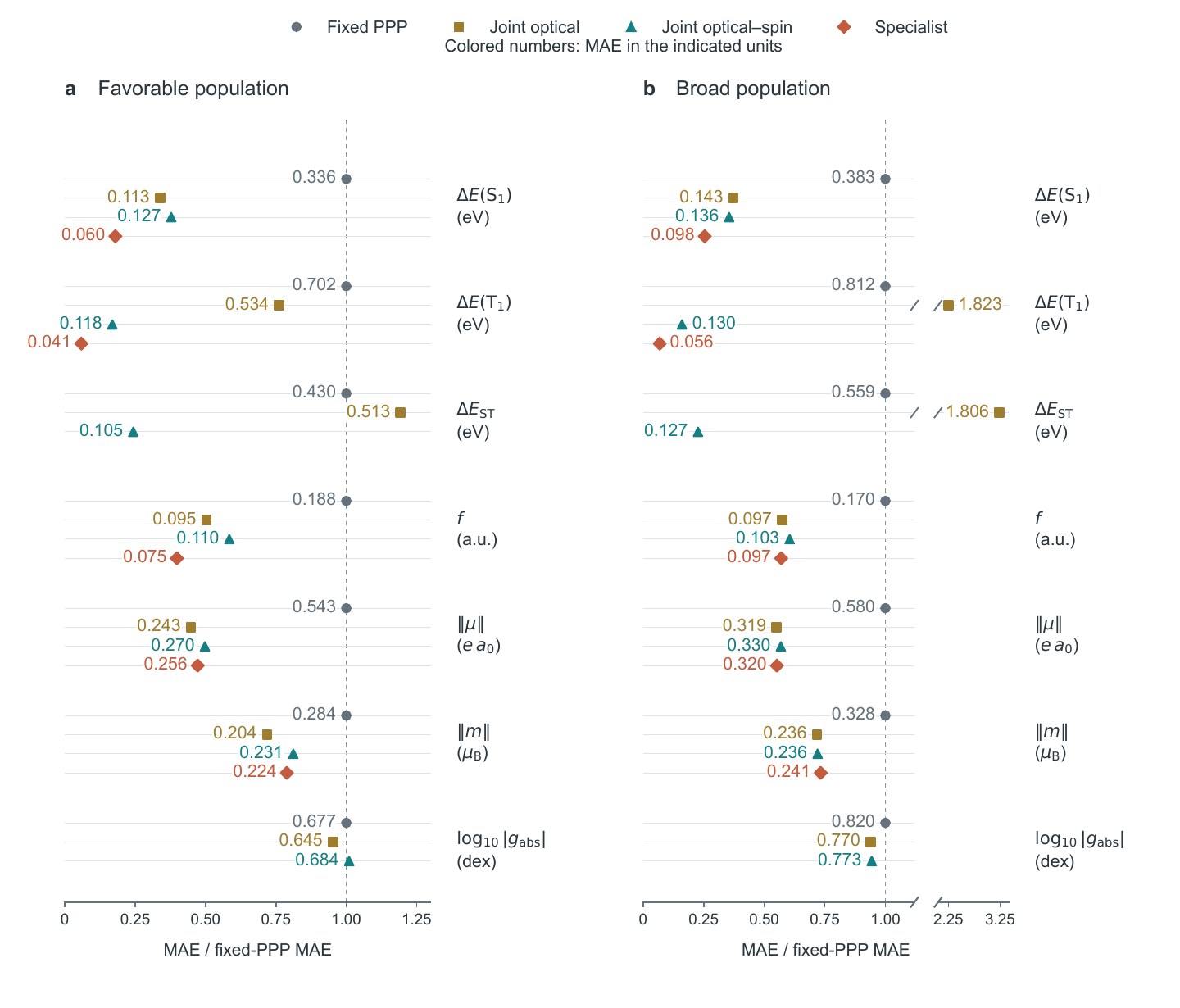}
\caption{Joint supervision balances target-specific accuracies in the favorable (a) and broad (b) populations. Errors are divided by fixed PPP's MAE on the same molecules: 1 denotes baseline accuracy, and values below 1 indicate improvement. {Each property uses separate horizontal lanes for fixed PPP, joint optical, joint optical--spin and specialist models. Colored numbers give MAE in the native units displayed with the property label at right; marker positions show the ratio to fixed PPP. Each specialist marker identifies the model trained on that property.} Comparisons use 884 favorable and 3,432 broad molecules, with 694 and 2,991, respectively, for \(\log_{10}|g_{\mathrm{abs}}|\).}
\label{fig:6}
\end{figure}

{The specialist trade-offs motivate{~a cross-examination with the }joint supervision of optical properties and \(\Delta E_{\mathrm{ST}}\) within one adapted Hamiltonian.~{{Consequently},~}Figure~\ref{fig:6}a compares the two joint models with fixed PPP and the specialist trained for each target. The joint optical--spin model reduces MAE relative to fixed PPP for \(\Delta E(\mathrm{S}_1)\), \(\Delta E(\mathrm{T}_1)\), \(f\), \(\lVert\boldsymbol{\mu}\rVert\) and \(\lVert\mathbf{m}\rVert\), reaching energy MAEs of 0.127, 0.118 and 0.105 eV for \(\Delta E(\mathrm{S}_1)\), \(\Delta E(\mathrm{T}_1)\) and \(\Delta E_{\mathrm{ST}}\), respectively. For each specialist target, its error lies between fixed PPP and the corresponding specialist. The remaining specialist advantage is small for the transition-moment magnitudes, larger for \(f\), and largest for the excitation energies.}

{Comparing the two joint objectives{{{,~}optical-spin and optical-only,}~}shows how this balance depends on{~the~}~supervision of \(\Delta E_{\mathrm{ST}}\). The optical-only model achieves a lower \(\Delta E(\mathrm{S}_1)\) error, but its \(\Delta E(\mathrm{T}_1)\) and \(\Delta E_{\mathrm{ST}}\) MAEs are 0.534 and 0.513 eV, compared with 0.118 and 0.105 eV for optical--spin training. Thus, optical-only training improves the triplet energy relative to fixed PPP but worsens the singlet--triplet gap. Neither favorable joint model produces a nonpositive \(\Delta E(\mathrm{T}_1)\) value in the common set.}

{The gains in energies and transition-moment magnitudes do not extend to dissymmetry: the joint optical--spin model's logarithmic dissymmetry MAE is slightly higher than fixed PPP, at 0.684 versus 0.677 dex. For a more detailed view, SI Figure~\ref{fig:S13} provides individual predictions and transition-moment direction comparisons, while Section~\ref{sec:si-optical-only-training} examines the effect of gap supervision. SI Tables~\ref{tab:S8}--\ref{tab:S9} provide MAEs and Pearson correlations for energies, moment magnitudes, \(f\) and \(\lvert g_{\mathrm{abs}}\rvert\) across all favorable models.}

\FloatBarrier
\subsection{\texorpdfstring{Broad Regime: Bright, Resolved S\textsubscript{1} Transitions}{Broad Regime: Bright, Resolved S1 Transitions}}\label{sec:broad-regime}

The broad population extends the evaluation to a wider range of reference S\textsubscript{1} local-\(\pi\) character. It retains the brightness and resolution criteria of Section~2.3 and includes molecules across the full range of character allowed by those criteria. Models for the broad population are trained separately on its training set and evaluated on its held-out test molecules. Applying the same evaluation protocol as in the favorable population retains 3,432 of the 3,436 broad test molecules (99.9\%). Of these, 2,991 pass the precision screen and are retained for dissymmetry comparisons.

\begin{figure}[!htbp]
\centering
\includegraphics[width=\linewidth,height=.68\textheight,keepaspectratio]{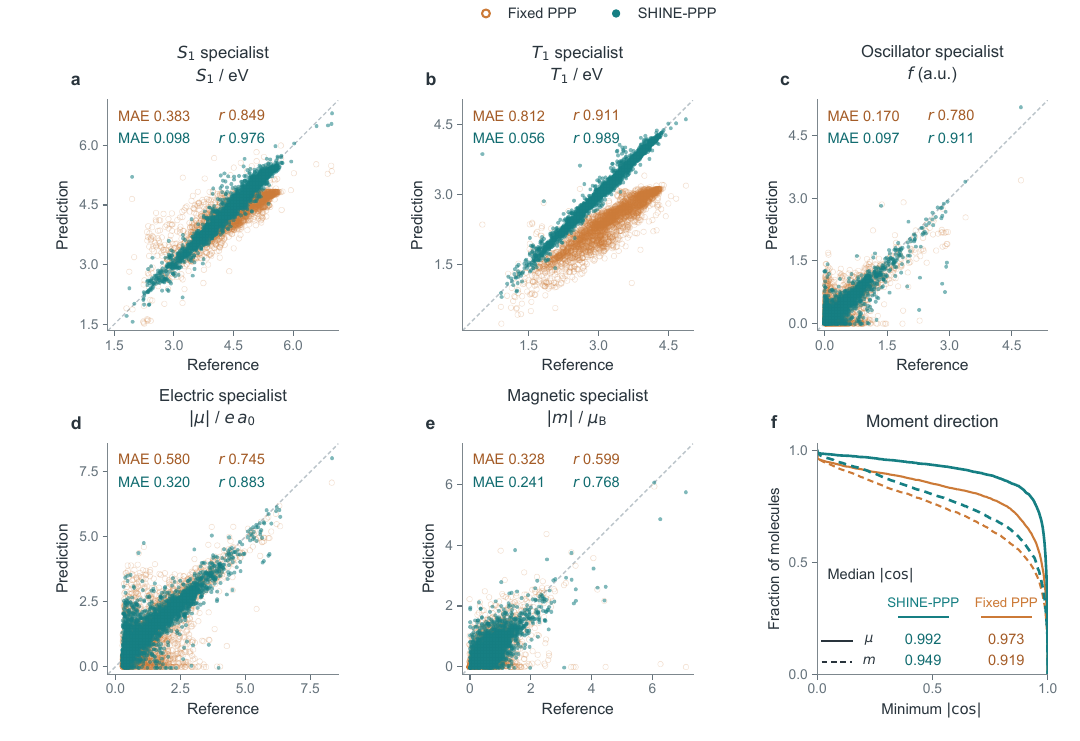}
\caption{Target-specific predictions of the five broad specialists. Panels show \(\Delta E(\mathrm{S}_1)\), \(\Delta E(\mathrm{T}_1)\), \(f\) and \(\lVert\boldsymbol{\mu}\rVert\) and \(\lVert\mathbf{m}\rVert\) against their references. Teal filled points show specialist predictions and orange open circles show fixed PPP on the same molecules. Matching scalar-panel annotations give each method's MAE and Pearson r; n is the evaluation count. {Panel f compares electric and magnetic direction agreement: teal curves show the specialists and orange curves show fixed PPP on the same molecules; solid lines indicate \(\boldsymbol{\mu}\) and dashed lines indicate \(\mathbf{m}\).} The horizontal axis is the minimum absolute direction cosine with the reference, and the vertical axis is the fraction of molecules reaching that value. The table gives specialist and fixed-PPP median absolute cosines. Scalar panels use the common 3,432-molecule evaluation set; direction curves retain 3,432 electric and 3,151 magnetic comparisons after the resolution masks.}
\label{fig:7}
\end{figure}

To test whether the gains from dedicated supervision extend to the broad population, we compare the five specialists' predictions with those of fixed PPP (Fig.~\ref{fig:7}). Each specialist reduces the MAE for its supervised target. As in the favorable population, the largest proportional reductions are for \(\Delta E\), and the smallest is for \(\lVert\mathbf{m}\rVert\).

\begin{figure}[!htbp]
\centering
\includegraphics[width=\linewidth,height=.68\textheight,keepaspectratio]{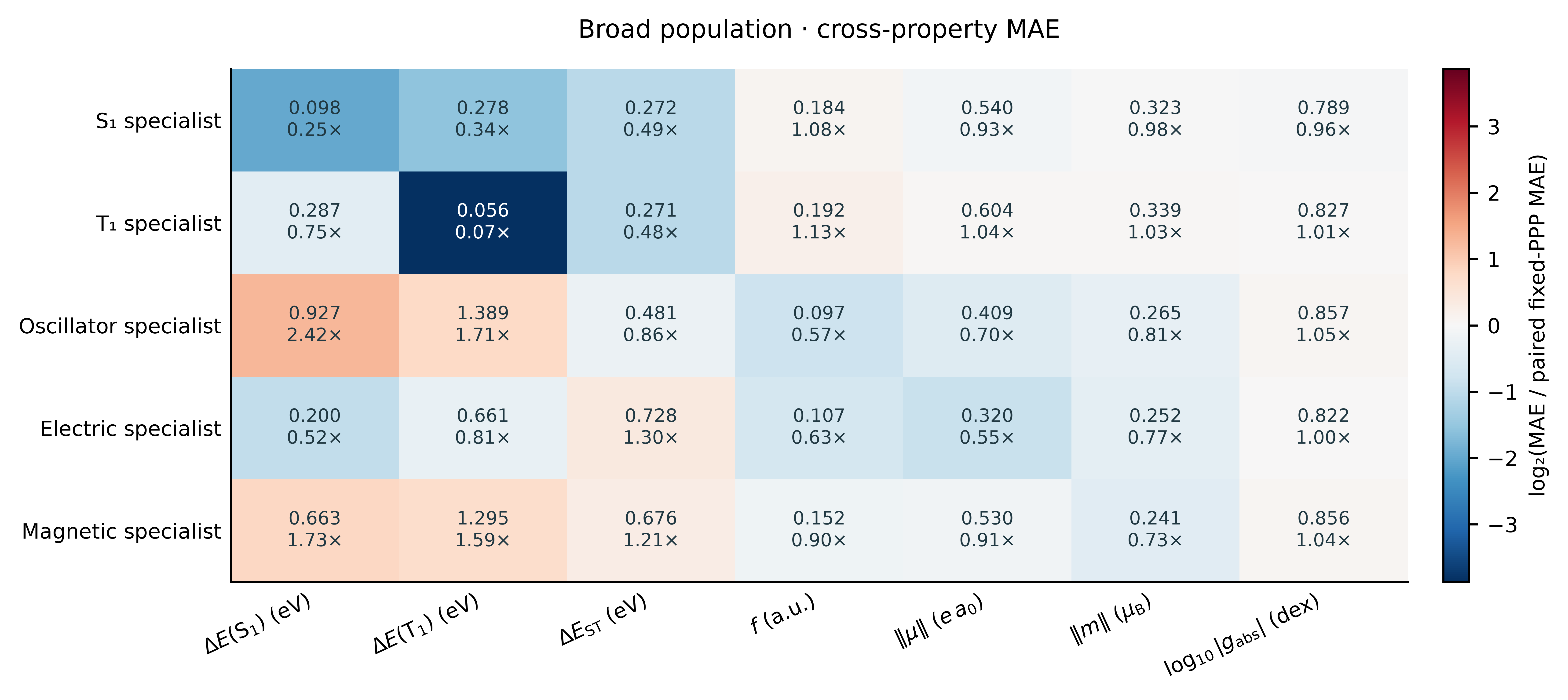}
\caption{Broad specialist transfer depends on the supervised observable. Rows show the five specialists and columns their supervised and unsupervised properties. Cells give MAE and its ratio to paired fixed PPP. All columns use the common 3,432-molecule broad evaluation set, except \(\log_{10}|g_{\mathrm{abs}}|\) (2,991 molecules). Accepted finite zero or negative triplet predictions remain included.}
\label{fig:8}
\end{figure}

{Cross-property transfer remains dependent on the target of supervision (Fig.~\ref{fig:8}). }The S\textsubscript{1} specialist slightly lowers \(\lVert\mathbf{m}\rVert\) MAE from 0.328 to 0.323 \(\mu\)B, whereas the T\textsubscript{1} specialist increases it to 0.339 \(\mu\)B; the magnetic specialist worsens accuracy for both \(\Delta E(\mathrm{S}_1)\) and \(\Delta E(\mathrm{T}_1)\). In contrast, the electric specialist improves both \(\Delta E\) and \(\lVert\mathbf{m}\rVert\): MAE in \(\Delta E(\mathrm{S}_1)\) falls from 0.383 to 0.200 eV, MAE in \(\Delta E(\mathrm{T}_1)\) from 0.812 to 0.661 eV, and \(\lVert\mathbf{m}\rVert\) MAE from 0.328 to 0.252 \(\mu\)B. However, its MAE in \(\Delta E_{\mathrm{ST}}\) rises from 0.559 to 0.728 eV, showing that lower errors in the individual energies can coexist with a less accurate \(\Delta E_{\mathrm{ST}}\).

{Comparing the broad joint models with the dedicated specialists shows how much target-specific accuracy joint supervision retains (Fig.~\ref{fig:6}b). The broad joint optical--spin model has lower errors than fixed PPP on all five specialist targets. Its \(\Delta E(\mathrm{S}_1)\), \(\Delta E(\mathrm{T}_1)\), \(\lVert\boldsymbol{\mu}\rVert\) and \(f\) errors remain above those of the corresponding specialists, while its \(\lVert\mathbf{m}\rVert\) error is slightly lower than that of the magnetic specialist (0.236 versus 0.241 \(\mu\)B).~{Notably, u}nlike in the favorable population, logarithmic dissymmetry also improves relative to fixed PPP, with MAE decreasing from 0.820 to 0.773 dex.}

{The importance of gap supervision becomes more pronounced in the broad population. The optical-only joint model has nearly the same error in \(\Delta E(\mathrm{S}_1)\) as the joint optical--spin model, but its error in \(\Delta E(\mathrm{T}_1)\) reaches 1.823 eV, compared with 0.130 eV for joint optical--spin training and 0.812 eV for fixed PPP. The optical-only model also produces 291 nonpositive \(\Delta E(\mathrm{T}_1)\) values in the common set, compared with none for the joint optical--spin model. These results show that similar singlet accuracy can conceal substantially different triplet predictions.}

{Individual joint optical--spin predictions and transition-moment direction comparisons are provided in SI Figure~\ref{fig:S14}, with MAEs and Pearson correlations across all models in SI Tables~\ref{tab:S10}--\ref{tab:S11}. Additional analysis of how prediction errors vary with reference S\textsubscript{1} local-\(\pi\) character is provided in SI Section~\ref{sec:si-screening-errors}.}

\FloatBarrier
\section{Discussion}\label{sec:discussion}

The results show that changing the training objective changes which properties the adapted Hamiltonian predicts most accurately. We next examine how the joint optical--spin model modifies the PPP interactions, before considering the limitations imposed by the electronic model.

\subsection{Environment Dependence of the Learned Corrections}\label{sec:learned-corrections}

The readouts allow the PPP corrections to depend on local atomic environments. This flexibility does not, by itself, establish whether the trained model distinguishes individual sites and pairs or mainly applies shared shifts within each molecule. We therefore examine the balance between molecular shifts and within-molecule variation in the joint optical--spin model. For example, C--C hopping could increase by approximately 10\% throughout molecule A and 20\% throughout molecule B, while varying little among pairs within either molecule. A wider distribution around each molecule’s average would indicate stronger differentiation among its local environments. We quantify the molecular shift by the mean fractional correction and the within-molecule variation by its population standard deviation.

The analysis uses the 884 favorable evaluation molecules and focuses on carbon sites and adjacent C--C pairs, which account for 83.0\% of sites and 76.5\% of adjacent pairs in this set. Adjacent pairs are those coupled by a nonzero baseline hopping. To compare C--C intersite corrections at the same distance, the correction map is evaluated at a common separation of 1.375 Å while holding the learned outputs fixed. This removes the explicit distance dependence of the map while retaining the environmental information encoded in those outputs.

\begin{figure}[!t]
\centering
\includegraphics[width=\linewidth,height=.76\textheight,keepaspectratio]{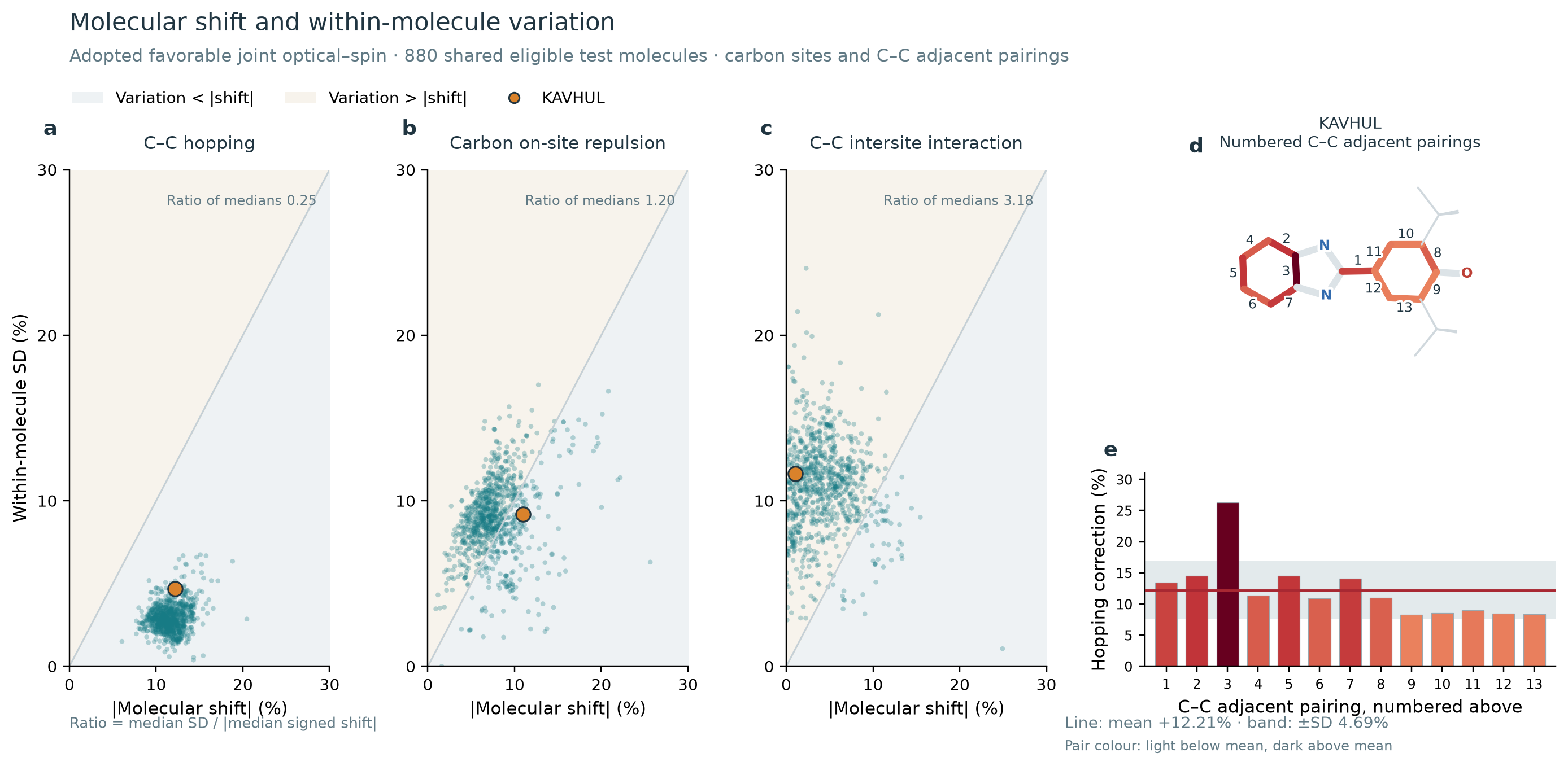}
\caption{Molecular shift and within-molecule variation in the joint optical--spin model. The 880 favorable test molecules retained for the parameter analysis have at least four carbon sites and three C--C adjacent pairings. Each point summarizes carbon on-site repulsion or C--C hopping/intersite corrections: absolute molecular mean on the horizontal axis and population standard deviation on the vertical axis. The diagonal marks equality. Orange points identify KAVHUL. Its C--C hopping bars have mean +12.21\% and SD 4.69\%; the line and band show these quantities. C--C intersite corrections use 1.375 Å with raw learned outputs held fixed.}
\label{fig:12}
\end{figure}

Figure~\ref{fig:12} compares molecular shift with within-molecule variation for the 880 molecules retained for the parameter analysis, each containing at least four carbon sites and three adjacent C--C pairs. Table~\ref{tab:2} summarizes the two quantities for each parameter type.

\begingroup
\footnotesize
\setlength{\tabcolsep}{3pt}
\renewcommand{\arraystretch}{1.18}
\begin{table}[!htbp]
\centering
\footnotesize
\caption{Molecular shift and within-molecule variation in the joint optical--spin model. Entries give separate medians over the 880 molecules. The final column divides median variation by the absolute median shift; it is not the median of per-molecule ratios. C--C intersite corrections are evaluated at 1.375 Å.}\label{tab:2}
\begin{tabular}{@{}>{\raggedright\arraybackslash}p{0.29\linewidth}>{\raggedright\arraybackslash}p{0.18\linewidth}>{\raggedright\arraybackslash}p{0.24\linewidth}>{\raggedright\arraybackslash}p{0.22\linewidth}@{}}
\toprule
Parameter & Molecular shift (\%) & Within-molecule variation (\%) & Variation / \textbar{}shift\textbar{} \\
\midrule
C--C hopping & +11.57 & 2.85 & 0.25 \\
Carbon on-site repulsion & -7.42 & 8.93 & 1.20 \\
C--C intersite interaction & +3.47 & 11.03 & 3.18 \\
\bottomrule
\end{tabular}
\end{table}
\endgroup

\Needspace{10\baselineskip}
Within-molecule variation is comparable to the magnitude of the molecular shift for carbon on-site repulsion and about three times larger for C--C intersite interactions. The learned corrections therefore do more than shift each molecule’s parameters together: they differentiate carbon sites and C--C pairs within the same molecule. C--C hopping shows a different balance. Its median molecular shift of +11.57\% dominates its median within-molecule variation of 2.85\%, indicating a predominantly shared increase with a smaller contribution from local differentiation. Taken together, these patterns show that the extent of local adaptation depends on the interaction being corrected. Section~\ref{sec:si-baseline-group-corrections} provides the distributions by baseline type.

\subsection{Limitations and Outlook}\label{sec:limitations}

The learned corrections operate within the existing electronic description of PPP. The benchmark measures agreement with TDA for bright, eligible transitions. It leaves open how the model performs relative to experimental observables or on weak and dark transitions. Within this benchmark, the orbital basis and excited-state configuration space also limit what parameter adaptation can achieve. These restrictions motivate extending the approach to a richer valence Hamiltonian and an electronic treatment that includes excitations beyond singles.

\mergedsubsectionanchor\label{sec:reduced-orbital-basis}

To test whether a larger valence basis improves the adapted predictions, SHINE could be coupled to a Hamiltonian such as INDO/S, the intermediate neglect of differential overlap method for spectroscopy, whose valence basis includes s and p orbitals on first-row heavy atoms and whose Hamiltonian retains one-center exchange integrals.\cite{ref22} Tight-binding approaches with excited-state treatments, such as TD-DFTB\cite{ref23} and simplified Tamm--Dancoff extended tight binding,\cite{ref24} offer additional potential hosts.

\mergedsubsectionanchor\label{sec:singles-only-treatment}

A larger valence basis would make more orbitals available for excitation, but capturing double and higher excitations would also require an excited-state treatment beyond CIS. These excitations can affect individual states and singlet--triplet ordering. Examples include the low-lying 2\textsuperscript{1}A\textsubscript{g} state of polyenes \cite{ref25} and the inverted \(\Delta E_{\mathrm{ST}}\).\cite{ref05} The corrections modify the Hamiltonian, allowing the same readouts, in principle, to be coupled to a correlated calculation in the same \ensuremath{\pi} space. Such an extension builds on correlated configuration-interaction treatments of PPP applied to low-lying singlet states and optical absorption in polyacenes.\cite{ref08} Bedogni et al.~\cite{bedogni2024} also used PPP to examine how excitations beyond singles affect singlet--triplet ordering in cycl{[}3.3.3{]}azine.{\cite{leupin1980cyclazine}} A correlated baseline could describe doubly excited states and correlation contributions to the \(\Delta E_{\mathrm{ST}}\) explicitly, reducing the need for parameter corrections to compensate for omitted correlation. However, the remaining \ensuremath{\pi}-space and Hamiltonian approximations could still require effective-parameter corrections. Furthermore, the cost of correlated methods grows steeply with the number of \ensuremath{\pi} sites, which bounds the systems that can be trained through them.

\section{Conclusion}\label{sec:conclusion}
SHINE {constructs rotation-invariant descriptors of local atomic environments from frozen MACE features and uses Kolmogorov--Arnold network readouts to map them to bounded corrections to the PPP Hamiltonian. During training, gradients from errors in the predicted excited-state properties pass through the RHF and CIS calculations to update the readouts.}

{On molecular scaffolds excluded from training, all five specialists improved their supervised scalar targets relative to fixed PPP in both FORMED populations, with the largest gains in \(\mathrm{S}_1\) and \(\mathrm{T}_1\) excitation energies. Transfer to other observables varied with the target. Joint optical--spin supervision lowered errors in excitation energies, the singlet--triplet gap, oscillator strength, and electric and magnetic transition moment magnitudes within one adapted Hamiltonian per population. In the broad population, including the gap in the training objective reduced the \(\mathrm{T}_1\) MAE from 1.823 to 0.130 eV compared with optical-only training. Logarithmic dissymmetry improved in the broad population and worsened slightly in the favorable population.}

{The learned corrections also varied by interaction. In the favorable joint model, C--C hopping showed a predominantly shared increase within each molecule, while carbon on-site repulsion and C--C intersite interactions showed appreciable variation among sites and pairs. These results establish the gains from adapting PPP for bright transitions in closed-shell conjugated molecules within the FORMED benchmark. A larger orbital basis and an excited-state treatment beyond CIS would test whether the same approach can improve predictions for a wider range of transitions.}

\section*{Supporting Information}
Descriptor and readout definitions, transition-moment conventions and precision criteria, dataset curation and scaffold splits, training and validation details, model comparisons and calculation failures, excitation-character analysis, and learned-parameter corrections (PDF).

\section*{Competing Interests}
The authors declare no competing interests.

\section*{Data and Code Availability}
The code, trained models and supporting numerical data will be made publicly available through a GitHub repository upon journal submission.

\section*{Acknowledgements}
I.A.E. and J.T.B. acknowledge the Alberta Machine Intelligence Institute (Amii) for support. J.T.B. acknowledges the Canadian Institute for Advanced Research for a CIFAR AI chair at Amii, the Natural Sciences and Engineering Research Council of Canada (RG-PIN 2026-06971 and DGECR 2026-00123), and the University of Alberta Faculty of Science. This research was enabled in part by computational support provided by the BC DRI Group and the Digital Research Alliance of Canada (alliancecan.ca).

\newpage
\section*{Declaration of AI Tools Usage}
The authors used Claude (Anthropic) and Codex (OpenAI) to assist with manuscript revision and the development of scripts used to generate figures. The authors reviewed the revised text and figures and take responsibility for the final manuscript.

\bibliography{references}
\end{document}


\begingroup
\setstretch{1.0}
\tableofcontents
\clearpage
\section*{List of Tables}
\makeatletter
\@starttoc{lot}
\makeatother
\clearpage
\section*{List of Figures}
\makeatletter
\@starttoc{lof}
\makeatother
\clearpage

\endgroup

\maketitle

{The Supporting Information provides readout definitions, centre-of-mass magnetic methods, population selection, reference-character analyses, training specifications, numerical failure criteria and full model comparisons in twelve sections, ten tables and thirteen figures.}

\FloatBarrier
\section{Readout Contractions}\label{sec:si-readout-contractions}

\begin{figure}[!htbp]
\centering
\includegraphics[width=\linewidth,height=.82\textheight,keepaspectratio]{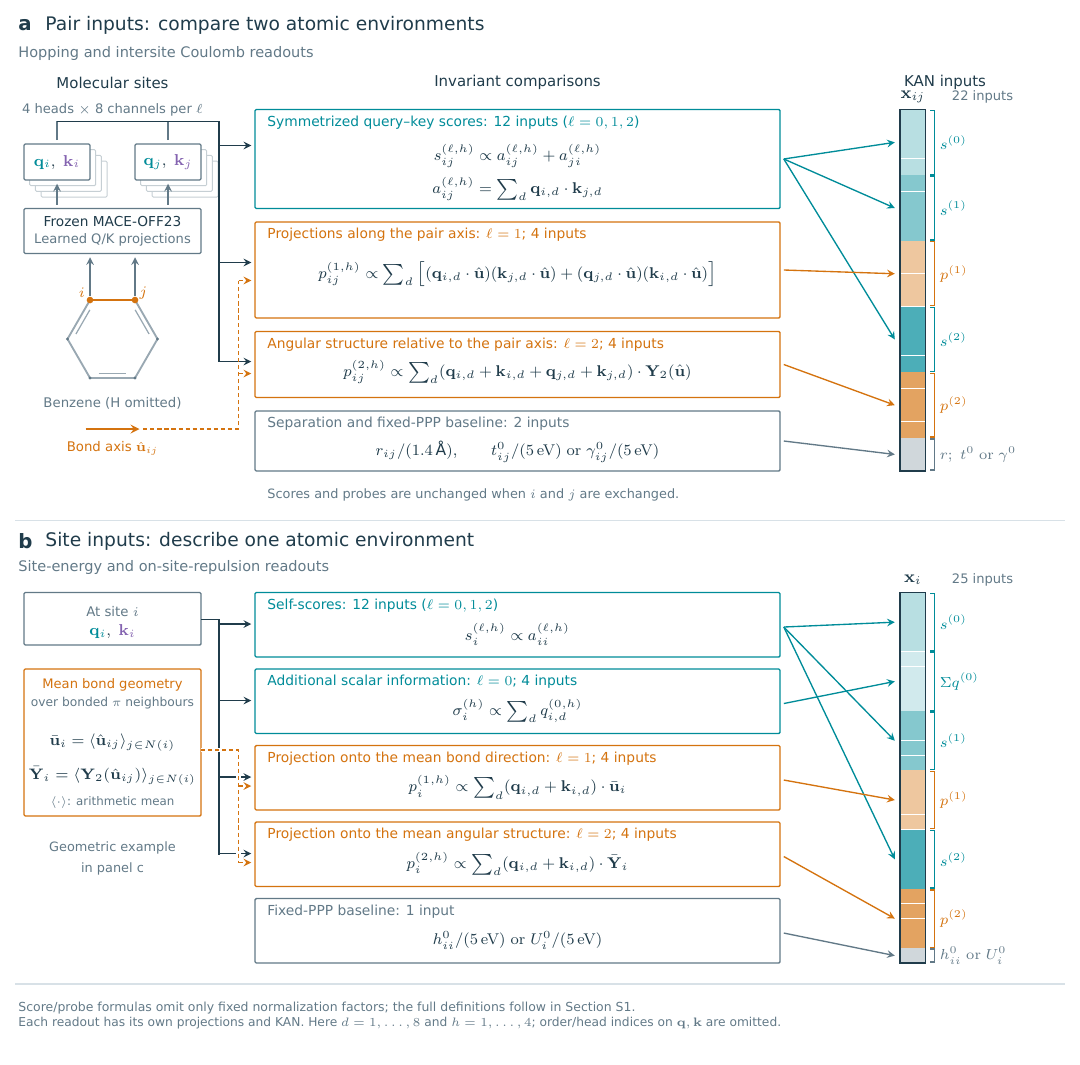}
\caption[Readout inputs]{Construction of the readout inputs. (a) Pair readouts combine query--key scores, pair-axis probes, separation and the baseline parameter into 22 invariant inputs. (b) Site readouts combine self-scores, scalar query sums, mean-geometry probes and the baseline parameter into 25 inputs. Bars show the KAN input order. Score and probe formulas omit fixed normalization factors; Tables~\ref{tab:S1}--\ref{tab:S2} and the equations below give the full definitions.}
\label{fig:S1}
\end{figure}

Fig.~\ref{fig:S1} shows how the four readouts turn atom-centred features into scalar inputs. Pair readouts compare two sites; site readouts summarize one site and its bonded \ensuremath{\pi} neighbours. We first define the learned projections, then construct the pair and site descriptors.

\begin{figure}[!htbp]
\centering
\includegraphics[width=\linewidth]{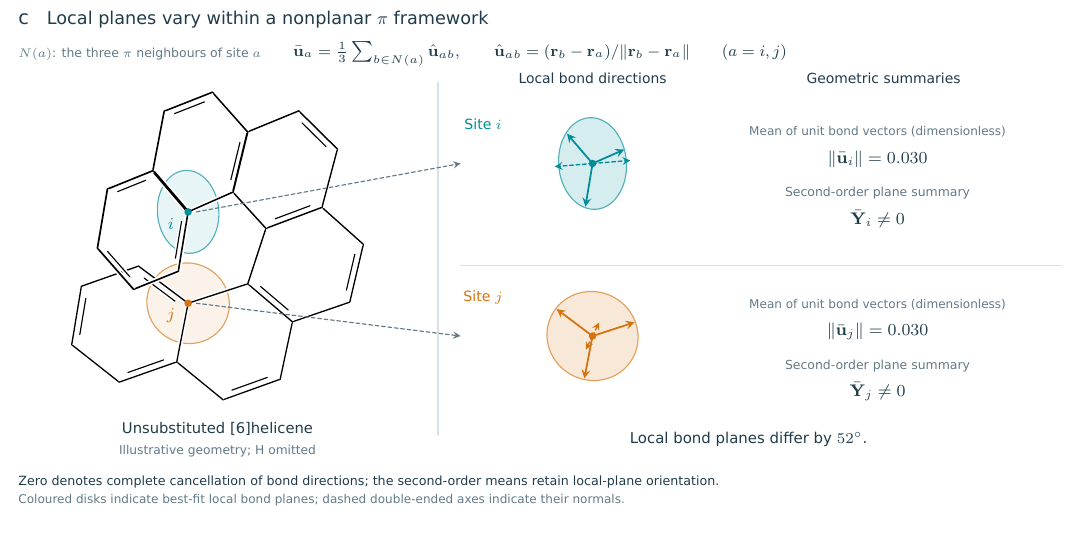}
\caption*{\textbf{Figure~\ref{fig:S1} (continued).} (c) In an illustrative [6]helicene geometry, two sites have small dimensionless mean bond-direction norms (0.030), while their local bond planes differ by $52^\circ$. Their second-order means retain the plane orientations. The geometry was optimized with MMFF94~\cite{halgren1996mmff94}; H atoms are omitted.}
\end{figure}

\subsection{Encoder Features and Learned Projections}\label{sec:si-encoder-projections}

We use the frozen MACE-OFF23 large encoder \cite{ref17} to describe each retained \ensuremath{\pi} site and its surrounding molecular environment within a 5 Å cutoff. From the first interaction layer, we extract 224 channels for each angular order, \ensuremath{\ell} = 0, 1, 2. These contain one, three and five components per channel, respectively, giving 2,016 feature components per atom. We retain the features at the PPP \ensuremath{\pi} sites. Before constructing the descriptors, we divide each feature channel by its root-mean-square (RMS) value over the training \ensuremath{\pi} sites and angular components, with a minimum scale of 10\textsuperscript{-8}. Each channel uses one scale shared across its 2\ensuremath{\ell} + 1 components, preserving its transformation under rotation. These feature-normalization scales are distinct from the target-error scales used in Section~\ref{sec:learning-objective}.

Each readout has its own query/key projection weights and its own KAN. The encoder supplies features F\textsuperscript{(\ensuremath{\ell})}\textsubscript{i,c}, with atom index i, channel c = 1, \ldots, 224, and angular order \ensuremath{\ell} = 0, 1, 2. Learned linear projections combine the encoder channels into queries and keys:

\begin{equation}
\mathbf q_{i,d}^{(\ell,h)}=\sum_{c=1}^{224}w_{dc}^{Q,\ell,h}\mathbf F_{i,c}^{(\ell)},\qquad\mathbf k_{i,d}^{(\ell,h)}=\sum_{c=1}^{224}w_{dc}^{K,\ell,h}\mathbf F_{i,c}^{(\ell)}.
\label{eq:source-19}
\end{equation}

\Needspace{4\baselineskip}
There are eight projected channels d = 1, \ldots, 8 in each of four heads h = 1, \ldots, 4. A query or key of order \ensuremath{\ell} has 2\ensuremath{\ell} + 1 components. The query and key weights are learned separately and shared across molecules.

Mixing channels within an angular order preserves their transformation under rotation. The contractions below therefore combine queries, keys and geometric probes into rotation-invariant scalar inputs.

\subsection{Pair Inputs}\label{sec:si-pair-inputs}

The hopping readout acts on bonded pairs of \ensuremath{\pi} sites, the only pairs with a baseline hopping t\textsuperscript{0}\textsubscript{ij}, and the screening readout on all pairs of \ensuremath{\pi} sites. Each receives the 22 inputs of Table~\ref{tab:S1}.

\begingroup
\footnotesize
\setlength{\tabcolsep}{3pt}
\renewcommand{\arraystretch}{1.18}
\begin{table}[!htbp]
\centering
\footnotesize
\caption{Pair inputs for atoms i and j.}\label{tab:S1}
\begin{tabular}{@{}>{\raggedright\arraybackslash}p{0.1063\linewidth}>{\raggedright\arraybackslash}p{0.1772\linewidth}>{\raggedright\arraybackslash}p{0.6025\linewidth}@{}}
\toprule
Inputs & Name & Description \\
\midrule

1--4 & \ensuremath{\ell} = 0 score & One symmetrized score per head for the scalar features: the average of atom i's query contracted with atom j's key and atom j's query contracted with atom i's key \\
5--8 & \ensuremath{\ell} = 1 score & The same score for the \ensuremath{\ell} = 1 features; it compares the directional environments of the two atoms, independently of the orientation of the pair \\
9--12 & \ensuremath{\ell} = 1 probe & The \ensuremath{\ell} = 1 queries of one atom and the keys of the other, each projected onto the pair axis, multiplied, and averaged over both orders \\
13--16 & \ensuremath{\ell} = 2 score & The same score for the \ensuremath{\ell} = 2 features \\
17--20 & \ensuremath{\ell} = 2 probe & The sum of the \ensuremath{\ell} = 2 queries and keys of both atoms, projected onto Y\textsubscript{2}(\ensuremath{\hat{u}}\textsubscript{ij}), the five real second-order spherical harmonics of the pair axis in the component order of the MACE features, scaled so that \textbar{}Y\textsubscript{2}(\ensuremath{\hat{u}})\textbar{} = 1 \\
21 & Separation & Interatomic distance r\textsubscript{ij} \\
22 & Baseline & Uncorrected hopping t\textsuperscript{0}\textsubscript{ij} or intersite interaction \ensuremath{\gamma}\textsuperscript{0}\textsubscript{ij} \\
\bottomrule
\end{tabular}
\end{table}
\endgroup

\Needspace{80pt}
To construct the scores, we first contract the projected features:

\begin{equation}
a_{ij}^{(\ell,h)}=\sum_{d=1}^{8}\mathbf q_{i,d}^{(\ell,h)}\cdot\mathbf k_{j,d}^{(\ell,h)}.
\label{eq:source-21}
\end{equation}

Each contraction sums over the angular components and eight projected channels. Queries and keys transform under the same orthogonal rotation, leaving their dot product unchanged.~\cite{fuchs2020se3}

The query and key weights differ, so a\textsubscript{ij} need not equal a\textsubscript{ji}. We average the two assignments to make the pair score independent of atom ordering:

\begin{equation}
s_{ij}^{(\ell,h)}=\frac{a_{ij}^{(\ell,h)}+a_{ji}^{(\ell,h)}}{2\sqrt{8(2\ell+1)}}.
\label{eq:source-22}
\end{equation}

The denominator averages the two atom assignments and normalizes for the channel and angular-component counts. Four heads at three angular orders provide twelve scores, used directly without softmax or attention weighting.

The probes supply the additional relationship to the pair axis. With \ensuremath{\hat{u}}\textsubscript{ij} the unit vector from atom i to atom j, the \ensuremath{\ell} = 1 probes are

\begin{equation}
\begin{aligned}
p_{ij}^{(1,h)} &= \frac{1}{2\sqrt8}\sum_{d=1}^{8}\Big[
  (\mathbf q_{i,d}^{(1,h)}\cdot\hat{\mathbf u}_{ij})(\mathbf k_{j,d}^{(1,h)}\cdot\hat{\mathbf u}_{ij}) \\
&\hphantom{{}=\frac{1}{2\sqrt8}\sum_{d=1}^{8}\Big[}
  + (\mathbf q_{j,d}^{(1,h)}\cdot\hat{\mathbf u}_{ij})(\mathbf k_{i,d}^{(1,h)}\cdot\hat{\mathbf u}_{ij})\Big],
\end{aligned}
\label{eq:source-23}
\end{equation}

and the \ensuremath{\ell} = 2 probes are

\begin{equation}
p_{ij}^{(2,h)}=\frac{1}{2\sqrt8}\sum_{d=1}^{8}\left(\mathbf q_{i,d}^{(2,h)}+\mathbf k_{i,d}^{(2,h)}+\mathbf q_{j,d}^{(2,h)}+\mathbf k_{j,d}^{(2,h)}\right)\cdot\mathbf Y_2(\hat{\mathbf u}_{ij}).
\label{eq:source-24}
\end{equation}

Both probes preserve atom-exchange symmetry: reversing the pair axis changes the sign of both factors in the \ensuremath{\ell} = 1 product, whereas Y\textsubscript{2} is even in its argument.

The separation and baseline, inputs 21 and 22, are divided by a typical bonded separation and by 5 eV:

\begin{equation}
\frac{r_{ij}}{1.4\,\text{\AA}},\qquad\frac{t_{ij}^{0}}{5\,\mathrm{eV}}\quad\mathrm{or}\quad\frac{\gamma_{ij}^{0}}{5\,\mathrm{eV}}.
\label{eq:source-25}
\end{equation}

\subsection{Site Inputs}\label{sec:si-site-inputs}

The site-energy and on-site-repulsion readouts receive the 25 inputs in Table~\ref{tab:S2}. Their probes use the mean bond direction and mean second-order harmonic over the \ensuremath{\pi} neighbours N(i) connected by baseline hopping. Bonds to atoms outside the \ensuremath{\pi} system are excluded. For a nonempty neighbour set,

\begin{equation}
\bar{\mathbf u}_i=\frac{1}{|N(i)|}\sum_{j\in N(i)}\hat{\mathbf u}_{ij},\qquad\bar{\mathbf Y}_i=\frac{1}{|N(i)|}\sum_{j\in N(i)}\mathbf Y_2(\hat{\mathbf u}_{ij}).
\label{eq:source-26}
\end{equation}

For a site with no \ensuremath{\pi} neighbour, both averages are zero.

\begingroup
\footnotesize
\setlength{\tabcolsep}{3pt}
\renewcommand{\arraystretch}{1.18}
\begin{table}[!htbp]
\centering
\footnotesize
\caption{Site inputs for atom i.}\label{tab:S2}
\begin{tabular}{@{}>{\raggedright\arraybackslash}p{0.2453\linewidth}>{\raggedright\arraybackslash}p{0.2558\linewidth}>{\raggedright\arraybackslash}p{0.3849\linewidth}@{}}
\toprule
Inputs & Name & Description \\
\midrule

1--4 & \ensuremath{\ell} = 0 self-score & Atom i's scalar query contracted with its own scalar key, one per head \\
5--8 & \ensuremath{\ell} = 0 query sum & The sum of atom i's eight scalar queries, one per head \\
9--12 & \ensuremath{\ell} = 1 self-score & The same self-score for the \ensuremath{\ell} = 1 features \\
13--16 & \ensuremath{\ell} = 1 probe & The sum of atom i's \ensuremath{\ell} = 1 queries and keys, projected onto the mean bond direction \ensuremath{\bar{u}}\textsubscript{i} \\
17--20 & \ensuremath{\ell} = 2 self-score & The same self-score for the \ensuremath{\ell} = 2 features \\
21--24 & \ensuremath{\ell} = 2 probe & The sum of atom i's \ensuremath{\ell} = 2 queries and keys, projected onto the mean harmonic \ensuremath{\bar{Y}}\textsubscript{i} \\
25 & Baseline & Uncorrected core diagonal h\textsuperscript{0}\textsubscript{ii} (site-energy readout) or on-site repulsion U\textsuperscript{0}\textsubscript{i} \\
\bottomrule
\end{tabular}
\end{table}
\endgroup

\FloatBarrier
The self-scores, inputs 1--4, 9--12 and 17--20, and query sums, inputs 5--8, are

\begin{equation}
s_i^{(\ell,h)}=\frac{a_{ii}^{(\ell,h)}}{\sqrt{8(2\ell+1)}},\qquad\sigma_i^{(h)}=\frac{1}{\sqrt8}\sum_{d=1}^{8}q_{i,d}^{(0,h)}.
\label{eq:source-27}
\end{equation}

The site probes need no pair symmetrization because the descriptor is centred on one atom. The probes, inputs 13--16 and 21--24, are

\begin{equation}
p_i^{(1,h)}=\frac{1}{2\sqrt8}\sum_{d=1}^{8}\bigl(\mathbf q_{i,d}^{(1,h)}+\mathbf k_{i,d}^{(1,h)}\bigr)\cdot\bar{\mathbf u}_i,
\label{eq:source-28}
\end{equation}

\begin{equation}
p_i^{(2,h)}=\frac{1}{2\sqrt8}\sum_{d=1}^{8}\left(\mathbf q_{i,d}^{(2,h)}+\mathbf k_{i,d}^{(2,h)}\right)\cdot\bar{\mathbf Y}_i.
\label{eq:source-29}
\end{equation}

The baseline, input 25, is

\begin{equation}
\frac{h_{ii}^{0}}{5\,\mathrm{eV}}\quad\mathrm{or}\quad\frac{U_i^{0}}{5\,\mathrm{eV}}.
\label{eq:source-30}
\end{equation}

The encoder features describe the local environment, while the probes supply its relationship to the bonded \ensuremath{\pi} neighbours. For three coplanar bonds separated by 120\ensuremath{{}^{\circ}}, the mean direction vanishes but the second-order harmonic mean does not. The scalar query sum is included only in site readouts.

\FloatBarrier
\section{Kolmogorov--Arnold Readout Construction}\label{sec:si-kan-construction}

\begin{figure}[!htbp]
\centering
\includegraphics[width=\linewidth,height=.76\textheight,keepaspectratio]{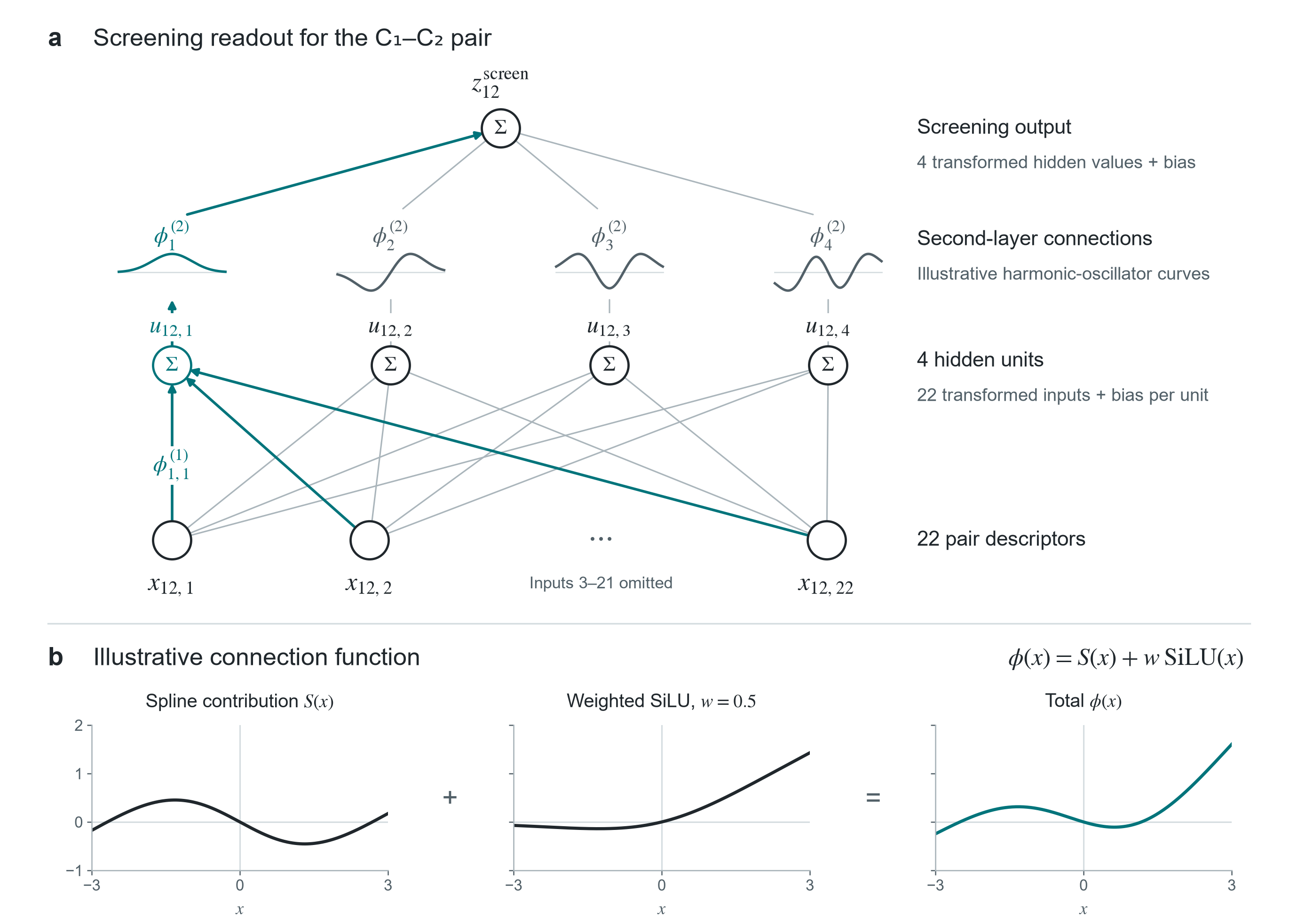}
\caption[KAN readout construction]{KAN mapping from a C--C pair descriptor to a screening output. The 22 inputs pass through four hidden units to one scalar output; inputs 3--21 are omitted from the diagram. Each connection applies its own learned cubic B-spline plus weighted-SiLU function, and each node sums its incoming values with a bias. Teal connections highlight one hidden unit. The second-layer curves illustrate harmonic-oscillator profiles. The lower strip illustrates the spline and SiLU contributions with w = 0.5; the curves are schematic, not fitted functions. The output enters the bounded Coulomb update of Section~\ref{sec:bounded-corrections}.}
\label{fig:S2}
\end{figure}

We expand the two KAN layers for a C--C screening readout (Fig.~\ref{fig:S2}), describe the spline implementation and initialization, and show how the screening output enters the intersite correction. Section~\ref{sec:si-readout-contractions} gives the descriptor definitions.

\Needspace{130pt}
\subsection{Learned Connection Functions}\label{sec:si-connection-functions}

Each KAN connection applies a learned scalar function \ensuremath{\phi}, consisting of a cubic B-spline expansion and a weighted SiLU term.~\cite{ref18} In the implementation, only the spline input is compressed:

\begin{equation}
\begin{aligned}\phi(x)&=\sum_k c_k B_k\!\left(g(x)\right)+w\,\mathrm{SiLU}(x),
\\ &\qquad  g(x)=3\tanh(x/3).\end{aligned}
\label{eq:source-31}
\end{equation}

The cubic B-spline basis functions B\textsubscript{k} are fixed, while their coefficients c\textsubscript{k} and the SiLU weight w are learned separately for every connection in each layer and readout. Biases are added when connection outputs are summed at a node. The same connection-function form is used in both KAN layers.

\subsection{From Descriptors to a Screening Output}\label{sec:si-screening-output}

For a C--C pair labelled C\textsubscript{1}--C\textsubscript{2}, let x\textsubscript{12,d} denote input d of the 22-component descriptor in Table~\ref{tab:S1}. Each of four hidden units sums the transformed inputs and a learned bias:

\begin{equation}
u_{12,a}=\sum_{d=1}^{22}\phi_{ad}^{(1,\mathrm{screen})}\!\left(x_{12,d}\right)+\beta_a^{(1,\mathrm{screen})},\qquad a=1,\ldots,4.
\label{eq:source-33}
\end{equation}

The second layer applies the same spline-plus-SiLU form, with its own coefficients and weights, to each hidden value. It then sums the results with an output bias:

\begin{equation}
z^{\mathrm{screen}}_{12}=\sum_{a=1}^{4}\phi^{(2,\mathrm{screen})}_{a}\bigl(u_{12,a}\bigr)+\beta^{(2,\mathrm{screen})}
\label{eq:source-34}
\end{equation}

The same learned mapping applies to every pair, while its descriptor determines the output. This output enters the bounded Coulomb correction in Section~\ref{sec:bounded-corrections} together with the adapted on-site repulsions and intersite distance.

\subsection{Spline Implementation and Initialization}\label{sec:si-spline-implementation}

Each readout passes its inputs through two KAN layers, from the inputs to four hidden units and from these to one output. In the connection function defined above, the spline terms act on the compressed input g(x) = 3 tanh(x/3), which lies between \ensuremath{-}3 and 3, and the SiLU term acts on x itself. The spline basis is eight cubic B-splines on twelve uniform knots 1.2 apart, from \ensuremath{-}6.6 to 6.6; five knot intervals lie across the compressed range, and the outer basis functions extend mostly beyond it. The second layer is initialized to zero, so an untrained readout returns zero and the baseline Hamiltonian.

\FloatBarrier
\Needspace{15\baselineskip}
\subsection[Rescaling and Screening of Intersite Coulomb Interactions]{Rescaling and Screening of Intersite Coulomb Interactions}\label{sec:si-intersite-rescaling}

We construct the provisional intersite interaction \(\widetilde{\gamma}_{ij}\) used in Section~\ref{sec:bounded-corrections} in two steps. First, we rescale the baseline interaction \(\gamma^0_{ij}\) by the ratio of the Beveridge--Hinze interactions~\citep{ref09} evaluated with the updated and baseline on-site repulsions at the same intersite distance \(r_{ij}\):

\begingroup
\begin{equation}
b_{ij}=\frac{\gamma^{\mathrm{BH}}(U_i^*,U_j^*,r_{ij})}{\gamma^{\mathrm{BH}}(U_i^0,U_j^0,r_{ij})}
\label{eq:source-08}
\end{equation}
\endgroup

We then apply the screening correction from the pair readout \(z_{ij}^{\mathrm{screen}}\):

\begingroup
\begin{equation}
\widetilde{\gamma}_{ij}=\gamma^0_{ij}\,b_{ij}\left[1+e^{-r_{ij}/\lambda}\tanh\left(0.1z^{\mathrm{screen}}_{ij}\right)\right],\qquad\lambda=3\,\text{\AA}
\label{eq:source-09}
\end{equation}
\endgroup

The exponential factor attenuates the learned screening term with distance. The final map in Section~\ref{sec:bounded-corrections} bounds the combined effect of rescaling and screening.

\FloatBarrier
\begingroup
Adapted Coulomb interactions also change the core diagonal through the PPP background-charge terms, so the total diagonal shift can exceed the \(\pm5\) eV bound on the explicit site-energy offset.

\section{Magnetic Transition Moments About the Molecular Centre of Mass}\label{sec:si-magnetic-moments}

Reference and PPP magnetic transition moments are compared about the full-molecule centre of mass, including hydrogens and atoms outside the PPP \ensuremath{\pi} space. We calculate it from the archived coordinates using atomic masses H 1.008, C 12.011, N 14.007, O 15.999, F 18.998, S 32.06 and Cl 35.45.

\subsection{Translation of Reference and PPP Moments}\label{sec:si-magnetic-translation}

The reference magnetic vector is translated algebraically from its archived origin using the printed Gaussian velocity-form transition vector. Let C be the displacement from the archived origin to the full-molecule centre of mass, expressed in bohr. In the real conventions used by the respective implementations,

\begin{equation}
\begin{aligned}
\mathbf m^{\mathrm{ref}}_{C} &= \mathbf m^{\mathrm{ref}}_{O} - \mathbf C \times \mathbf v^{\mathrm{printed}},\\
\mathbf m^{\mathrm{PPP}}_{C} &= \mathbf m^{\mathrm{PPP}}_{O} + \mathbf C \times \mathbf v^{\mathrm{red}}.
\end{aligned}
\label{eq:source-35}
\end{equation}

Magnetic components are expressed in \ensuremath{\mu}B and velocity coefficients in atomic units. The opposite signs reflect the Gaussian gradient-form and PPP reduced physical-velocity conventions. We use the velocity-form transition vector printed by Gaussian to translate the reference magnetic moment. The length-form electric transition dipole is independent of the choice of origin.

The archived Gaussian calculations used NoSymm, and we checked that the reported geometries match the input coordinate frame. We take the origin of that frame as the origin of the printed magnetic transition moments. 

For both fixed PPP and SHINE, we calculate the magnetic transition moment directly using site coordinates measured from the full molecule's centre of mass. 

We use the magnetic moments about this common origin to compute magnetic magnitudes, rotatory strengths, and dissymmetries as defined in Section~\ref{sec:optical-spin-conventions}.

\subsection{Resolution of Magnetic Directions and Dissymmetry}\label{sec:si-magnetic-resolution}

To decide whether a magnetic reference direction is resolved, we account for rounding in both vectors used in the translation. Gaussian prints the magnetic and velocity components to four decimal places, giving a maximum rounding error of \(5\times10^{-5}\) per component. We set the archived magnetic magnitude threshold \(q_0 = 3\sqrt{3}\times 5\times10^{-5}\,\mu_{\mathrm{B}}\) at three times the rounding bound for the archived magnetic vector. After translation to the centre of mass (COM), the rounding bound for each molecule is

\begin{equation}
\delta_m
=
\underbrace{\frac{q_0}{3}}_{\text{archived magnetic rounding}}
+
\underbrace{\frac{q_0}{3}\,d}_{\text{propagated velocity rounding}}.
\label{eq:source-39}
\end{equation}

Here \(d\) is the distance from the archived origin to the COM, expressed numerically in bohr. We treat the reference direction as resolved when the archived magnetic magnitude exceeds \(q_0\) and the COM magnetic magnitude exceeds \(3\delta_m\). If either condition fails, the magnetic specialist uses the magnitude error alone at full weight. For resolved references, the direction term remains in the loss even if the predicted magnetic moment is close to zero.

Direction comparisons use resolved reference axes and finite saved stabilized unsigned cosines, without a predicted-magnitude cutoff. All eight models have finite scores on the same reference sets; Table~\ref{tab:S7b} gives the counts for electric and magnetic directions.

For the dissymmetry score, we estimate the rounding uncertainty in the reference magnitude as the root-mean-square change in the unsigned value across 2,048 independent draws from uniform half-last-digit rounding intervals for the printed electric, magnetic and velocity components (seed 1729). Each draw propagates velocity rounding through the center-of-mass translation. For reporting, we retain reference magnitudes at least three times this estimated uncertainty. This criterion concerns printed precision, not electronic-structure accuracy.

\begin{figure}[!htbp]
\centering
\includegraphics[width=.96\linewidth]{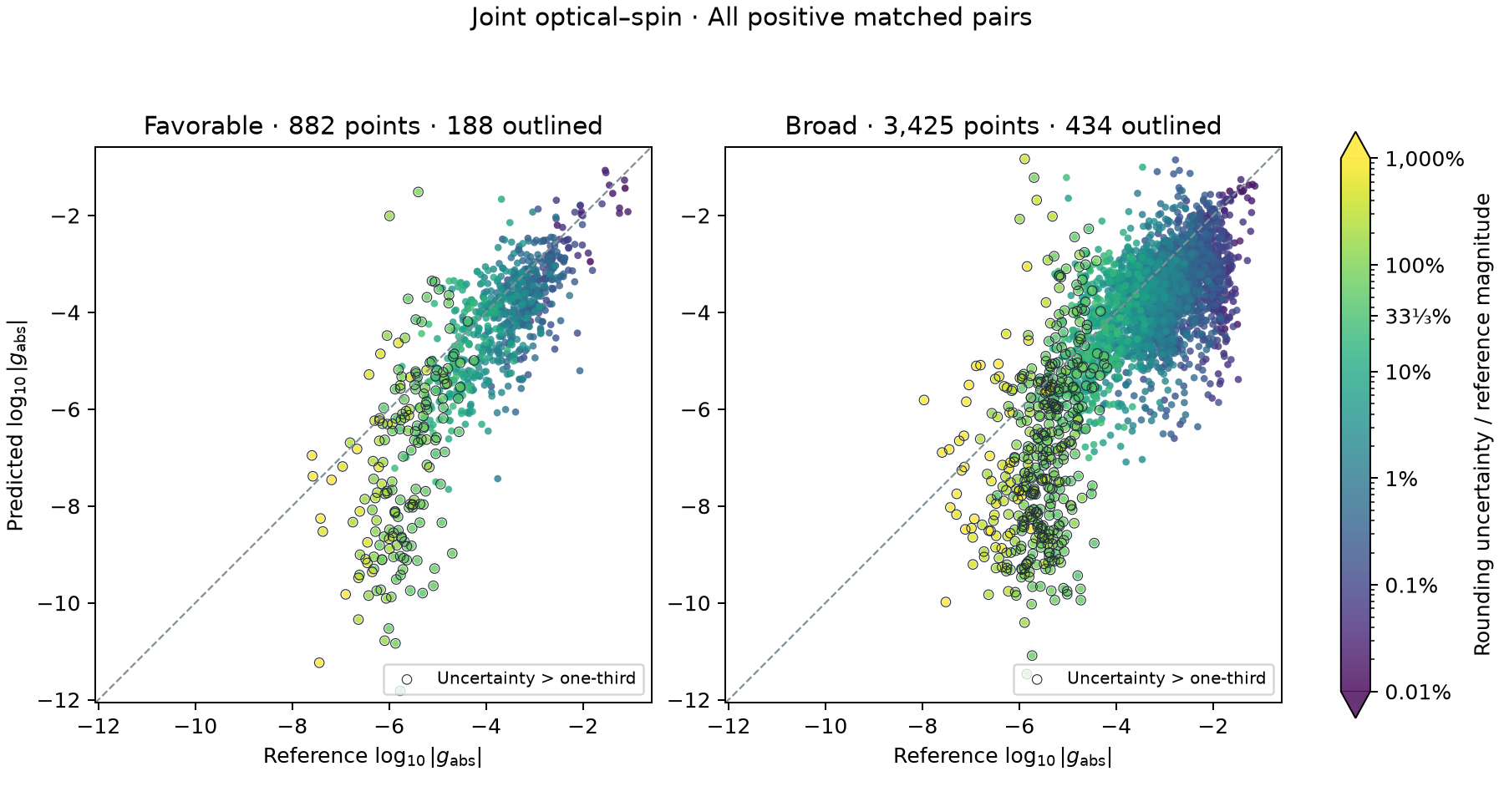}
\caption[Prescreening dissymmetry predictions]{Prescreening dissymmetry predictions for the joint optical--spin models. Points show the 882 favorable and 3,425 broad positive matched pairs. Colour gives the ratio of estimated rounding uncertainty to the reference magnitude. Outlines mark the 188 favorable and 434 broad pairs excluded by the one-third rule. Dashed lines show exact agreement.}\label{fig:S3}
\end{figure}

\FloatBarrier
\endgroup
\section{\texorpdfstring{The S\textsubscript{1} Local-\ensuremath{\pi} Character Score}{The S1 Local-pi Character Score}}\label{sec:si-local-pi-score}

The reference S\textsubscript{1} local-\ensuremath{\pi} score, \(\chi\), estimates the contribution of p orbitals perpendicular to local molecular planes from the archived FORMED~\cite{ref16} \ensuremath{\omega}B97X-D/6-31G(d) Tamm--Dancoff calculation. Its construction is independent of PPP and follows the four steps below; Table~\ref{tab:S3} gives two worked examples. The top 15\% of the 37,519 calculation-eligible molecules (\(\chi\geq89.3\%\)) form the favorable pool, whose reference-qualified members define the favorable population (Table~\ref{tab:1}).

Step 1: a local normal for each atom. The covalent radii used here match those tabulated by Cordero et al.~\citep{cordero2008} (H 0.31, C 0.76, N 0.71, O 0.66, F 0.57, S 1.05 and Cl 1.02 Å), with 0.76 Å used for all carbon atoms. We treat two atoms, hydrogens included, as bonded if their distance is less than 1.3 times the sum of their radii. For an atom A with two bonded neighbours, the normal is the normalized cross product of the two unit bond vectors \textbf{\ensuremath{\hat{b}}}\textsubscript{1} and \textbf{\ensuremath{\hat{b}}}\textsubscript{2}:

\begin{equation}
\mathbf{n}_A=\hat{\mathbf{b}}_1\times\hat{\mathbf{b}}_2\,/\,|\hat{\mathbf{b}}_1\times\hat{\mathbf{b}}_2|
\label{eq:source-40}
\end{equation}

The normal is undefined when the cross-product norm is below 0.15. With three or more neighbours, it is the right singular vector of the uncentred matrix of unit bond vectors with the smallest singular value; it is undefined if the second singular value is below 0.15 or the smallest exceeds 0.30. Atoms with one neighbour receive no normal for this score. This differs from the PPP site-axis construction, which assigns a terminal \ensuremath{\pi} site its neighbour's normal projected perpendicular to the bond.

Step 2: the \ensuremath{\pi} fraction of each orbital. For molecular orbital \ensuremath{\phi}, c\textsubscript{\ensuremath{\phi}\ensuremath{\mu}} denotes its coefficient on atomic basis function \ensuremath{\mu}. For each p shell s on atom A, the printed Cartesian coefficients form a vector c\textsubscript{\ensuremath{\phi},A,s}. Its projection onto the local normal gives the perpendicular p contribution. We sum the squared projections over atoms with defined normals and divide by the orbital's total squared coefficient weight:

\begin{equation}
f_{\phi}=\sum_{A\in\mathcal{N}}\sum_{s}\left(\mathbf{c}_{\phi,A,s}\cdot\mathbf{n}_A\right)^{2}\,/\,\sum_{\mu}c_{\phi\mu}^{2}
\label{eq:source-41}
\end{equation}

where \ensuremath{\mathcal{N}} is the set of atoms with a normal. Everything else counts only in the denominator: in-plane p weight, s and d weight, and all weight on atoms without a normal. The fraction is an index of squared coefficients, not an orbital population, since no overlap matrix enters.

Step 3: the dominant configurations. We rank the printed occupied-to-virtual S\textsubscript{1} configurations by squared amplitude and retain three to eight, stopping when they account for at least 75\% of the printed weight. The retained share has a median of 89\% across the 37,519 calculated molecules and reaches 75\% in 97.8\% of them. In 49 cases the eight-configuration limit leaves less than half of the printed weight, with a minimum of 29.5\%.

Step 4: the score. Each kept configuration k, an excitation from orbital i\textsubscript{k} to orbital a\textsubscript{k}, contributes the mean \ensuremath{\pi} fraction of its two orbitals, weighted by its squared amplitude:

\begin{equation}
\chi=\sum_{k\in K}w_{k}\,\left(f_{i_k}+f_{a_k}\right)/2,\qquad w_{k}=c_{k}^{2}\,/\sum_{k'\in K}c_{k'}^{2}
\label{eq:source-42}
\end{equation}

Here K is the retained configuration set and w\textsubscript{k} is each configuration's normalized share of its squared-amplitude weight. The score is reported as a percentage. Table~\ref{tab:S3} gives two worked examples.

\begingroup
\scriptsize
\setlength{\tabcolsep}{3pt}
\renewcommand{\arraystretch}{1.18}
\begin{table}[!htbp]
\centering
\scriptsize
\caption{The score of two FORMED molecules, configuration by configuration. Kept S\textsubscript{1} configurations in order of weight; c\textsuperscript{2} is the printed squared amplitude, the share is c\textsuperscript{2} over the kept total, and the contribution is c\textsuperscript{2}(f\textsubscript{occ} + f\textsubscript{virt})/2. The score is the sum of contributions divided by the sum of c\textsuperscript{2}. The kept configurations carry 92.4\% (KAVHUL) and 95.8\% (KEQHET) of the printed weight. The \ensuremath{\pi} fractions are rounded to three decimals, so recomputed contributions can differ in the last digit.}\label{tab:S3}
\begin{tabular}{@{}>{\raggedright\arraybackslash}p{0.1110\linewidth}>{\raggedright\arraybackslash}p{0.1110\linewidth}>{\raggedright\arraybackslash}p{0.1110\linewidth}>{\raggedright\arraybackslash}p{0.1339\linewidth}>{\raggedright\arraybackslash}p{0.1110\linewidth}>{\raggedright\arraybackslash}p{0.1110\linewidth}>{\raggedright\arraybackslash}p{0.1250\linewidth}@{}}
\toprule
Molecule & Configuration & c\textsuperscript{2} & Share of kept weight & f\textsubscript{occ} & f\textsubscript{virt} & Contribution \\
\midrule

KAVHUL & HOMO \ensuremath{\to} LUMO & 0.4364 & 94.6\% & 0.973 & 0.899 & 0.4086 \\
 & HOMO\ensuremath{-}1 \ensuremath{\to} LUMO+3 & 0.0159 & 3.4\% & 0.993 & 0.948 & 0.0154 \\
 & HOMO\ensuremath{-}1 \ensuremath{\to} LUMO & 0.0089 & 1.9\% & 0.993 & 0.899 & 0.0084 \\
 & Sum and score & 0.4612 & 100\% &  &  & 0.4324 \ensuremath{\to} \ensuremath{\chi} = 93.8\% \\
KEQHET & HOMO\ensuremath{-}1 \ensuremath{\to} LUMO & 0.4086 & 85.4\% & 0.000 & 0.675 & 0.1380 \\
 & HOMO \ensuremath{\to} LUMO & 0.0417 & 8.7\% & 0.728 & 0.675 & 0.0292 \\
 & HOMO\ensuremath{-}1 \ensuremath{\to} LUMO+1 & 0.0283 & 5.9\% & 0.000 & 0.724 & 0.0103 \\
 & Sum and score & 0.4785 & 100\% &  &  & 0.1775 \ensuremath{\to} \ensuremath{\chi} = 37.1\% \\
\bottomrule
\end{tabular}
\end{table}
\endgroup

The score describes perpendicular-p coefficient weight in the retained configurations, not excitation delocalization or an overlap-weighted orbital population. Weight on terminal atoms and d functions enters only the denominator. For KEQHET, including the terminal oxygen\textquotesingle s \ensuremath{\pi}* contribution raises \(\chi\) from 37\% to 44\%, still below the favorable threshold. Section~\ref{sec:si-screening-errors} examines prediction errors across the score.

\FloatBarrier
\section{\texorpdfstring{Structural Coverage and Reference S\textsubscript{1} Local-\ensuremath{\pi} Character}{Structural Coverage and Reference S1 Local-pi Character}}\label{sec:si-structural-coverage}

\begin{figure}[!htbp]
\centering
\includegraphics[width=\linewidth,height=.76\textheight,keepaspectratio]{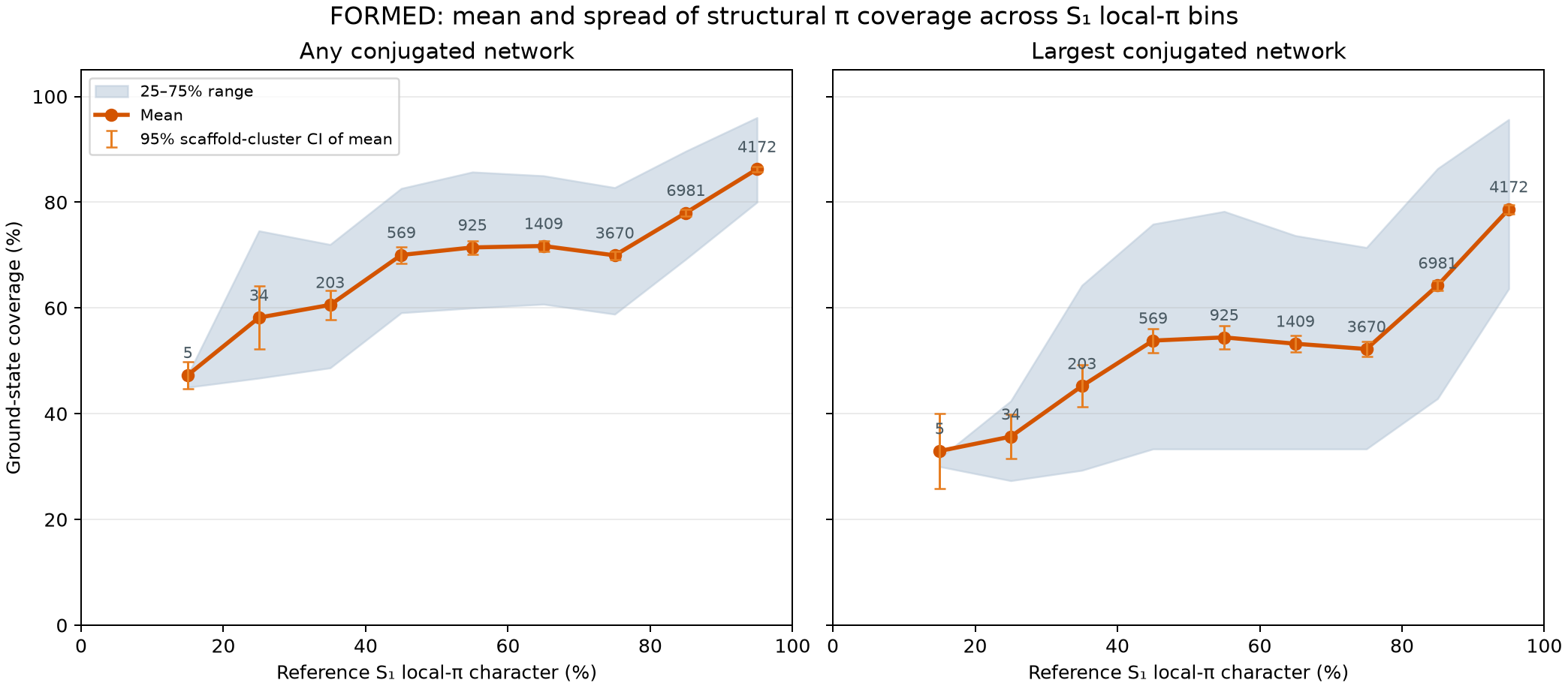}
\caption[Structural coverage and excitation character]{Structural coverage tracks S\textsubscript{1} character on average, but not molecule by molecule. Mean heavy-atom coverage of any conjugated network (left) and of the largest conjugated network (right) is shown across bins of reference S\textsubscript{1} local-\ensuremath{\pi} character. The analysis uses the 17,968 molecules that pass the first three reference gates, before the triplet gate removes seven. Orange points give molecule-weighted bin means, error bars give 95\% confidence intervals calculated as the mean \ensuremath{\pm}1.96 standard errors, with standard errors clustered by Bemis--Murcko scaffold,~\cite{bemis1996} and shading spans the middle 50\% of molecules; counts accompany each bin. Coverage is available without an excited-state calculation, but its wide within-bin spread limits its interpretation for any individual excitation.}
\label{fig:S4}
\end{figure}

Any-network coverage is the fraction of heavy atoms belonging to at least one conjugated network; largest-network coverage is the fraction belonging to the largest connected network. The former measures conjugation across the molecule, and the latter its concentration in one framework. Both are obtained from molecular structure without an excited-state calculation. Figure~\ref{fig:S4} compares them with reference S\textsubscript{1} local-\ensuremath{\pi} character.

Both measures tend to rise with S\textsubscript{1} local-\ensuremath{\pi} character, particularly at high scores, but their broad within-bin spread limits their use for identifying the character of individual excitations.

\FloatBarrier
\section{Population Selection and Reference Criteria}\label{sec:si-population-selection}

\begin{figure}[htbp]
\centering
\includegraphics[width=\linewidth,height=.76\textheight,keepaspectratio]{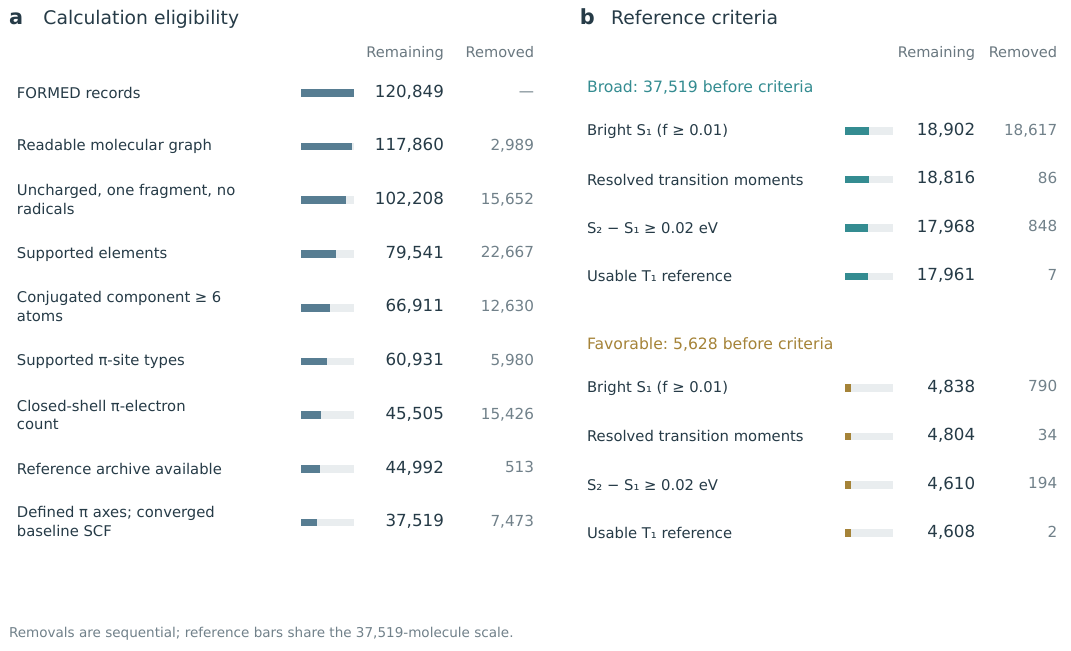}
\caption[Population-selection counts]{Population-selection counts. (a) Calculation eligibility reduces 120,849 FORMED records to 37,519 calculated cases. (b) Reference criteria are applied separately to this pool and its 5,628 highest-character molecules, yielding 17,961 broad and 4,608 favorable molecules. Removals are counted as sequential first failures; the broad and favorable reference bars share one scale. Favorable molecules also belong to the broad population. No gate uses PPP prediction error.}
\label{fig:S6}
\end{figure}

\begin{figure}[!htbp]
\centering
\includegraphics[width=\linewidth,height=.76\textheight,keepaspectratio]{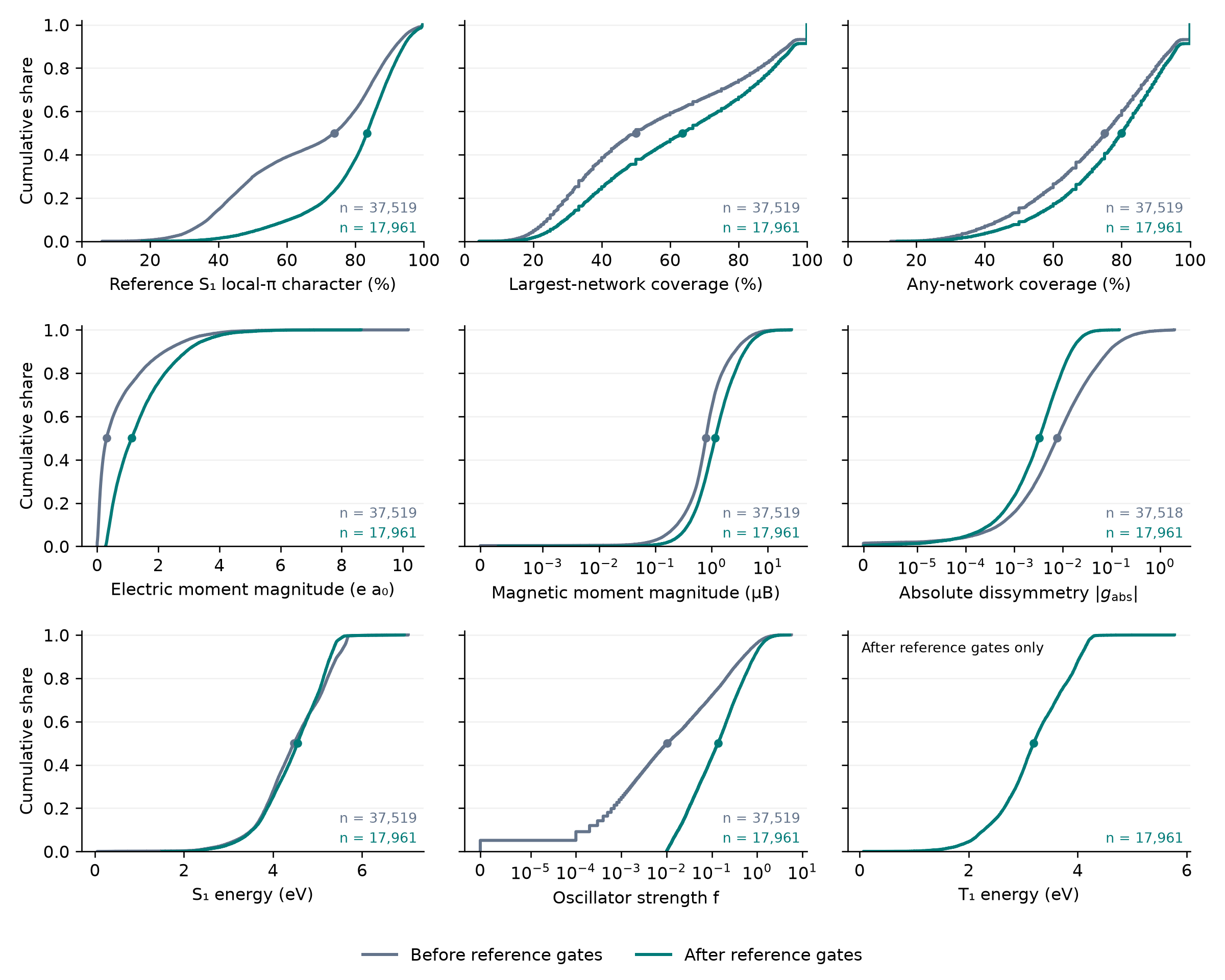}
\caption[Effects of the reference criteria]{Effect of the reference criteria on cohort composition. Each panel shows the cumulative distribution of available reference values. The comparison uses 37,519 calculation-eligible molecules before the reference criteria and the 17,961 molecules in the broad population afterward. Dots mark medians, and counts give the number of finite values. Coverage measures are fractions of heavy atoms. The T\textsubscript{1} panel shows values after the reference gates. The archived reference table does not cover the pre-gate cohort.}
\label{fig:S7}
\end{figure}

\siadded{Figure~\ref{fig:S6} records the selection counts, and Fig.~\ref{fig:S7} shows how the reference criteria change the population distributions. Eligibility requires defined local \(\pi\) axes, not a common molecular plane. Among the 37,519 calculation-eligible molecules, the reference criteria raise median S\textsubscript{1} local-\(\pi\) character from 73.8\% to 83.4\%, largest-network coverage from 50.0\% to 63.6\%, and any-network coverage from 75.0\% to 80.0\%.}

\FloatBarrier
\Needspace{7\baselineskip}
\siadded{At the closed-shell \(\pi\)-electron-count filter (Fig.~\ref{fig:S6}a), the screening code missed hydrogens when assigning atom types, undercounting \(\pi\) electrons on some nitrogen and oxygen sites. This incorrectly excluded some otherwise eligible closed-shell molecules whose assigned electron count became odd. The error restricted benchmark coverage; production calculations used explicit-hydrogen geometries for atom typing, so the defect concerns selection rather than the calculations on included molecules.}

\FloatBarrier
\section{Training and Checkpoint Selection}\label{sec:si-training-specification}

\subsection{Loss Normalization}\label{sec:si-loss-normalization}

\Needspace{7\baselineskip}
\siadded{Each target is normalized by fixed PPP's RMS error on the training partition:}

\begin{equation}
s_k=\max\left(\sqrt{\frac{1}{N_{\mathrm{train},k}}\sum_{n\in\mathrm{train}_k}\left(y^{\mathrm{PPP}}_{n,k}-y_{n,k}\right)^2},\;\varepsilon_k\right)
\label{eq:source-43}
\end{equation}

\siadded{Here, \(N_{\mathrm{train},k}\) counts training molecules with finite reference and fixed-PPP values for target \(k\); direction scales additionally require resolved reference axes. Scales are calculated separately for each population and target and remain fixed throughout training and validation. Validation and test data do not enter their calculation.}

\siadded{The floors \(\varepsilon_k\) are 0.05 eV for S\textsubscript{1}, 0.1 eV for T\textsubscript{1} and the signed gap, 10\textsuperscript{-3} for oscillator strength, 0.02 \(e a_0\) for electric magnitude, 0.02 \(\mu_{\mathrm B}\) for magnetic magnitude, 10\textsuperscript{-5} \((e a_0)^2\) for unsigned rotatory strength, 10\textsuperscript{-4} for unsigned dissymmetry, and 0.02 for direction cosines. All training-derived scales exceed these floors, so none is active.}

\siadded{The specialist and joint objectives follow Section~\ref{sec:learning-objective}. For magnetic directions, the training scales are 0.3532 and 0.4262 in the favorable and broad populations, respectively, calculated from 2,572 and 10,802 resolved reference axes. The remaining 402 and 725 training molecules retain magnitude supervision alone.}

\siadded{Every training objective includes an additional penalty of \(10^{-3}\langle z^2\rangle\), where the average is over scalar readout outputs. Since zero outputs recover fixed PPP, this penalty discourages unnecessary departures from the baseline. This penalty contributes to the training loss but not to the validation score used to select the checkpoint.}

\subsection{Optimization}\label{sec:si-differentiation-optimization}

\siadded{Only the feature projections and KAN readouts are optimized. The implicit RHF response is obtained through an adjoint equation solved by GMRES~\cite{saad1986gmres}; differentiation through CIS and the observables follows Section~\ref{sec:learning-objective}.}

\begingroup
\footnotesize
\setlength{\tabcolsep}{3pt}
\renewcommand{\arraystretch}{1.18}
\begin{table}[!htbp]
\centering
\footnotesize
\caption{\siadded{Shared optimization settings for all fourteen models.}}\label{tab:S5}
\begin{tabular}{@{}>{\raggedright\arraybackslash}p{0.5521\linewidth}>{\raggedright\arraybackslash}p{0.3519\linewidth}@{}}
\toprule
Setting & Value \\
\midrule

Optimizer & Adam \\
Learning rate & Constant, 10\textsuperscript{-3} \\
Training duration & 20 epochs \\
Batch size & 12 \\
Random seed & 17 \\
Gradient-norm clipping & 10 \\
\bottomrule
\end{tabular}
\end{table}
\endgroup

\siadded{The Adam optimizer~\cite{kingma2015adam} receives no update from a batch if its forward or response calculation fails, or if its gradients are nonfinite. Failed validation and test calculations are handled separately.}

\subsection{Checkpoint Selection}\label{sec:si-validation-trajectories}

\siadded{Checkpoints are selected using the validation criterion described in Section~\ref{sec:learning-objective}. For joint models, the logarithmic dissymmetry residual uses training-derived scales of 0.724 dex for the favorable population and 0.991 dex for the broad population, calculated with the same 10\textsuperscript{-5} floor applied during validation.}

\siadded{Each failed validation calculation contributes a score of 10 and remains in the full validation denominator. The readout-output penalty is excluded from this score. Section~\ref{sec:si-failed-calculations} gives the test-calculation rejection criteria and resulting comparison sets.}

\FloatBarrier
\section{Failed Test Calculations and Comparison Sizes}\label{sec:si-failed-calculations}

The comparisons in Sections~\ref{sec:favorable-regime}--\ref{sec:broad-regime} require accepted predictions from every model on the same molecules. Table~\ref{tab:S7} records calculation failures and the resulting scalar, direction and dissymmetry comparison sets.

\subsection{Failures and Comparison Sizes}\label{sec:si-comparison-sizes}

RHF is attempted with density-damping factors of 0.25, 0.65 and 0.85 in sequence, allowing up to 1,000 iterations per attempt and stopping at the first successful convergence. A calculation is classified as a self-consistent-field (SCF) failure only if all three attempts fail to converge. Test calculations are also rejected if the two lowest CIS roots within either spin sector differ by less than 10\textsuperscript{-8} eV. Table~\ref{tab:S7a} reports these failures by model; Table~\ref{tab:S7b} gives the common evaluation sets used in Sections~\ref{sec:favorable-regime} and \ref{sec:broad-regime}.

All learned-model test failures arise from lowest-root degeneracy and occur among the fixed-PPP exclusions in Table~\ref{tab:S7a}. No additional learned-model SCF failures occur in these evaluations.

\begingroup
\footnotesize
\setlength{\tabcolsep}{3pt}
\renewcommand{\arraystretch}{1.18}
\begin{table}[!htbp]
\centering
\footnotesize
\begin{tabular}{@{}>{\raggedright\arraybackslash}p{0.1357\linewidth}>{\raggedright\arraybackslash}p{0.1689\linewidth}>{\raggedright\arraybackslash}p{0.1583\linewidth}>{\raggedright\arraybackslash}p{0.1741\linewidth}>{\raggedright\arraybackslash}p{0.2130\linewidth}@{}}
\toprule
Population & Model & Failed / registered & Failure type & Failed molecules \\
\midrule

Favorable & Fixed PPP & 2 / 886 & Triplet degeneracy & CEXQOL, HAVSIH \\
Broad & Fixed PPP & 4 / 3,436 & Lowest-root degeneracy & DOBVAR, JOFHUI, SAXYIZ, VEXGUA \\
Favorable & S\textsubscript{1} specialist & 1 / 886 & Triplet degeneracy & CEXQOL \\
Favorable & T\textsubscript{1} specialist & 1 / 886 & Triplet degeneracy & CEXQOL \\
Favorable & Oscillator specialist & 0 / 886 & None & --- \\
Favorable & Electric specialist & 0 / 886 & None & --- \\
Favorable & Magnetic specialist & 0 / 886 & None & --- \\
Favorable & Joint optical & 1 / 886 & Triplet degeneracy & CEXQOL \\
Favorable & Joint optical--spin & 1 / 886 & Triplet degeneracy & CEXQOL \\
Broad & S\textsubscript{1} specialist & 2 / 3,436 & Triplet degeneracy & JOFHUI, SAXYIZ \\
Broad & T\textsubscript{1} specialist & 2 / 3,436 & Triplet degeneracy & JOFHUI, SAXYIZ \\
Broad & Oscillator specialist & 2 / 3,436 & Triplet degeneracy & JOFHUI, SAXYIZ \\
Broad & Electric specialist & 2 / 3,436 & Triplet degeneracy & JOFHUI, SAXYIZ \\
Broad & Magnetic specialist & 2 / 3,436 & Triplet degeneracy & JOFHUI, SAXYIZ \\
Broad & Joint optical & 2 / 3,436 & Triplet degeneracy & JOFHUI, SAXYIZ \\
Broad & Joint optical--spin & 2 / 3,436 & Triplet degeneracy & JOFHUI, SAXYIZ \\
\bottomrule
\end{tabular}
\refstepcounter{table}\label{tab:S7}
\addcontentsline{lot}{table}{\protect\numberline{\thetable}Failed test calculations and evaluation populations (parts a--b)}
\refstepcounter{tablepart}\label{tab:S7a}
\caption*{Table~\ref{tab:S7a}. Failed test calculations by model. Failures are counted against the registered favorable and broad test partitions. The affected molecular structures appear in Fig.~\ref{fig:S12}. Fixed-PPP entries refer to the joint spin calculation; optical predictions can remain available. DOBVAR is rejected for singlet-root degeneracy; the other fixed-PPP cases have triplet-root degeneracy.}
\end{table}
\endgroup

The common scalar set requires accepted finite predictions from fixed PPP, all five specialists and both joint models. A failure in any required calculation excludes that molecule from every scalar model comparison, without changing the registered test partition. Finite nonpositive T\textsubscript{1} predictions remain in the error statistics. Direction and dissymmetry comparisons apply the additional reference-resolution criteria of Section~\ref{sec:si-magnetic-resolution}.

The rounding screen retains 694 of the 884 favorable molecules and 2,991 of the 3,432 broad molecules for both native and logarithmic dissymmetry comparisons. It excludes 190 favorable and 441 broad cases, including two and seven zero references, respectively. These exclusions are separate from calculation failures.

\begingroup
\footnotesize
\setlength{\tabcolsep}{3pt}
\renewcommand{\arraystretch}{1.18}
\begin{table}[!htbp]
\centering
\footnotesize
\begin{tabular}{@{}>{\raggedright\arraybackslash}p{0.3722\linewidth}>{\raggedright\arraybackslash}p{0.2766\linewidth}>{\raggedright\arraybackslash}p{0.2372\linewidth}@{}}
\toprule
Properties evaluated & Favorable molecules & Broad molecules \\
\midrule

Energies, transition magnitudes, oscillator strength & 884 & 3,432 \\
Native and logarithmic unsigned dissymmetry & 694 & 2,991 \\
Resolved electric reference axes within the common scalar set & 884 & 3,432 \\
Resolved COM magnetic reference axes within the common scalar set & 771 & 3,151 \\
\bottomrule
\end{tabular}
\refstepcounter{tablepart}\label{tab:S7b}
\caption*{Table~\ref{tab:S7b}. Evaluation populations used throughout the model comparisons. The common scalar sets retain 99.8\% of the favorable and 99.9\% of the broad registered partitions. The dissymmetry row gives counts after applying the same rounding screen to all models. Reference-axis rows give the eligible references for the saved stabilized direction scores. The registered test partitions contain 773 favorable and 3,153 broad resolved magnetic reference axes. Within the common scalar sets, 113 favorable and 281 broad magnetic reference axes are unresolved; their scalar predictions remain included. Calculation failures overlap across models, so Table~\ref{tab:S7a} must not be summed to obtain exclusions.}
\end{table}
\endgroup

\FloatBarrier

\subsection{Molecular Structures of Failed Cases}\label{sec:si-failed-case-structures}

\begin{figure}[!htbp]
\centering
\includegraphics[width=\linewidth,height=.76\textheight,keepaspectratio]{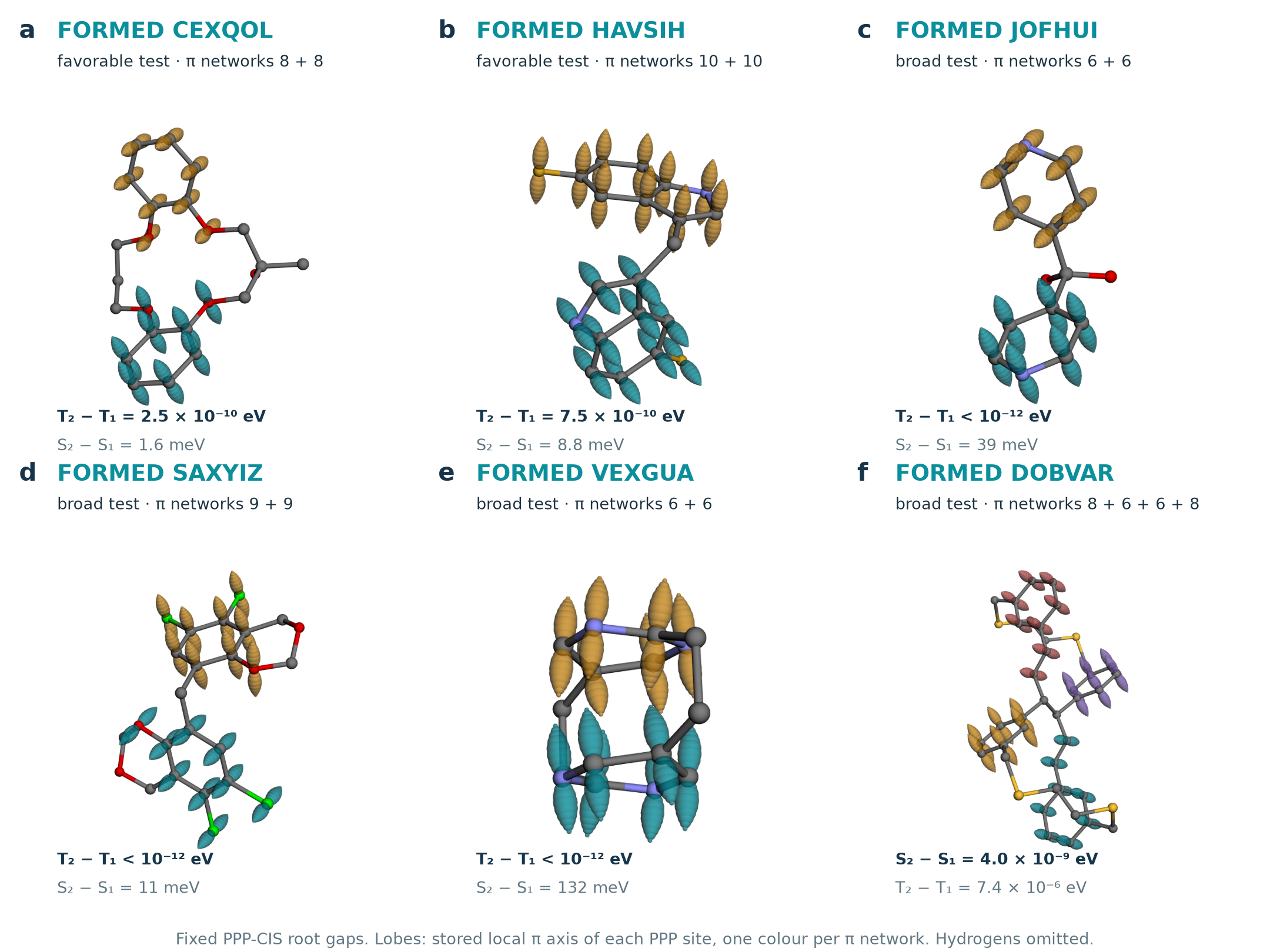}
\caption[Structures of degenerate test cases]{Test molecules excluded by the fixed-PPP lowest-root degeneracy criterion. Heavy atoms are shown with two-lobed glyphs along the local \ensuremath{\pi} axes of PPP sites; colours distinguish \ensuremath{\pi} networks. Bold annotations give the gap between the nearly degenerate roots; grey annotations give the corresponding gap in the other spin sector. Panels (a--b) show favorable test molecules and (c--f) broad test molecules. The degeneracy occurs in the singlet sector for DOBVAR (f) and in the triplet sector for the other five molecules.}
\label{fig:S12}
\end{figure}

The six fixed-PPP degeneracy cases in Fig.~\ref{fig:S12} are listed in Table~\ref{tab:S7a}. Each contains repeated identical \ensuremath{\pi} systems that are not conjugated with one another. Within PPP, these equivalent fragments can produce degenerate excited states; the rejected roots are triplets in five molecules and singlets in DOBVAR.

The learned-model failures are confined to three of these molecules: CEXQOL, JOFHUI and SAXYIZ. Table~\ref{tab:S7a} identifies the affected models.

\FloatBarrier
\section{Joint-Model Predictions and the Role of Gap Supervision}\label{sec:si-optical-only-training}

{The joint optical and joint optical--spin models use the same training protocol, with checkpoints selected separately by their validation objectives. The optical-only objective averages the five optical groups, while the optical--spin objective also includes the signed singlet--triplet gap.}

\begin{figure}[!htbp]
\centering
\includegraphics[width=\linewidth,height=.67\textheight,keepaspectratio]{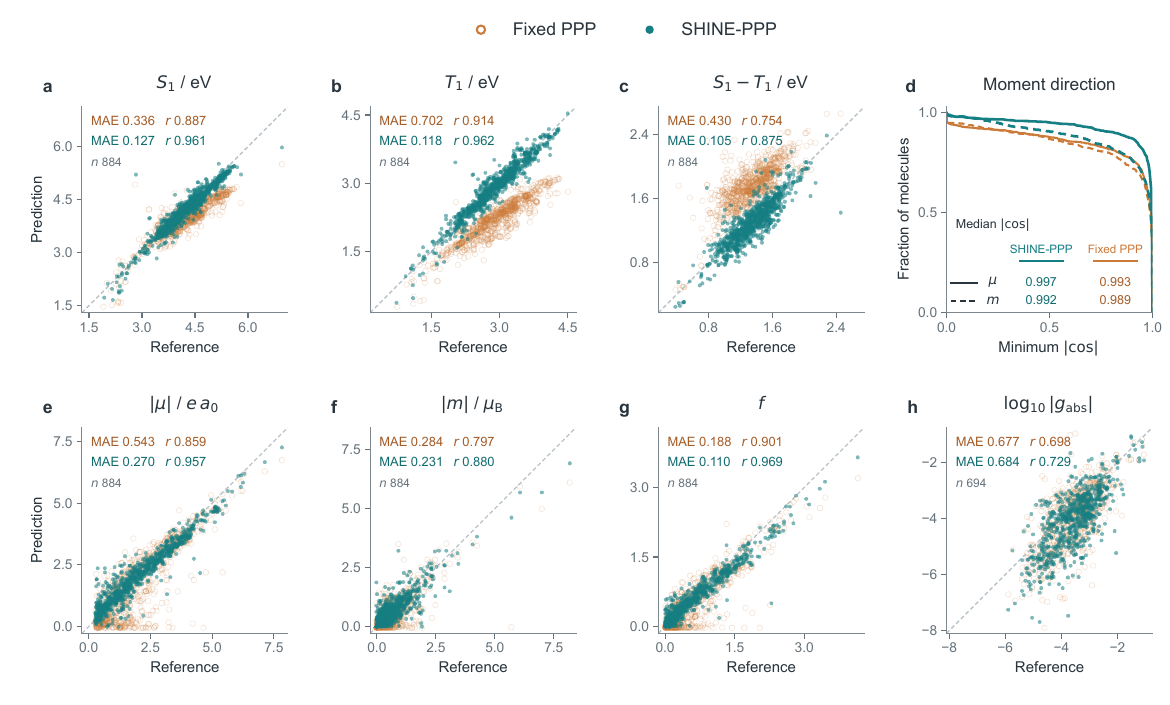}
\caption[Favorable joint optical–spin predictions]{{Favorable joint optical--spin predictions. Panels compare \(\Delta E(\mathrm{S}_1)\), \(\Delta E(\mathrm{T}_1)\) and \(\Delta E_{\mathrm{ST}}\) (a--c), and \(\lVert\boldsymbol{\mu}\rVert\), \(\lVert\mathbf{m}\rVert\), \(f\) and \(\log_{10}|g_{\mathrm{abs}}|\) (e--h), with their references. Teal filled points show SHINE-PPP and orange open circles show fixed PPP on the same molecules. Matching annotations give each method's MAE and Pearson \(r\); \(n\) is the comparison count, and dashed diagonals mark exact agreement. Panel d compares unsigned direction agreement: teal curves show SHINE-PPP and orange curves show fixed PPP; solid lines indicate \(\boldsymbol{\mu}\) and dashed lines indicate \(\mathbf{m}\). Its table gives the median absolute direction cosines. Scalar comparisons use 884 molecules, except logarithmic unsigned dissymmetry, which uses the 694 passing the reference-rounding screen. Direction curves use 884 electric and 771 magnetic references whose axes are resolved.}}
\label{fig:S13}
\end{figure}

\begin{figure}[!htbp]
\centering
\includegraphics[width=\linewidth,height=.67\textheight,keepaspectratio]{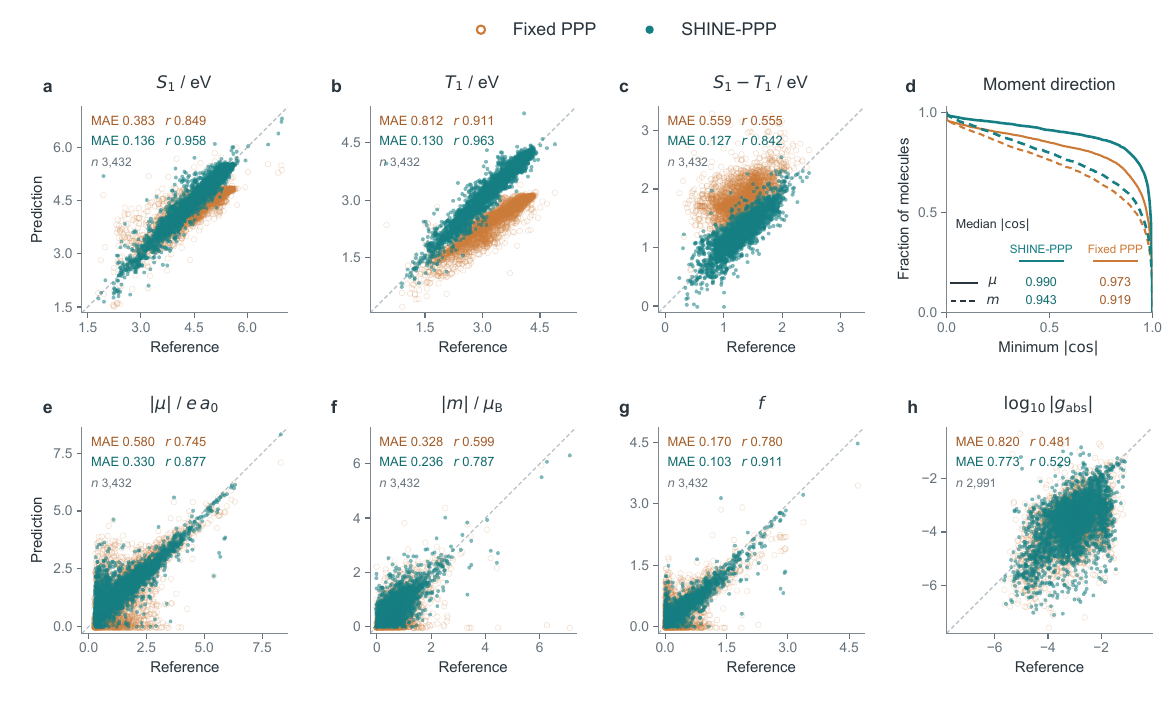}
\caption[Broad joint optical–spin predictions]{{Broad joint optical--spin predictions. Panels compare \(\Delta E(\mathrm{S}_1)\), \(\Delta E(\mathrm{T}_1)\) and \(\Delta E_{\mathrm{ST}}\) (a--c), and \(\lVert\boldsymbol{\mu}\rVert\), \(\lVert\mathbf{m}\rVert\), \(f\) and \(\log_{10}|g_{\mathrm{abs}}|\) (e--h), with their references. Teal filled points show SHINE-PPP and orange open circles show fixed PPP on the same molecules. Matching annotations give each method's MAE and Pearson \(r\); \(n\) is the comparison count, and dashed diagonals mark exact agreement. Panel d compares unsigned direction agreement: teal curves show SHINE-PPP and orange curves show fixed PPP; solid lines indicate \(\boldsymbol{\mu}\) and dashed lines indicate \(\mathbf{m}\). Its table gives the median absolute direction cosines. Scalar comparisons use 3,432 molecules, except logarithmic unsigned dissymmetry, which uses the 2,991 passing the reference-rounding screen. Direction curves use 3,432 electric and 3,151 magnetic references whose axes are resolved.}}
\label{fig:S14}
\end{figure}

{Figures~\ref{fig:S13}--\ref{fig:S14} show the individual joint optical--spin predictions for the favorable and broad populations, including transition-moment direction agreement. Among resolved comparisons, the median unsigned direction cosines are 0.997 for \(\boldsymbol{\mu}\) and 0.992 for \(\mathbf{m}\) in the favorable population, and 0.990 and 0.943, respectively, in the broad population. These comparisons measure alignment with the reference axes without distinguishing parallel from antiparallel vectors.}

\begin{figure}[!htbp]
\centering
\includegraphics[width=\linewidth,height=.67\textheight,keepaspectratio]{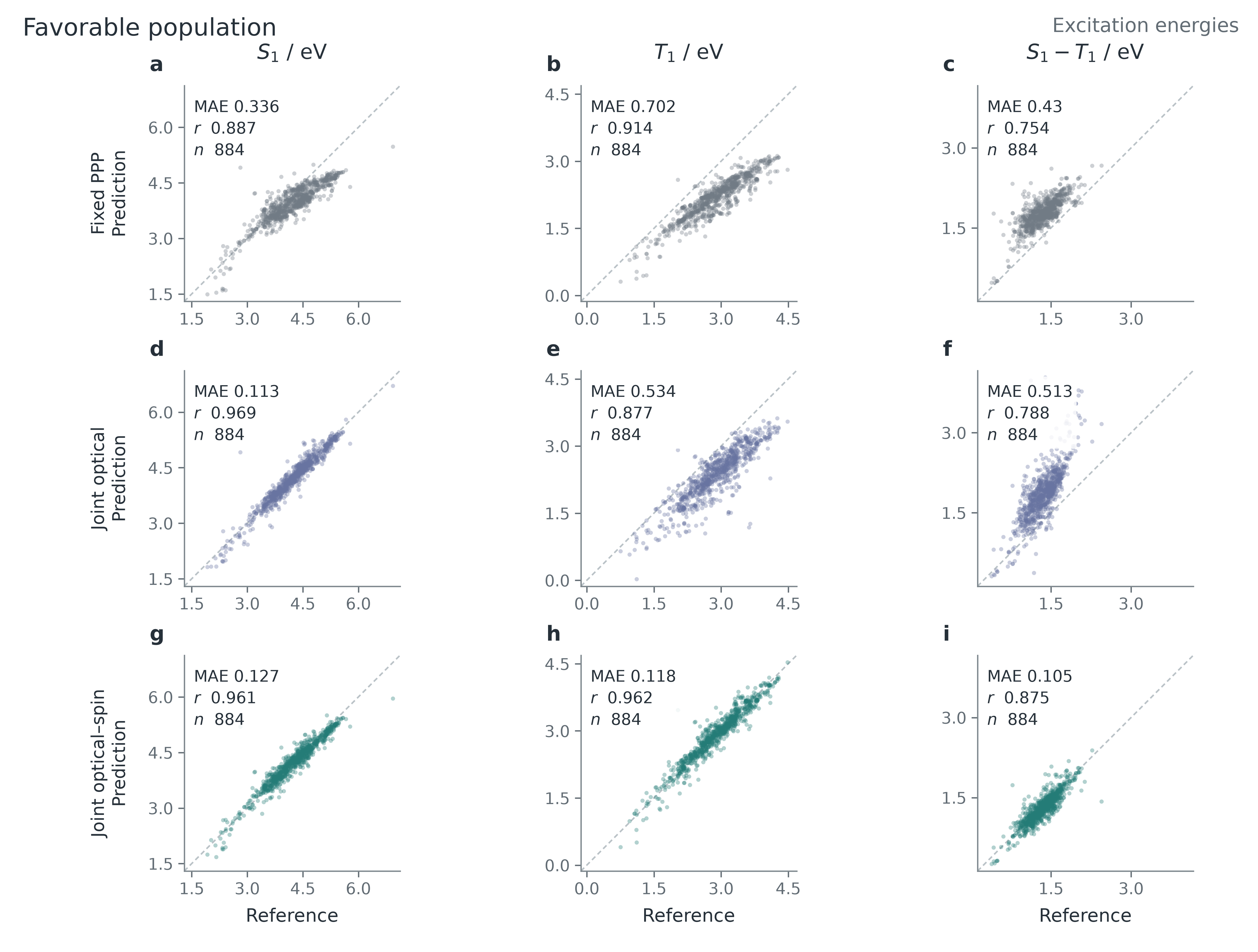}
\caption[Favorable energy predictions]{{Effect of gap supervision on favorable energy predictions. Rows compare fixed PPP, joint optical and joint optical--spin predictions for \(\Delta E(\mathrm{S}_1)\), \(\Delta E(\mathrm{T}_1)\) and \(\Delta E_{\mathrm{ST}}\) on the common 884-molecule evaluation set. Axes span the full prediction range and are shared within each property. Neither learned joint model produces a nonpositive T\textsubscript{1} prediction in this set.}}
\label{fig:S15}
\end{figure}

\begin{figure}[!htbp]
\centering
\includegraphics[width=\linewidth,height=.67\textheight,keepaspectratio]{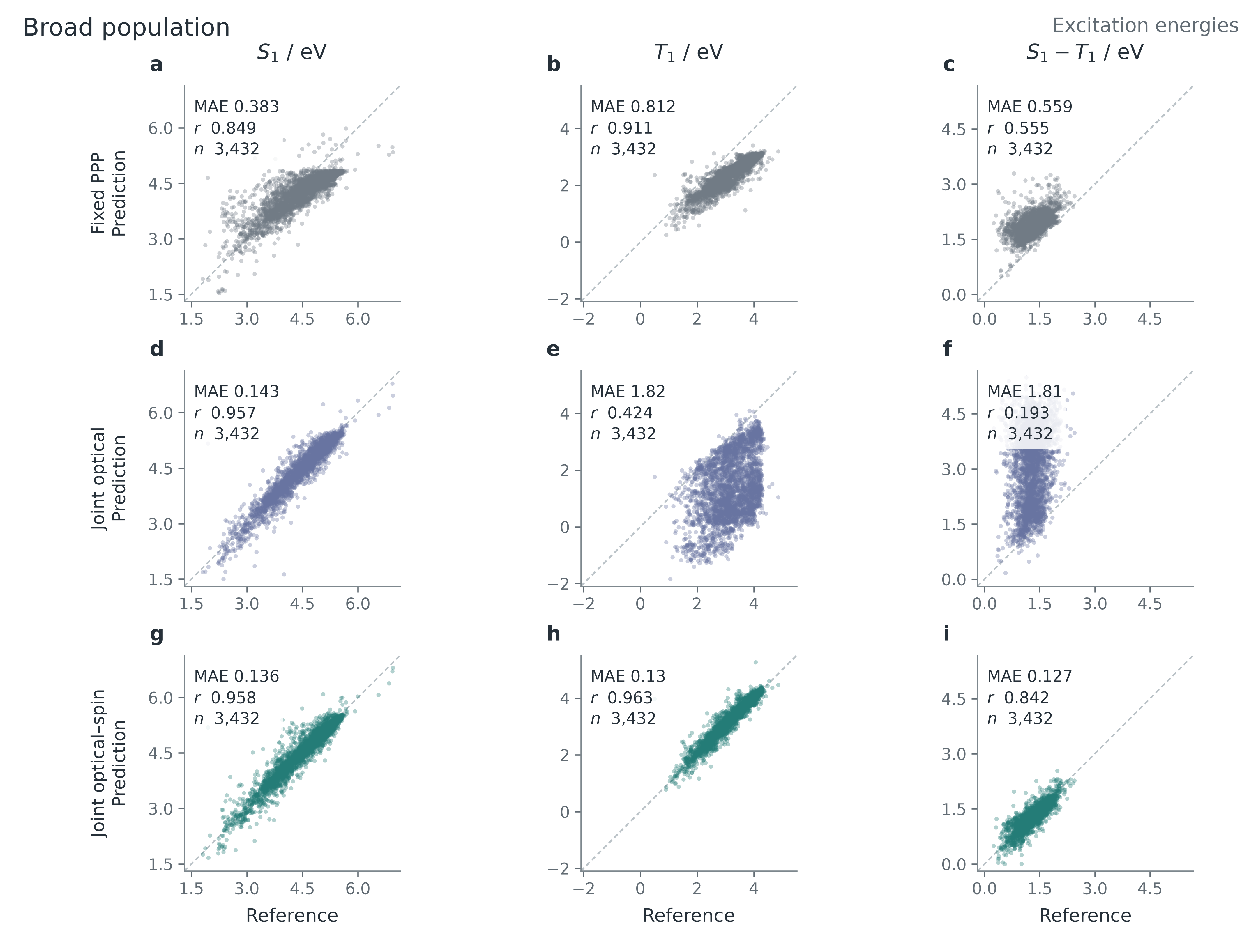}
\caption[Broad energy predictions]{{Effect of gap supervision on broad energy predictions. Rows compare fixed PPP and the two joint models on the common 3,432-molecule evaluation set. The optical-only model's 291 nonpositive T\textsubscript{1} predictions are retained in the plotted points and errors; the optical--spin model produces none in this set. Axes are shared within each property.}}
\label{fig:S16}
\end{figure}

{The strongest effect of gap supervision is on triplet energetics (Figs.~\ref{fig:S15}--\ref{fig:S16}). T\textsubscript{1} MAE decreases from 0.534 to 0.118 eV in the favorable population and from 1.823 to 0.130 eV in the broad population; the corresponding signed-gap errors decrease from 0.513 to 0.105 eV and from 1.806 to 0.127 eV. The broad optical-only model produces 291 nonpositive T\textsubscript{1} predictions, compared with none for optical--spin training. Neither favorable joint model produces a nonpositive T\textsubscript{1} prediction. These finite predictions remain included in the plots and error calculations; they are not calculation failures.}

\FloatBarrier

{The following section provides the full comparison across fixed PPP, both joint models and all five specialists (Tables~\ref{tab:S8}--\ref{tab:S11}).}

\FloatBarrier
\section{Full Cross-Property Comparisons}\label{sec:si-cross-property-comparisons}

These tables bring together the scalar-property MAEs and Pearson correlations for fixed PPP, both joint models and all five specialists, supporting the accuracy and cross-property comparisons in Sections~\ref{sec:favorable-regime} and \ref{sec:broad-regime}.

\begingroup
\scriptsize
\setlength{\tabcolsep}{3pt}
\renewcommand{\arraystretch}{1.18}
\begin{table}[!htbp]
\centering
\scriptsize
\caption{Favorable all-model mean absolute errors. All models use the same 884 molecules, except native and logarithmic unsigned dissymmetry (694 molecules). Magnetic and derived quantities use the full-molecule centre of mass. Bold entries identify the lowest MAE in each column, using unrounded values.}\label{tab:S8}
\begin{tabular}{@{}>{\raggedright\arraybackslash}p{0.1276\linewidth}>{\raggedright\arraybackslash}p{0.0813\linewidth}>{\raggedright\arraybackslash}p{0.0813\linewidth}>{\raggedright\arraybackslash}p{0.0813\linewidth}>{\raggedright\arraybackslash}p{0.0813\linewidth}>{\raggedright\arraybackslash}p{0.0813\linewidth}>{\raggedright\arraybackslash}p{0.0813\linewidth}>{\raggedright\arraybackslash}p{0.0813\linewidth}>{\raggedright\arraybackslash}p{0.0813\linewidth}@{}}
\toprule
Model & S\textsubscript{1} (eV) & T\textsubscript{1} (eV) & Gap (eV) & Electric (e a\textsubscript{0}) & Magnetic (\ensuremath{\mu}B) & f & Native \textbar{}g\textbar{} & log\textsubscript{10} \textbar{}g\textbar{} (dex) \\
\midrule

Fixed PPP & 0.336 & 0.702 & 0.430 & 0.543 & 0.284 & 0.188 & 0.00138 & 0.677 \\
Joint optical & 0.113 & 0.534 & 0.513 & \textbf{0.243} & \textbf{0.204} & 0.095 & 0.00123 & \textbf{0.645} \\
Joint optical--spin & 0.127 & 0.118 & 0.105 & 0.270 & 0.231 & 0.110 & 0.00123 & 0.684 \\
S\textsubscript{1} specialist & \textbf{0.060} & 0.121 & 0.122 & 0.380 & 0.303 & 0.156 & 0.00134 & 0.658 \\
T\textsubscript{1} specialist & 0.109 & \textbf{0.041} & \textbf{0.098} & 0.409 & 0.319 & 0.164 & 0.00161 & 0.687 \\
Oscillator specialist & 0.584 & 0.744 & 0.217 & 0.271 & 0.226 & \textbf{0.075} & \textbf{0.00109} & 0.689 \\
Electric + direction specialist & 0.377 & 0.716 & 0.360 & 0.256 & 0.233 & 0.122 & 0.00221 & 0.712 \\
Magnetic + direction specialist & 0.483 & 1.099 & 0.635 & 0.369 & 0.224 & 0.156 & 0.00179 & 0.733 \\
\bottomrule
\end{tabular}
\end{table}
\endgroup

\begingroup
\scriptsize
\setlength{\tabcolsep}{3pt}
\renewcommand{\arraystretch}{1.18}
\begin{table}[!htbp]
\centering
\scriptsize
\caption{Favorable all-model Pearson correlations. All models use the same 884 molecules, except native and logarithmic unsigned dissymmetry (694 molecules). Magnetic and derived quantities use the full-molecule centre of mass. Bold entries identify the highest correlation in each column, using unrounded values.}\label{tab:S9}
\begin{tabular}{@{}>{\raggedright\arraybackslash}p{0.1276\linewidth}>{\raggedright\arraybackslash}p{0.0813\linewidth}>{\raggedright\arraybackslash}p{0.0813\linewidth}>{\raggedright\arraybackslash}p{0.0813\linewidth}>{\raggedright\arraybackslash}p{0.0813\linewidth}>{\raggedright\arraybackslash}p{0.0813\linewidth}>{\raggedright\arraybackslash}p{0.0813\linewidth}>{\raggedright\arraybackslash}p{0.0813\linewidth}>{\raggedright\arraybackslash}p{0.0813\linewidth}@{}}
\toprule
Model & S\textsubscript{1} (eV) & T\textsubscript{1} (eV) & Gap (eV) & Electric (e a\textsubscript{0}) & Magnetic (\ensuremath{\mu}B) & f & Native \textbar{}g\textbar{} & log\textsubscript{10} \textbar{}g\textbar{} (dex) \\
\midrule

Fixed PPP & 0.887 & 0.914 & 0.754 & 0.859 & 0.797 & 0.901 & 0.486 & 0.698 \\
Joint optical & 0.969 & 0.877 & 0.788 & 0.962 & \textbf{0.895} & 0.973 & 0.475 & \textbf{0.747} \\
Joint optical--spin & 0.961 & 0.962 & 0.875 & 0.957 & 0.880 & 0.969 & 0.646 & 0.729 \\
S\textsubscript{1} specialist & \textbf{0.990} & 0.965 & \textbf{0.897} & 0.952 & 0.857 & 0.966 & 0.509 & 0.737 \\
T\textsubscript{1} specialist & 0.965 & \textbf{0.994} & 0.869 & 0.936 & 0.828 & 0.954 & 0.230 & 0.717 \\
Oscillator specialist & 0.969 & 0.908 & 0.807 & \textbf{0.971} & 0.872 & \textbf{0.983} & \textbf{0.708} & 0.742 \\
Electric + direction specialist & 0.949 & 0.854 & 0.757 & 0.960 & 0.863 & 0.968 & 0.228 & 0.725 \\
Magnetic + direction specialist & 0.895 & 0.572 & 0.578 & 0.909 & 0.869 & 0.929 & 0.164 & 0.680 \\
\bottomrule
\end{tabular}
\end{table}
\endgroup

\begingroup
\scriptsize
\setlength{\tabcolsep}{3pt}
\renewcommand{\arraystretch}{1.18}
\begin{table}[!htbp]
\centering
\scriptsize
\caption{Broad all-model mean absolute errors. All models use the same 3,432 molecules, except native and logarithmic unsigned dissymmetry (2,991 molecules). Magnetic and derived quantities use the full-molecule centre of mass. Bold entries identify the lowest MAE in each column, using unrounded values.}\label{tab:S10}
\begin{tabular}{@{}>{\raggedright\arraybackslash}p{0.1276\linewidth}>{\raggedright\arraybackslash}p{0.0813\linewidth}>{\raggedright\arraybackslash}p{0.0813\linewidth}>{\raggedright\arraybackslash}p{0.0813\linewidth}>{\raggedright\arraybackslash}p{0.0813\linewidth}>{\raggedright\arraybackslash}p{0.0813\linewidth}>{\raggedright\arraybackslash}p{0.0813\linewidth}>{\raggedright\arraybackslash}p{0.0813\linewidth}>{\raggedright\arraybackslash}p{0.0813\linewidth}@{}}
\toprule
Model & S\textsubscript{1} (eV) & T\textsubscript{1} (eV) & Gap (eV) & Electric (e a\textsubscript{0}) & Magnetic (\ensuremath{\mu}B) & f & Native \textbar{}g\textbar{} & log\textsubscript{10} \textbar{}g\textbar{} (dex) \\
\midrule

Fixed PPP & 0.383 & 0.812 & 0.559 & 0.580 & 0.328 & 0.170 & 0.00305 & 0.820 \\
Joint optical & 0.143 & 1.823 & 1.806 & \textbf{0.319} & \textbf{0.236} & 0.097 & 0.00303 & \textbf{0.770} \\
Joint optical--spin & 0.136 & 0.130 & \textbf{0.127} & 0.330 & 0.236 & 0.103 & \textbf{0.00270} & 0.773 \\
S\textsubscript{1} specialist & \textbf{0.098} & 0.278 & 0.272 & 0.540 & 0.323 & 0.184 & 0.00326 & 0.789 \\
T\textsubscript{1} specialist & 0.287 & \textbf{0.056} & 0.271 & 0.604 & 0.339 & 0.192 & 0.00357 & 0.827 \\
Oscillator specialist & 0.927 & 1.389 & 0.481 & 0.409 & 0.265 & \textbf{0.097} & 0.00315 & 0.857 \\
Electric + direction specialist & 0.200 & 0.661 & 0.728 & 0.320 & 0.252 & 0.107 & 0.00292 & 0.822 \\
Magnetic + direction specialist & 0.663 & 1.295 & 0.676 & 0.530 & 0.241 & 0.152 & 0.00300 & 0.856 \\
\bottomrule
\end{tabular}
\end{table}
\endgroup

\begingroup
\scriptsize
\setlength{\tabcolsep}{3pt}
\renewcommand{\arraystretch}{1.18}
\begin{table}[!htbp]
\centering
\scriptsize
\caption{Broad all-model Pearson correlations. All models use the same 3,432 molecules, except native and logarithmic unsigned dissymmetry (2,991 molecules). Magnetic and derived quantities use the full-molecule centre of mass. Bold entries identify the highest correlation in each column, using unrounded values.}\label{tab:S11}
\begin{tabular}{@{}>{\raggedright\arraybackslash}p{0.1276\linewidth}>{\raggedright\arraybackslash}p{0.0813\linewidth}>{\raggedright\arraybackslash}p{0.0813\linewidth}>{\raggedright\arraybackslash}p{0.0813\linewidth}>{\raggedright\arraybackslash}p{0.0813\linewidth}>{\raggedright\arraybackslash}p{0.0813\linewidth}>{\raggedright\arraybackslash}p{0.0813\linewidth}>{\raggedright\arraybackslash}p{0.0813\linewidth}>{\raggedright\arraybackslash}p{0.0813\linewidth}@{}}
\toprule
Model & S\textsubscript{1} (eV) & T\textsubscript{1} (eV) & Gap (eV) & Electric (e a\textsubscript{0}) & Magnetic (\ensuremath{\mu}B) & f & Native \textbar{}g\textbar{} & log\textsubscript{10} \textbar{}g\textbar{} (dex) \\
\midrule

Fixed PPP & 0.849 & 0.911 & 0.555 & 0.745 & 0.599 & 0.780 & 0.124 & 0.481 \\
Joint optical & 0.957 & 0.424 & 0.193 & \textbf{0.884} & \textbf{0.791} & \textbf{0.917} & 0.167 & \textbf{0.540} \\
Joint optical--spin & 0.958 & 0.963 & \textbf{0.842} & 0.877 & 0.787 & 0.911 & \textbf{0.289} & 0.529 \\
S\textsubscript{1} specialist & \textbf{0.976} & 0.908 & 0.516 & 0.830 & 0.706 & 0.872 & 0.127 & 0.523 \\
T\textsubscript{1} specialist & 0.919 & \textbf{0.989} & 0.583 & 0.822 & 0.677 & 0.846 & 0.058 & 0.504 \\
Oscillator specialist & 0.916 & 0.789 & 0.560 & 0.882 & 0.709 & 0.911 & 0.083 & 0.508 \\
Electric + direction specialist & 0.913 & 0.888 & 0.607 & 0.883 & 0.758 & 0.903 & 0.143 & 0.504 \\
Magnetic + direction specialist & 0.836 & 0.771 & 0.377 & 0.818 & 0.768 & 0.837 & 0.155 & 0.515 \\
\bottomrule
\end{tabular}
\end{table}
\endgroup

\FloatBarrier
{The full comparison shows smaller, property-dependent differences in optical accuracy between the two joint objectives. Optical-only training gives lower electric-magnitude and oscillator-strength MAEs in both populations, and lower magnetic-magnitude MAE in the favorable population. Both broad models give a magnetic-magnitude MAE of 0.236 \(\mu\)B at the displayed precision. Logarithmic dissymmetry errors are also lower for optical-only training: 0.645 versus 0.684 dex in the favorable population and 0.770 versus 0.773 dex in the broad population (Tables~\ref{tab:S8}--\ref{tab:S11}).}

\FloatBarrier
\section{Prediction Errors Across Excitation Character}\label{sec:si-screening-errors}

\begin{figure}[!htbp]
\centering
\includegraphics[width=.85\linewidth,keepaspectratio]{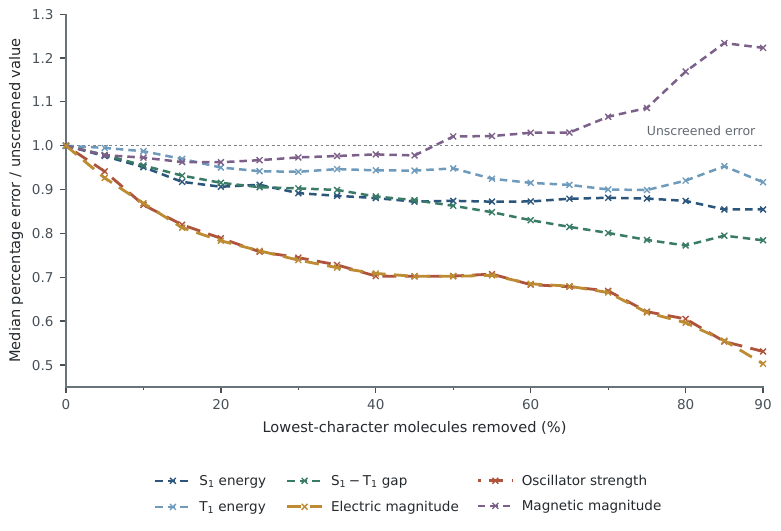}
\caption[Errors across excitation character]{Errors under reference-character screening. Each curve is the retained subset\textquotesingle s median of \(100|y_{\mathrm{pred}}-y_{\mathrm{ref}}|/|y_{\mathrm{ref}}|\), divided by its unscreened value for the same property. Unity marks the unscreened error; 0.8 denotes a 20\% reduction. Crosses mark 5\% steps; dashed lines guide the eye.}
\label{fig:S17}
\end{figure}

The 3,432 common broad-test molecules are ranked by reference S\textsubscript{1} local-\ensuremath{\pi} character, and the lowest-ranked are removed in 5\% steps using the saved tie ordering. Figure~\ref{fig:S17} compares each retained subset\textquotesingle s median percentage error with its unscreened value. The predictions remain fixed throughout.

Higher-character subsets show lower percentage errors for excitation energies, the signed gap, electric magnitude and oscillator strength. Magnetic-magnitude error initially decreases slightly, then rises. Because percentage errors depend on reference magnitude, these associations do not establish lower native-unit errors or performance beyond the reference-filtered population.

\FloatBarrier
\section{Learned Corrections by Baseline Group}\label{sec:si-baseline-group-corrections}

\begin{figure}[!htbp]
\centering
\includegraphics[width=\linewidth,height=.76\textheight,keepaspectratio]{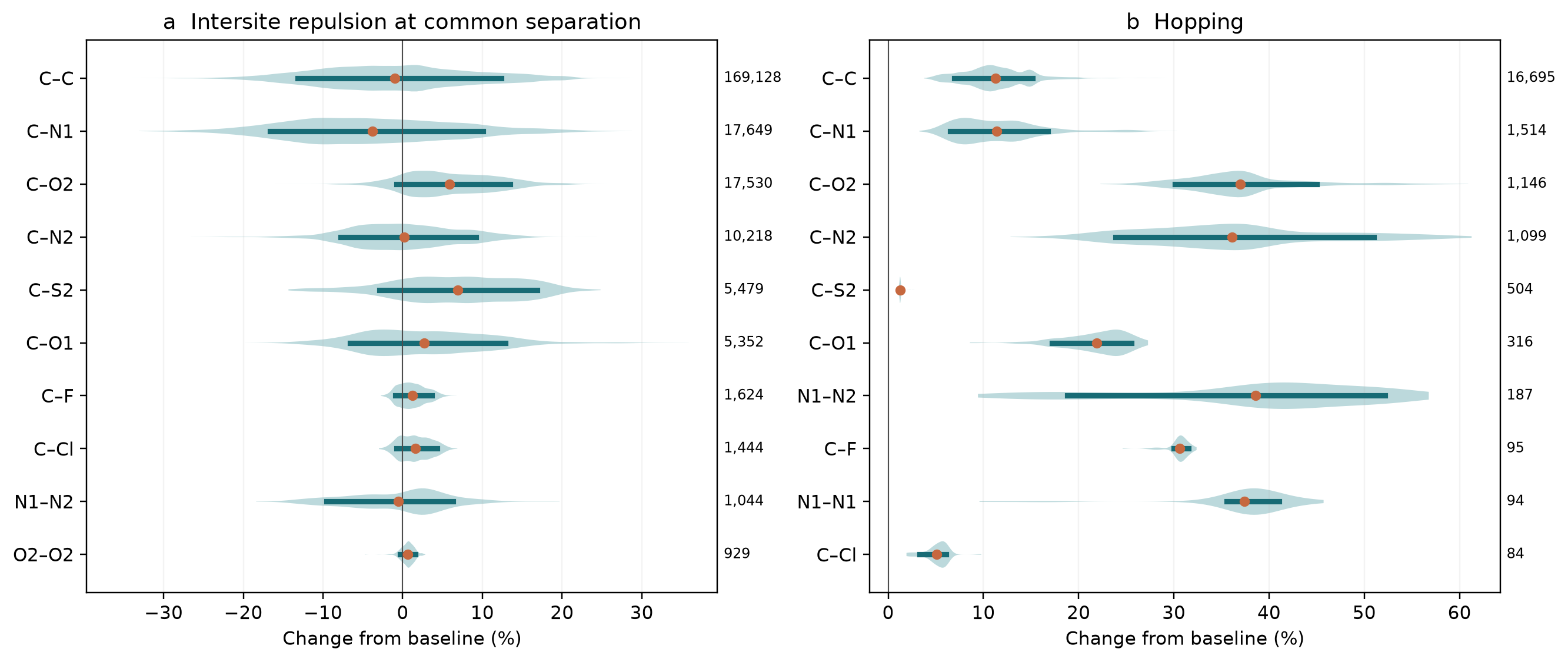}
\caption[Pair corrections by endpoint group]{Pair corrections by endpoint group. Corrections from the epoch-5 favorable COM joint optical--spin model on 884 evaluation molecules. Intersite corrections use each group's modal-distance reference (0.05 Å bins, midpoint, lowest bin for ties), holding learned outputs fixed; all pairs are included. Hopping uses adjacent pairings. Both fractional corrections have \ensuremath{\pm}100\% bounds. Shading is the distribution, orange points the mean, and bars the 10th--90th percentiles; counts are pair observations. Each endpoint group is evaluated at its own reference separation, so differences in distribution width reflect both endpoint type and the chosen separation.}
\label{fig:S22}
\end{figure}

To see whether environment-dependent corrections differ among sites and pairs with the same baseline parameters, we group the favorable joint optical--spin results by site type and pair endpoints. Table~\ref{tab:S12} and Fig.~\ref{fig:S22} show the distributions within each group. For the selected epoch-5 COM model, C--C hopping magnitude increases by 11.30\% on average, with a 10th--90th percentile range of 6.98--15.17\%. Carbon on-site repulsion decreases by 0.88 eV on average, while the explicit carbon site-energy offset increases by 0.47 eV. These pooled distributions combine differences between molecules with differences within each molecule; main Fig.~\ref{fig:12} separates those two contributions.

\begingroup
\scriptsize
\setlength{\tabcolsep}{3pt}
\renewcommand{\arraystretch}{1.18}
\begin{table}[!htbp]
\centering
\scriptsize
\caption{Single-site corrections by baseline site type. Changes in on-site repulsion and explicit site energy across 884 favorable molecules evaluated with the epoch-5 COM joint optical--spin model. Each site contributes one observation. Entries give the mean and 10th--90th percentiles. Site-type numerals denote \ensuremath{\pi}-electron contributions. The explicit site offset is bounded at \ensuremath{\pm}5 eV; the on-site fractional correction is bounded at \ensuremath{\pm}100\%.}\label{tab:S12}
\begin{tabular}{@{}>{\raggedright\arraybackslash}p{0.1123\linewidth}>{\raggedright\arraybackslash}p{0.1123\linewidth}>{\raggedright\arraybackslash}p{0.1123\linewidth}>{\raggedright\arraybackslash}p{0.1123\linewidth}>{\raggedright\arraybackslash}p{0.1264\linewidth}>{\raggedright\arraybackslash}p{0.1123\linewidth}>{\raggedright\arraybackslash}p{0.1264\linewidth}@{}}
\toprule
Site type & Sites & Molecules & \ensuremath{\Delta}U, mean (eV) & \ensuremath{\Delta}U, 10th--90th (eV) & \ensuremath{\delta}\ensuremath{\varepsilon}, mean (eV) & \ensuremath{\delta}\ensuremath{\varepsilon}, 10th--90th (eV) \\
\midrule

C & 16,281 & 884 & -0.88 & -2.34 to +0.31 & +0.47 & +0.27 to +0.71 \\
O2 & 1,002 & 482 & +2.79 & +2.26 to +3.43 & +0.49 & +0.42 to +0.56 \\
N1 & 995 & 492 & -1.39 & -2.52 to -0.14 & +0.31 & +0.12 to +0.53 \\
N2 & 582 & 412 & +1.14 & +0.56 to +1.73 & +1.00 & +0.61 to +1.32 \\
O1 & 316 & 230 & -0.98 & -1.19 to -0.70 & -0.75 & -0.84 to -0.64 \\
S2 & 267 & 149 & +2.66 & +2.26 to +3.22 & +0.18 & +0.18 to +0.18 \\
F & 95 & 47 & +0.54 & +0.40 to +0.78 & +0.27 & +0.26 to +0.30 \\
Cl & 84 & 55 & -2.02 & -2.16 to -1.93 & -0.11 & -0.16 to -0.06 \\
\bottomrule
\end{tabular}
\end{table}
\endgroup

\FloatBarrier
\clearpage
\bibliography{references}